\documentclass[pdflatex,sn-mathphys-ay]{sn-jnl}

\usepackage{graphicx}%
\usepackage{multirow}%
\usepackage{amsmath,amssymb,amsfonts}%
\usepackage{amsthm}%
\usepackage{mathrsfs}%
\usepackage[title]{appendix}%
\usepackage{xcolor}%
\usepackage{textcomp}%
\usepackage{manyfoot}%
\usepackage{booktabs}%
\usepackage{algorithm}%
\usepackage{algorithmicx}%
\usepackage{algpseudocode}%
\usepackage{listings}%

\usepackage{subfigure}
\usepackage{bm}
\usepackage{natbib}
\DeclareMathOperator{\argmax}{argmax}

\theoremstyle{thmstyleone}%
\theoremstyle{thmstyletwo}%

\theoremstyle{thmstylethree}%

\begin{document}

\title[Article Title]{Kriging for large datasets via penalized neighbor selection}


\author*[1]{\fnm{Francisco} \sur{Cuevas-Pacheco}}\email{francisco.cuevas@usm.cl}

\author[2]{\fnm{Jonathan} \sur{Acosta}}\email{jonathan.acosta@uc.cl}
\equalcont{These authors contributed equally to this work.}

\affil*[1]{\orgdiv{Departamento de Matemática}, \orgname{Universidad Técnica Federico Santa María}, \orgaddress{\street{Avenida Espa\~na 1680}, \city{Valparaíso}, \postcode{2340000}, \state{Valparaíso}, \country{Chile}}}

\affil[2]{\orgdiv{Departamento de Estadística}, \orgname{Pontificia Universidad Católica de Chile}, \orgaddress{\street{Avenida Vicuña Mackenna 4860}, \city{Macul}, \postcode{7820436}, \state{Santiago de Chile}, \country{Chile}}}


\abstract{Kriging is a fundamental tool for spatial prediction, but its computational complexity of $O(N^3)$ becomes prohibitive for large datasets. While local kriging using $K$-nearest neighbors addresses this issue, the selection of $K$ typically relies on ad-hoc criteria that fail to account for spatial correlation structure. We propose a penalized kriging framework that incorporates LASSO-type penalties directly into the kriging equations to achieve automatic, data-driven neighbor selection.  We further extend this to adaptive LASSO, using data-driven penalty weights that account for the spatial correlation structure. Our method determines which observations contribute non-zero weights through $\ell_1$ regularization, with the penalty parameter selected via a novel criterion based on effective sample size that balances prediction accuracy against information redundancy. Numerical experiments demonstrate that penalized kriging automatically adapts neighborhood structure to the underlying spatial correlation, selecting fewer neighbors for smoother processes and more for highly variable fields, while maintaining prediction accuracy comparable to global kriging at substantially reduced computational cost.}





\maketitle

\section{Introduction}

Spatial prediction through kriging has become a fundamental tool in geostatistics for interpolating and predicting spatial phenomena across diverse scientific fields \citep{cressie1993statistics}. However, the standard kriging equations require the inversion of an $N \times N$ covariance matrix, resulting in $O(N^3)$ computational complexity \citep{cressie2008fixed}, which becomes prohibitive as dataset sizes grow into the thousands or millions of observations. Indeed, modern applications routinely generate massive spatial datasets: satellite remote sensing platforms such as MODIS and Landsat produce multi-spectral imagery at millions of locations, climate reanalysis products like ERA5 provide gridded fields at hundreds of thousands of spatial points, and LiDAR surveys generate point clouds with millions of elevation measurements. Oceanographic monitoring systems exemplify this scale: the COBE Sea Surface Temperature dataset alone contains over 43,000 spatial locations at 1° × 1° resolution globally \citep{ishii2005objective}, while modern satellite altimetry and Argo float networks produce even larger datasets.

To address this computational challenge, several strategies have been developed within the geostatistical community. Among the most widely adopted approaches is the use of local neighborhoods, where kriging predictions are computed using only a subset of $K$ nearest neighbors rather than the full dataset \citep{gramacy2015local}. Alternative strategies include covariance tapering to induce sparsity in the covariance matrix \citep{furrer2006covariance} and low-rank approximations \citep{cressie2008fixed}. A comprehensive comparison of these methods by \cite{heaton2019comparison} demonstrates that while neighborhood-based approaches offer computational efficiency, their performance depends critically on the choice of neighborhood size and configuration.

Despite the widespread use of $K$-nearest neighbor kriging, the selection of neighborhood size $K$ remains problematic. The choice typically relies on user-specified criteria or cross-validation procedures, both of which are computationally expensive for large datasets. \cite{emery2009kriging} demonstrated that fixed neighborhood selection strategies may not be optimal in the sense of minimizing kriging variance among all possible selections with the same number of observations. The $K$-nearest neighbor approach fails to account for spatial correlation structure and information redundancy: in regions with strong spatial correlation, a smaller neighborhood may suffice, while areas with high spatial variability may require more neighbors to achieve comparable prediction accuracy.

Penalized regression methods, particularly the Least Absolute Shrinkage and Selection Operator (LASSO) introduced by \cite{tibshirani1996regression}, provide a principled framework for automatic variable selection through $\ell_1$ regularization. 

The LASSO penalty induces sparsity in regression coefficients, effectively performing variable selection while simultaneously estimating model parameters. Several authors have extended this approach to the kriging context: \cite{hung2011penalized} introduced penalized blind kriging for automatic mean function selection in universal kriging with oracle properties, while \cite{park2021lasso} applied LASSO regularization to kriging for global trend model selection, demonstrating improved predictive accuracy and computational efficiency. However, these methods focus primarily on mean function specification rather than neighbor selection for computational efficiency.

Building on these developments, we propose a penalized kriging framework that incorporates LASSO-type penalties directly into the kriging equations to achieve sparse neighbor selection. Unlike traditional $K$-nearest neighbor approaches where the number and identity of neighbors are fixed a priori, our method automatically determines which observations contribute non-zero weights to the kriging predictor at each prediction location. The sparsity level is controlled by a tuning parameter, $\eta$, which we select through a criterion based on the effective sample size concept that explicitly accounts for spatial correlation structure. This criterion balances prediction accuracy against information redundancy in the spatial data, providing an alternative to computationally expensive cross-validation.

Our main contributions are threefold. First, we reformulate the kriging problem to incorporate $\ell_1$ penalties that induce sparsity in the kriging weights while maintaining the unbiasedness constraints inherent to kriging. Second, we develop an effective sample size-based criterion for automatic selection of the penalty parameter that explicitly accounts for spatial correlation and information redundancy, eliminating the need for computationally expensive cross-validation. Third, through numerical experiments with both simulated and real data, we demonstrate that our penalized kriging approach automatically adapts the neighborhood structure to the underlying spatial correlation: selecting fewer neighbors for smoother processes and more neighbors in regions of high spatial variability, while maintaining prediction accuracy comparable to global kriging at substantially reduced computational cost.

The remainder of this paper is organized as follows. Section \ref{section:kriging_equations} reviews the kriging equations and establishes the connection between nugget effects and Ridge penalization, motivating our penalized approach. Section \ref{section::penalized_kriging} presents our LASSO-penalized kriging formulation, introduces the adaptive LASSO extension, and analyzes the solution path as a function of the penalty parameter. Section \ref{section:tuning_parameter} develops the effective sample size criterion for selecting the tuning parameter. Section \ref{section:simulations} presents numerical experiments examining how neighbor selection adapts to different covariance structures and spatial configurations. Section \ref{section:datanalysis} demonstrates the methodology on two real datasets spanning different spatial scales: the Jura heavy metals dataset and the large-scale COBE Sea Surface Temperature dataset. Finally, Section \ref{section:conclusions} concludes with a discussion of implications and directions for future research.

\section{Kriging equations}\label{section:kriging_equations}

We begin by establishing the notation and reviewing the kriging framework that forms the basis for our penalized approach. This section also demonstrates how penalization naturally arises in kriging through the nugget effect, providing theoretical justification for our LASSO extension.

Let $\{ \bm{s}_{i} \}_{i=1}^{N}  \subset \mathbb{R}^{d}$ be a set of $N$ spatial locations, and let $\bm{Z} = (Z(\bm{s}_{1}), \ldots, Z(\bm{s}_{N}))^{\top}$ be the vector of observed values, in the respective locations, of the Gaussian random field, $Z$, with mean function $\mu(\bm{s}) = \mathbb{E}[Z(\bm{s})]$ and covariance function $C(\bm{s}, \bm{s}^{\prime}) = {\rm{cov}}(Z(\bm{s}), Z(\bm{s}^{\prime}))$. We are interested in predicting $Z$ at an unobserved location $\bm{s}_{0}$ using the information in the vector of observed values $\bm{Z}$. We introduce the notation $\Sigma = {\rm{Var}}[\bm{Z}]$, $\bm{c}_{0} = {\rm{cov}}(Z(\bm{s}_{0}), \bm{Z})=(C(\bm{s}_0,\bm{s}_1),\dots,C(\bm{s}_0,\bm{s}_N))^{\top}$, and $\sigma^2_0={\rm{Var}}[Z(\bm{s}_0)]$. 

Spatial prediction is done by proposing linear combinations of the observed variable at different spatial locations. This procedure is known as kriging. Formally, we consider the following predictor
\begin{equation*}
    P(\bm{Z},\bm{s}_{0}) = \lambda_0+\sum_{i = 1}^{N} \lambda_{i} Z(\bm{s}_{i}) = \lambda_0+\bm{\lambda}^{\top}\bm{Z},
\end{equation*}
where $\lambda_{0}$ is a real number used as an intercept and $\bm{\lambda} = (\lambda_{1}, \ldots, \lambda_{N})^{\top}$ is a vector of real numbers used as weights. The predictor $P(\bm{Z},\bm{s}_{0})$ seeks to optimally combine observed values to predict the unobserved value at $\bm{s}_0$, with weights $\bm{\lambda}$ chosen to minimize prediction error while ensuring unbiasedness.

Commonly, the vector of coefficients $\bm{\lambda}$ is obtained by minimizing the mean squared error (MSE) function:
\begin{equation}\label{eq:simple_kriging}
    {\rm{MSE}}(P(\bm{Z},\bm{s}_{0})):=\mathbb{E}\left[ \left(Z(\bm{s}_0)-P(\bm{Z},\bm{s}_0)\right)^{2} \right]
    = \bm{\lambda}^{\top} \Sigma \bm{\lambda} - 2 \bm{\lambda}^{\top}\bm{c}_{0} + \sigma^{2}_0,
\end{equation}
however, the ${\rm{MSE}}(P(\bm{Z},\bm{s}_{0}))$ does not depend on $\lambda_0$. Therefore, it is necessary to impose the so-called unbiasedness condition, that is, $\mathbb{E}[P(\bm{Z},\bm{s}_0)]=\mathbb{E}[Z(\bm{s}_0)]$. Thus, the weights for the kriging predictor are obtained via solving the restricted minimization problem
\begin{align} 
    \min_{\bm{\lambda} \in \mathbb{R}^{N}} ~&~\bm{\lambda}^{\top} \Sigma \bm{\lambda} - 2 \bm{\lambda}^{\top}\bm{c}_{0} + \sigma^{2}_0 \label{eq:original_problem} \\ \nonumber
    \text{s.t.}~&~\mathbb{E}[P(\bm{Z},\bm{s}_0)]=\mathbb{E}[Z(\bm{s}_0)]
\end{align}
According to the imposed assumptions on the mean function, different versions of kriging are obtained: simple, ordinary, and universal \citep{schabenberger2005}.

For the assumption that the mean function is known, it is straightforward to see that the solution of \eqref{eq:original_problem} is $\widehat{\bm{\lambda}} = \Sigma^{-1}\bm{c}_{0}$, and  $\widehat{\lambda}_{0} = \mu(\bm{s}_0)-\widehat{\bm{\lambda}}^{\top}\bm{\mu}$, where $\bm{\mu}=(\mu(\bm{s}_1),\dots,\mu(\bm{s}_N))^{\top}$. Under this assumption, we obtain the simple kriging estimator, 
\begin{equation*}
  \widehat{P}_{\text{sk}}(\bm{Z},\bm{s}_{0}) = \mu(\bm{s}_{0}) + \bm{c}_{0}^{\top}\Sigma^{-1}(\bm{Z}-\bm{\mu}).
\end{equation*}
This predictor adjusts the known mean $\mu(\bm{s}_0)$ by a weighted combination of deviations from the mean at observed locations, with weights determined by the spatial covariance structure. Of course, the assumption of known mean is not realistic in practice. The classical way to deal with this assumption is via statistical modelling and estimation of the mean. Therefore, the proposed mean structure leads to a new set of restrictions that must be included in the minimization problem \citep{gaetan2010spatial}. For example, the predictor is called ordinary kriging when we assume that the mean is constant and unknown, while universal kriging assumes that the mean is a linear combination of known covariates and unknown coefficients. 

For the assumption that the mean function follows a linear structure, 
$\mu(\bm{s})=\bm{x}^{\top}(\bm{s})\bm{\beta}$, where 
$\bm{x}(\bm{s})=(x_1(\bm{s}),\dots,x_p(\bm{s}))^{\top}\in\mathbb{R}^p$ 
contains $p$ covariate functions and 
$\bm{\beta}=(\beta_1,\dots,\beta_p)^{\top}\in\mathbb{R}^p$ contains the 
corresponding coefficients. The unbiasedness constraint becomes 
$\mathbb{E}[Z(\bm{s}_0)]=\bm{x}_0^{\top}\bm{\beta}$ and 
$\mathbb{E}[P(\bm{Z},\bm{s}_0)]=\bm{\lambda}^{\top}\bm{X}\bm{\beta}$, where 
$\bm{x}_0=\bm{x}(\bm{s}_0)$, and $\bm{X}$ is the $N\times p$ design matrix 
with element $X_{ij}=x_j(\bm{s}_i)$ for $i=1,\ldots,N$ and $j=1,\ldots,p$. 
Equivalently, the $i$-th row of $\bm{X}$ is $\bm{x}^{\top}(\bm{s}_i)$. Typically, $x_1(\bm{s})=1$ for all $\bm{s}$, and so, the condition of unbiasedness becomes $\widehat{\lambda}_0=0$ and $\bm{X}^{\top}\bm{\lambda}=\bm{x}_0$, that is, the condition is equivalent to $p$ linear constraints on the parameters $\sum_{i=1}^{N}\lambda_ix_{j}(\bm{s}_i)=x_j(\bm{s}_0)$ for $j=1,\dots,p$. In this case, the solution of \eqref{eq:original_problem} is
\begin{equation*}
\widehat{\bm{\lambda}}=\Sigma^{-1}\left[\bm{c}_{0} +\bm{X}(\bm{X}^{\top}\Sigma^{-1}\bm{X})^{-1}(\bm{x}_0-\bm{X}^{\top}\Sigma^{-1}\bm{c}_{0})\right],
\end{equation*}
and so yields the universal kriging estimator, 
\begin{equation*}
  \widehat{P}_{\text{uk}}(\bm{Z},\bm{s}_{0}) = \bm{x}_0^{\top}\widehat{\bm{\beta}}+\bm{c}_0^{\top}\Sigma^{-1}(\bm{Z}-\bm{X}\widehat{\bm{\beta}}),\quad\text{where}\quad \widehat{\bm{\beta}}=(\bm{X}^{\top}\Sigma^{-1}\bm{X})^{-1}\bm{X}^{\top}\Sigma^{-1}\bm{Z}.
\end{equation*}

Note that, if $x_1(\bm{s})=1$ then the first restriction is $\sum_{i=1}^{N}\lambda_i=1$. Moreover, if this is the only covariate, that is, $p=1$ and $x_1(\bm{s})=1$, the predictor is known as ordinary kriging, and the solution for the optimal estimator is given by
\begin{equation*}
  \widehat{P}_{\text{ok}}(\bm{Z},\bm{s}_{0}) = \left[\bm{c}_0^{\top}+\dfrac{1-\bm{c}_0^{\top}\Sigma^{-1}\bm{1}_{N}}{\bm{1}_N^{\top}\Sigma^{-1}\bm{1}_N}\bm{1}_N^{\top}\right]\Sigma^{-1}\bm{Z}.
\end{equation*}

The obtained predictor also allows the computation of the prediction variance, which depends on the type of restrictions imposed. The simple kriging variance is
\begin{eqnarray*}
    \sigma^2_{\text{sk}}(\bm{s}_0)&=&{\rm{MSE}}(\widehat{P}_{\text{sk}}(\bm{Z},\bm{s}_{0}))\\
  &=&\sigma^2_0-\bm{c}_{0}^{\top}\Sigma^{-1}\bm{c}_{0},
\end{eqnarray*}
and the universal kriging variance is
\begin{eqnarray*}
    \sigma^2_{\text{uk}}(\bm{s}_0)&=&{\rm{MSE}}(\widehat{P}_{\text{uk}}(\bm{Z},\bm{s}_{0}))\\
  &=&\sigma^2_0-\bm{c}_{0}^{\top}\Sigma^{-1}\bm{c}_{0}+(\bm{x}_0-\bm{X}^{\top}\Sigma^{-1}\bm{c}_{0})^{\top} (\bm{X}^{\top}\Sigma^{-1}\bm{X})^{-1} (\bm{x}_0-\bm{X}^{\top}\Sigma^{-1}\bm{c}_{0}).
\end{eqnarray*}

\subsection{Connection to Penalized Regression}\label{subsec:ridge_connection}

A key insight motivating our penalized kriging approach is that penalization already appears implicitly in standard kriging when the covariance exhibits a nugget effect. This connection has not been widely recognized in the geostatistical literature but provides strong theoretical justification for our LASSO extension.

Consider kriging with nugget effect, where ${\rm{Var}}[\bm{Z}] = \Sigma_{\tau}$ with $\Sigma_{\tau} = \Sigma + \tau^{2}\bm{I}$, and ${\rm{Var}}[Z(\bm{s}_0)]=\sigma^2_0+\tau^2$. Remarkably, the kriging variance remains unchanged compared to the nugget-free case, but the objective function of Problem \eqref{eq:original_problem} can be written as
\begin{align*}
     {\rm{MSE}}(P(\bm{Z}, \bm{s}_{0})) &= \bm{\lambda}^{\top} \Sigma_{\tau} \bm{\lambda} - 2 \bm{\lambda}^{\top}\bm{c}_{0} + \sigma^{2}_0 + \tau^{2} \\
    &= \bm{\lambda}^{\top} (\Sigma + \tau^{2}\bm{I}) \bm{\lambda} - 2\bm{\lambda}^{\top}\bm{c}_{0} + \sigma^{2}_0 +\tau^{2}  \\
    &= (\bm{\lambda}^{\top} \Sigma\bm{\lambda} - 2\bm{\lambda}^{\top}\bm{c}_{0} + \sigma^{2}_0) +\tau^{2}(1 + \|\bm{\lambda}\|_2^{2}),
\end{align*}
where $\|\cdot\|_2$ is the Euclidean norm. Therefore, the kriging problem with nugget effect is equivalent to solving the following problem

\begin{align}
    \min_{\bm{\lambda} \in \mathbb{R}^{N}} ~&~\bm{\lambda}^{\top} \Sigma \bm{\lambda} - 2 \bm{\lambda}^{\top}\bm{c}_{0} + \sigma^{2}_0 + \tau^{2}(1+\| \bm{\lambda} \|_2^{2})\label{eq:problema_ridge} \\ \nonumber
    \text{s.t.}~&~\mathbb{E}[Z(\bm{s}_0)]=\mathbb{E}[P(\bm{Z},\bm{s}_0)]
\end{align}
Problem \eqref{eq:problema_ridge} is a penalized version of problem \eqref{eq:original_problem} with a Ridge ($\ell_2$) penalty where the penalty parameter $\tau^2$ is obtained from the data via maximum likelihood or weighted least squares. This reveals that Ridge penalization is already implicit in standard geostatistical practice whenever a nugget effect is present.

\subsection{Local Kriging}

The computation of $\widehat{\bm{\lambda}}$ in \eqref{eq:original_problem} requires $O(N^3)$ operations to invert the $N \times N$ covariance matrix $\Sigma$, which becomes computationally prohibitive for large $N$ and may lead to numerical instabilities. Local kriging addresses this issue by using only a subset of $K$ observations from the available data, applying the same optimization problem \eqref{eq:original_problem} to the subset $\bm{Z}_K$. While various selection criteria exist for choosing these $K$ observations, the most common approach is to select the $K$ nearest neighbors to the prediction location $\bm{s}_0$. Without loss of generality, we order the observation locations by increasing Euclidean distance to the prediction location $\bm{s}_0$, so that the $K$ nearest neighbors correspond to the first $K$ ordered locations. This leads to the linear predictor
\begin{equation*}
    P(\bm{Z}, \bm{s}_{0}, K) =\lambda_0+\sum_{i = 1}^{K} \lambda_{i} Z(\bm{s}_{(i)}) = \lambda_0 + \bm{\lambda}^{\top} \bm{Z}_{K},
\end{equation*}
where $\bm{s}_{(i)}$ denotes the $i$-th location in the selected subset. By defining $\bm{Z}_{K} = (Z(\bm{s}_{(1)}), \ldots, Z(\bm{s}_{(K)}))^{\top}$, ${\rm{Var}}[\bm{Z}_{K}] = \Sigma_{K}$, and $\bm{c}_{0K} = {\rm{Cov}}( \bm{Z}_{K}, Z(\bm{s}_{0}))$, the local kriging weights are obtained by solving the problem 
\begin{align} 
    \min_{\bm{\lambda} \in \mathbb{R}^{K}} ~&~\bm{\lambda}^{\top} \Sigma_{K} \bm{\lambda} - 2 \bm{\lambda}^{\top}\bm{c}_{0K} + \sigma^{2}_0 \label{eq:original_local_kriging} \\ \nonumber
    \text{s.t.}~&~\mathbb{E}[Z(\bm{s}_0)]=\mathbb{E}[P(\bm{Z},\bm{s}_0, K)]
\end{align}

Local kriging produces a sparse predictor by using only $K$ of the $N$ available observations. The key limitation is that $K$ must be chosen a priori without accounting for the spatial correlation structure or information redundancy in the data. This motivates the inclusion of a LASSO-type penalty to obtain data-driven sparse solutions where the sparsity level is automatically determined by the penalty parameter rather than fixed by the user.

\section{Penalized Kriging equations} \label{section::penalized_kriging}

We now present our main methodological contribution: a LASSO-penalized kriging framework that automatically selects neighbors through sparse regularization. Recall from Section \ref{section:kriging_equations} that we ordered the observation locations by increasing Euclidean distance to the prediction location $\bm{s}_0$. This ordering ensures that the first $p$ locations are the nearest neighbors, which will be preserved in the solution due to the structure of the penalty we introduce below.

The motivation for sparse solutions becomes clear when we recognize that for a given 
number of neighbors $K$, solving the local kriging problem \eqref{eq:original_local_kriging} 
is equivalent to 
\begin{align} 
    \min_{\bm{\lambda} \in \mathbb{R}^{N}} ~&~\bm{\lambda}^{\top} \Sigma \bm{\lambda} - 2 \bm{\lambda}^{\top}\bm{c}_{0} + \sigma^{2}_0 \label{eq:local_kriging_equivalency} \\ \nonumber
    \text{s.t.}~&~\mathbb{E}[Z(\bm{s}_0)]=\mathbb{E}[P(\bm{Z},\bm{s}_0)] \\ 
        ~&~\bm{\lambda}_{-K} = \bm{0}_{N-K} \nonumber
\end{align}
where $\bm{0}_{m}$ is the $m$-dimensional zero vector and $\bm{\lambda}_{-K} = [\lambda_{K+1}, \ldots, \lambda_{N}]^{\top}$ is the vector of the last $(N-K)$ coefficients. The restriction $\bm{\lambda}_{-K} = \bm{0}_{N-K}$ implies that the solution of \eqref{eq:local_kriging_equivalency} has the form $\widehat{\bm{\lambda}}_{N} = (\widehat{\bm{\lambda}}_{K},\bm{0}_{N-K})$, where $\widehat{\bm{\lambda}}_{K}$ is the solution of \eqref{eq:original_local_kriging}.

Following \citet{emery2009kriging}, the $K$ nearest neighbors are not necessarily optimal for minimizing kriging variance among all possible selections with the same number of observations. To achieve sparsity while preserving the statistical properties of the kriging predictor, we incorporate a LASSO penalty into the kriging equations. A natural first approach would be to add an $\ell_1$ penalty to all weights:

\begin{align} 
    \min_{\bm{\lambda} \in \mathbb{R}^{N}} &\bm{\lambda}^{\top} \Sigma \bm{\lambda} - 2 \bm{\lambda}^{\top}\bm{c}_{0} + \sigma^{2}_0 + \eta \|\bm{\lambda}\|_1 \label{eq:proposed_lasso_problem_original}\\
    \text{s.t.} & \mathbb{E}[Z(\bm{s}_0)]=\mathbb{E}[P(\bm{Z},\bm{s}_0)] \notag
\end{align}

where $\eta$ is a non-negative constant, and $\|\cdot\|_1$ is the $\ell_1$-norm. Problem \eqref{eq:proposed_lasso_problem_original} is equivalent to a constrained LASSO \citep{gaines2018algorithms}. However, this formulation has a critical limitation. The existence and uniqueness of solutions depend strongly on the interplay between the penalty parameter $\eta$ and the feasible region defined by the unbiasedness constraint. For certain values of $\eta$, the constrained optimization problem may have no solution or multiple solutions, particularly when the constraint set intersects the boundary of the $\ell_1$ ball at non-differentiable points. This motivates a modification of Problem \eqref{eq:proposed_lasso_problem_original}.

To ensure existence and uniqueness of solutions for all $\eta \geq 0$, we modify the penalty structure. The key insight is that the unbiasedness constraints allow us to express $p$ of the kriging weights as functions of the remaining weights. Recall from Section \ref{section:kriging_equations} that under a linear mean structure 
$\mu(\bm{s}) = \bm{x}^{\top}(\bm{s})\bm{\beta}$ with $p$ covariates, the unbiasedness constraint becomes $\bm{X}^{\top}\bm{\lambda}=\bm{x}_0$. We partition the design matrix as $\bm{X}=[\bm{X}_p^{\top}, \bm{X}_{-p}^{\top}]^{\top}$, where $\bm{X}_p\in\mathbb{R}^{p\times p}$ contains the first $p$ rows 
(corresponding to the $p$ nearest neighbors) and is assumed invertible, and 
$\bm{X}_{-p}\in\mathbb{R}^{(N-p)\times p}$ contains the remaining rows. We can isolate $p$ components of $\bm{\lambda}$ via the relationship

$$\bm{\lambda}_{p} = \bm{X}_{p}^{-1}(\bm{x}_{0} - \bm{X}_{-p}\bm{\lambda}_{-p}),$$

\noindent where $\bm{\lambda}_{p} = [\lambda_{1}, \ldots, \lambda_{p}]^{\top}$ and $\bm{\lambda}_{-p} = [\lambda_{p+1}, \ldots, \lambda_{N}]^{\top}$ form a partition of $\bm{\lambda}$. By penalizing only $\bm{\lambda}_{-p}$, we ensure that the first $p$ neighbors (the nearest neighbors under our ordering assumption from Section \ref{section:kriging_equations}) always receive non-zero weights, while more distant neighbors may be shrunk to zero. This leads to our proposed problem:

\begin{align} \label{eq:proposed_lasso_problem_mod}
    \min_{\bm{\lambda} \in \mathbb{R}^{N}} &\bm{\lambda}^{\top} \Sigma \bm{\lambda} - 2 \bm{\lambda}^{\top}\bm{c}_{0} + \sigma^{2}_0 + \eta \|\bm{\lambda}_{-p}\|_{1} \\
    \text{s.t.} & \quad \bm{X} \bm{\lambda} = \bm{x}_{0} \notag
\end{align}

for a fixed positive $\eta$. Observe that we only need to solve for $\bm{\lambda}_{-p}$, reducing the dimensionality of the problem and ensuring the existence of the solution for all $\eta\geq 0$. Let $\widehat{\bm{\lambda}}(\eta)$ be the solution of \eqref{eq:proposed_lasso_problem_mod} for fixed $\eta$, then the sparse kriging predictor is given by
\begin{equation*}
    \widehat{P}_{\eta}(\bm{Z}, \bm{s}_{0}) = \widehat{\bm{\lambda}}(\eta)^{\top} \bm{Z}.
\end{equation*}

To understand the behavior of the solution with respect to $\eta$, observe that $\widehat{\bm{\lambda}}(0)$ coincides with the standard kriging predictor from \eqref{eq:original_problem}, while for sufficiently large $\eta$ there exists a value $M$ such that the solution vector becomes fully sparse: $\widehat{\bm{\lambda}}(\eta) = [\bm{X}_{p}^{-1}\bm{x}_{0}, \bm{0}_{N-p}]^{\top}$ for all $\eta>M$. This means that $\widehat{\bm{\lambda}}(\eta)$ provides a continuous connection between the kriging predictor and a predictor using only the $p$-nearest neighbors. The parameter $\eta$ controls the sparsity-accuracy trade-off: small $\eta$ yields solutions close to standard kriging (using many neighbors), while large $\eta$ produces highly sparse solutions (using few neighbors). The solution path $\widehat{\bm{\lambda}}(\eta)$ is piecewise linear in $\eta$, with breakpoints occurring when neighbors enter or leave the active set.

For the ordinary kriging case, we have $\bm{X}_{p} = 1$ and $\bm{x}_{0} = 1$, and so $\bm{X}_{p}^{-1}\bm{x}_{0}=1$. The restriction in \eqref{eq:proposed_lasso_problem_mod} can be written by isolating the first weight as $\lambda_1=1-\sum_{i=2}^{N}\lambda_i$. 
Consequently, $\widehat{\bm{\lambda}}(M) = [1, \bm{0}_{N - 1}]^{\top}$, meaning that $\widehat{P}_{M}(\bm{Z}, \bm{s}_{0})$ is simply the nearest neighbor predictor.

\begin{figure}
    \subfigure[LASSO surface problem]{\includegraphics[width=0.45\linewidth]{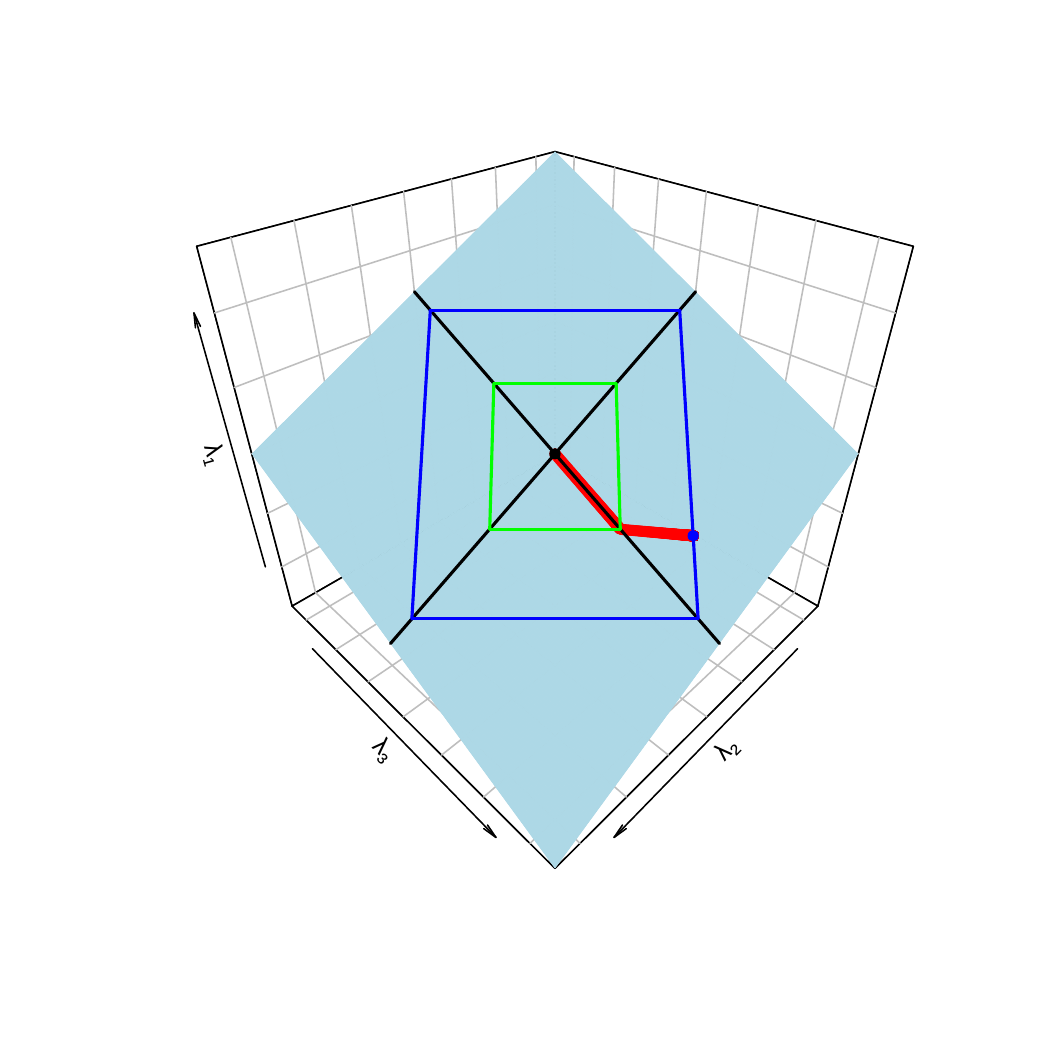}\label{fig:lasso_diamond:a}}
    \subfigure[Trace plot]{\includegraphics[width=0.45\linewidth]{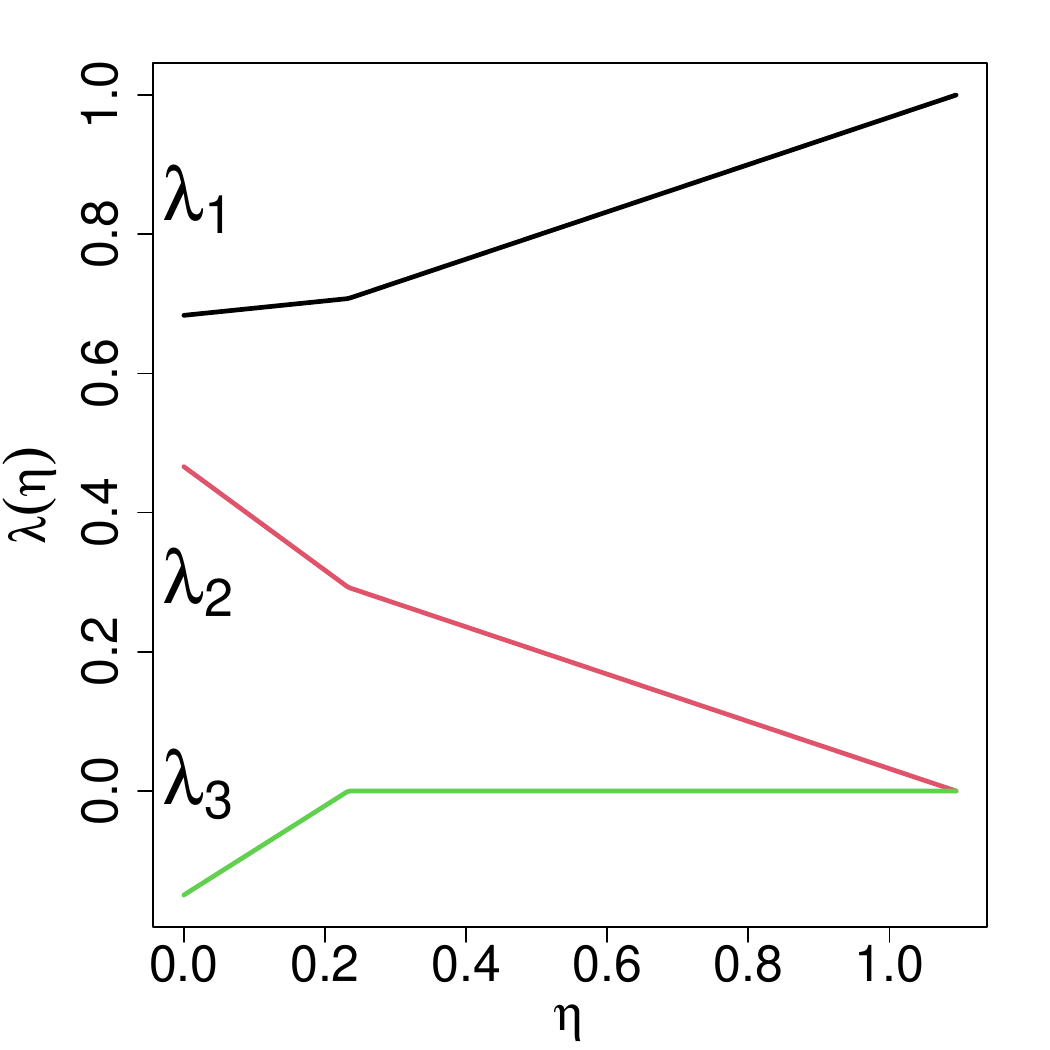}\label{fig:lasso_diamond:b}}
    \caption{Left: surface $\lambda_{1} = 1 - \lambda_{2} - \lambda_{3}$ obtained from Problem \ref{eq:proposed_lasso_problem_mod} under the assumption that the mean is unknown (ordinary kriging setup). The green and the blue curves are the intersection of the LASSO restrictions $|\bm{\lambda}_{-1}| = 0.291$ (OK with two locations) and $|\bm{\lambda}_{-1}| = 0.615$ (exact OK solution) with the restriction plane respectively. The red curve is the solution path $\bm{\lambda}(\eta)$, that connects the solution $\bm{\lambda}(M) = [1,0,0]^{\top}$ and the ordinary kriging solution $\widehat{\bm{\lambda}}(0) = [0.683, 0.466, -0.149]^{\top}$. Right: Trace plot created with the solution path $\widehat{\bm{\lambda}}(\eta)$, showing how the method assigns zero weight to the farthest location.}
    \label{fig:lasso_diamond}
\end{figure}

We illustrate this solution path behavior with a simple example. Suppose we want to predict at the unobserved location $\bm{s}_{0} = (0.4, 0)$ given observations at $\bm{s}_{1} = (0,0)$, $\bm{s}_{2} = (1,0)$, and $\bm{s}_{3} = (2,0)$. Assuming unknown constant mean (ordinary kriging) and exponential covariance $C(\| \bm{s}_{i} - \bm{s}_{j} \|) = \exp( -\| \bm{s}_{i} - \bm{s}_{j} \|/0.75)$, the solution must satisfy $\lambda_{1} + \lambda_2 + \lambda_{3} = 1$, equivalently $\lambda_{1} = 1 - \lambda_{2} - \lambda_{3}$. Figure \ref{fig:lasso_diamond:a} gives a geometrical interpretation of the problem, where we can see the restriction set, and the penalty term. Figure \ref{fig:lasso_diamond:b} depicts the solution path generated by $\bm{\lambda}(\eta)$.

An important feature of our framework is that it naturally accommodates nugget effects in the covariance structure. Indeed, by the same arguments used in Section \ref{subsec:ridge_connection}, Problem \eqref{eq:proposed_lasso_problem_mod} with nugget effect $\tau^2$ yields
\begin{align} \label{eq:proposed_lasso_problem_mod_en}
    \min_{\bm{\lambda} \in \mathbb{R}^{N}} &\bm{\lambda}^{\top} \Sigma \bm{\lambda} - 2 \bm{\lambda}^{\top}\bm{c}_{0} + \sigma^{2}_0 + \eta \|\bm{\lambda}_{-p}\|_1 + \tau^{2}(1+\| \bm{\lambda}_{p} \|^{2}_2 + \| \bm{\lambda}_{-p} \|^{2}_2) \\
    \text{s.t.} & \quad \bm{X} \bm{\lambda} = \bm{x}_{0}. \notag
\end{align}
Since $\bm{\lambda}_{p} = \bm{X}_{p}^{-1}(\bm{x}_{0} - \bm{X}_{-p}\bm{\lambda}_{-p})$, Problem \eqref{eq:proposed_lasso_problem_mod_en} corresponds to a weighted elastic-net problem \citep{zou2005regularization}. For both problems \eqref{eq:proposed_lasso_problem_mod} and \eqref{eq:proposed_lasso_problem_mod_en}, the selection of $\eta$ is crucial for balancing sparsity and statistical efficiency. This is addressed in Section \ref{section:tuning_parameter}.

\subsection{Adaptive LASSO Formulation}\label{sec:adaptive_lasso}

While the standard LASSO penalty in \eqref{eq:proposed_lasso_problem_mod} provides automatic neighbor selection, it treats all distant neighbors equally regardless of their spatial relationship with the prediction location. In practice, some distant neighbors may be more informative than others due to spatial correlation patterns. For instance, a neighbor at moderate distance in a direction of strong spatial correlation may be more valuable than a closer neighbor in a direction of weak correlation. This motivates an adaptive weighting scheme that accounts for the spatial covariance structure.

Problem \eqref{eq:proposed_lasso_problem_mod} applies uniform penalization $\eta|\lambda_i|$ to all weights $\lambda_i$ for $i > p$, treating all distant neighbors equally. This ignores their relative importance for prediction at $\bm{s}_0$ because neighbors that contribute strongly to the prediction should receive less penalization than those that contribute weakly. Following \cite{zou2006adaptive}, we propose a two-stage procedure where the second stage uses data-driven weights based on an initial estimate.

The adaptive LASSO employs a two-stage procedure. In the first stage, we obtain an initial estimate $\tilde{\bm{\lambda}}$ via standard kriging using all $N$ observations. In the second stage, we define adaptive weights based on the magnitude of $\tilde{\lambda}_i$:

\begin{equation}\label{eq:adaptive_weights}
    w_i = \frac{1}{|\tilde{\lambda}_i|}, \quad i > p.
\end{equation}
Neighbors with large $|\tilde{\lambda}_i|$ (indicating strong contribution to prediction) receive small weights $w_i$ (less penalization), while those with small $|\tilde{\lambda}_i|$ 
receive large weights (more aggressive shrinkage). The adaptive LASSO kriging weights are 
obtained by solving
\begin{align} \label{eq:adaptive_lasso_problem}
    \min_{\bm{\lambda} \in \mathbb{R}^{N}} &\quad \bm{\lambda}^{\top} \Sigma \bm{\lambda} - 2 \bm{\lambda}^{\top}\bm{c}_{0} + \sigma^{2}_0 + \eta \sum_{\substack{i > p}} w_i|\lambda_i| \\
    \text{s.t.} &\quad \bm{X} \bm{\lambda} = \bm{x}_{0}. \notag
\end{align}
Setting $w_i = 1$ for all $i$ recovers the standard LASSO problem \eqref{eq:proposed_lasso_problem_mod}.

Problem \eqref{eq:adaptive_lasso_problem} is solved using a modified gradient descent approach with weighted soft-thresholding, detailed in Appendix \ref{app:gradient_descent} (Algorithm \ref{alg:GD}). 
The initial estimate $\tilde{\bm{\lambda}}$ also provides an excellent starting point for the gradient descent optimization, reducing the number of gradient descent iterations by 40-50\% compared to cold start initialization. When the initial estimate uses all $N$ observations, the total complexity is $O(N^3 + TN)$ where $T$ is the number of gradient descent iterations (typically $T < 100$). When local kriging with $K$ neighbors is used in Stage 1, the complexity reduces to $O(K^3 + TN)$. For large datasets where $K \ll N$, this represents a substantial computational saving compared to $O(N^3)$ for standard kriging. 
In the kriging context, adaptive LASSO identifies minimal neighborhoods for accurate prediction while maintaining prediction variance close to standard kriging, as demonstrated in Section \ref{section:datanalysis}.

\section{Selection of the Tuning Parameter}\label{section:tuning_parameter} 

We now address the crucial question of selecting the penalty parameter $\eta$. This parameter 
controls the trade-off between prediction accuracy and neighborhood sparsity, where small 
values yield dense solutions similar to standard kriging, while large values produce highly 
sparse solutions that may sacrifice prediction quality.

For independent data in standard regression settings, tuning parameter selection typically 
relies on cross-validation or information criteria such as AIC or BIC \citep{tibshirani1996regression, 
hastie2009elements, friedman2010regularization}. However, these standard approaches face two 
fundamental challenges in the kriging context. First, they are computationally expensive. 
Unlike standard regression where a single penalty parameter is selected for the entire dataset, 
our penalized kriging framework requires selecting a penalty parameter $\eta$ at each prediction location to adapt to local spatial correlation structure. 
For a dataset with $M$ prediction  locations, evaluating $G$ candidate $\eta$ values at each location requires solving $O(MG)$ penalized kriging problems, each involving matrix operations of cost $O(N^3)$ for standard  kriging or $O(K^3)$ when using $K$ neighbors, making cross-validation prohibitively expensive.

Second, and more fundamentally, kriging variance does not depend on the observed data values  but rather on the spatial configuration of observations and the covariance structure \citep{heuvelink2002ordinary}. This contrasts with prediction error in standard regression. 
This makes data-splitting approaches less informative about prediction uncertainty, as the kriging variance at a held-out location depends only on its spatial relationship to the 
training locations, not on the actual observed values. Consequently, cross-validation may select $\eta$ that minimizes empirical prediction error without properly accounting for 
spatial correlation structure and information redundancy. These limitations motivate the need for a different approach that explicitly accounts for spatial correlation and can be 
computed efficiently.

\subsection{Effective Sample Size and Information Redundancy}

The key insight underlying our approach is that positively spatially correlated observations carry redundant information. In the extreme case of perfect correlation, a sample of $N$ observations provides no more information than a single observation. Conversely, under independence, all $N$ observations contribute unique information. For intermediate levels of spatial correlation, the effective amount of information lies somewhere between $1$ and $N$.

This concept is formalized through the notion of \emph{effective sample size} (ESS), which quantifies the number of independent observations that would be equivalent to a given spatially correlated sample \cite[][p.~15]{Cressie1993}. In the kriging context, ESS provides a natural criterion for neighbor selection because we seek the sparsity level (controlled by $\eta$) that removes redundant neighbors while preserving the effective information content needed for accurate prediction.

\cite{Cressie1993} derived an expression for ESS by comparing 
the variance of the sample average $\bar{Z}$ for both uncorrelated and spatially dependent 
observations, yielding:
\begin{equation}\label{eq:ess_cressie} 
 n^{\text{ESS}}= \frac{N^2}{\bm 1^{\top}_{N}\bm R\bm 1_{N}} = \frac{N^2}{\displaystyle\sum_{i=1}^N\sum_{j=1}^N\rho(\bm h_{ij})}, \quad \bm h_{ij}=\bm s_i-\bm s_j,
\end{equation}
where $\rho(\cdot)$ is a valid correlation function, $\bm R$ is the correlation matrix with entries $\bm R_{ij} := \rho(\bm h_{ij})$, and $\bm 1_{N}$ is a column vector of ones of size $N$. 



As an example, consider the equicorrelation structure where $\text{cov}(Z_i,Z_j)=\sigma^2\rho$ 
for $i\neq j$ with $\sigma^2=\text{var}(Z_i)$ and $\rho>0$. Then $n^{\text{ESS}}=N/(1+(N-1)\rho)$, 
so if $\rho=0$ then $n^{\text{ESS}}=N$ (independence preserves full sample size), and if 
$\rho=1$ then $n^{\text{ESS}}=1$ (perfect correlation yields no additional information). 
This extreme behavior highlights the impact of spatial correlation on effective information: 
strong positive correlation dramatically reduces the effective sample size, while independence 
preserves it. For realistic spatial processes, ESS typically lies between these extremes, 
depending on the range and strength of spatial correlation. These properties hold not only 
for the equicorrelation structure but also for any valid correlation structure 
\citep[see][for additional ESS properties]{Acosta2018}.

For neighbor selection, we focus on the complementary quantity $r^{\text{ESS}}=N-n^{\text{ESS}}$, which measures the redundant information in a sample of size $N$. This redundancy quantifies how many observations could be removed without substantially reducing the effective information content. Note that $r^{\text{ESS}} = 0$ when observations are independent (no redundancy), while $r^{\text{ESS}} = N-1$ when observations are perfectly correlated (maximal redundancy). 
In general, $0 \leq r^{\text{ESS}} \leq N-1$.

The screening effect \citep{stein2002screening} provides intuition for why redundancy matters in kriging: when observations are highly correlated, nearby neighbors provide overlapping information, leading to unstable kriging weights that oscillate around zero despite satisfying the unbiasedness constraints. The LASSO penalty addresses this by explicitly removing redundant 
neighbors.

\subsection{ESS for Local and Penalized Kriging}

Recall that global kriging uses all $N$ spatial locations, while local kriging uses $K \leq N$ spatial locations. Let $\bm e_{N,K}$ be the $N$-dimensional binary vector where the $i$-th entry is $1$ if the location $\bm s_{i}$ is used for local kriging and $0$ otherwise, then 
$\| \bm e_{N,K} \|_{0} = \sum_{i = 1}^{N} \mathbb{I}(e_{N,K,i} \neq 0) = K$, where $\mathbb{I}(\cdot)$ is the indicator function  and $\|\cdot\|_0$ is the $\ell_0$-norm. 
The constraint $\|\bm e_{N,K}\|_0 = K$ ensures exactly $K$ neighbors are selected. The effective sample size for this subset is
$$n^{\text{ESS}}_{\text{loc}}= \frac{K^2}{\bm e_{N,K}^{\top}\bm R\bm e_{N,K}},$$
which accounts for the correlation structure within the selected neighborhood, and the redundancy is $r^{\text{ESS}}_{\text{loc}} = K - n^{\text{ESS}}_{\text{loc}}$.

Now, recall that $\widehat{P}_{\eta}(\bm{Z}, \bm s_{0}) = \widehat{\bm\lambda}(\eta)^{\top}\bm Z$ 
is the predictor obtained by solving \eqref{eq:proposed_lasso_problem_mod}. The sparsity of 
$\widehat{\bm\lambda}(\eta)$ depends on $\eta$, and consequently so does the number of selected 
locations and their effective sample size. We define the binary indicator vector $\bm e_{N,\eta}$ 
with $e_{N,\eta,i} = \mathbb{I}(\widehat{\lambda}_i(\eta) \neq 0)$, which identifies the active 
set of neighbors at penalty level $\eta$. The effective sample size as a function of $\eta$ 
is then

$$n_{\eta}^{\text{ESS}} = \frac{\|\widehat{\bm\lambda}(\eta)\|_0^2}{\bm e_{N,\eta}^{\top}\bm R\bm e_{N,\eta}},$$

\noindent and the redundancy is $r_{\eta}^{\text{ESS}} = \|\widehat{\bm\lambda}(\eta)\|_0 - n_{\eta}^{\text{ESS}}$. 
Note that as $\eta$ increases, $\|\widehat{\bm\lambda}(\eta)\|_0$ decreases (fewer neighbors), 
but the relationship between $\eta$ and $n_{\eta}^{\text{ESS}}$ is more complex: removing 
highly correlated neighbors may increase the effective sample size per observation, while 
removing informative neighbors decreases overall information content.

Since both information and redundancy depend on the correlation structure, direct comparison 
across different $\eta$ values requires normalization. We propose the normalized redundancy 
measure 
$$s_{\eta} = \frac{r_{\eta}^{\text{ESS}}}{r_{0}^{\text{ESS}}} = \frac{\|\widehat{\bm\lambda}(\eta)\|_0 - n_{\eta}^{\text{ESS}}}{N - n_{0}^{\text{ESS}}},$$
which expresses the current redundancy as a fraction of the total redundancy in the global 
kriging solution ($\eta = 0$). This normalization ensures $0 \leq s_{\eta} \leq 1$, where 
$s_\eta = 1$ corresponds to global kriging (all redundancy present) and $s_\eta \to 0$ as 
$\eta \to \infty$ (redundancy removed through sparsification).

Importantly, $s_\eta$ is not guaranteed to be monotonically decreasing in $\eta$. While the 
number of neighbors $\|\widehat{\bm\lambda}(\eta)\|_0$ decreases monotonically, the effective 
sample size $n_\eta^{\text{ESS}}$ may increase when highly redundant neighbors are removed, 
leading to non-monotonic behavior in $s_\eta$ with discontinuous jumps at values of $\eta$ 
where neighbors enter or leave the active set.

\subsection{Optimal selection of $\eta$}

By itself, $s_\eta$ cannot determine an optimal value of $\eta$ unless a specific quantity is fixed—for example, $s_\eta = 0.5$, which represents a $50\%$ relative redundancy. To identify $\eta$, we need a measure that quantifies the quality of the fit. In this sense, as discussed in Section \ref{section:kriging_equations}, the prediction variance $\sigma_{\eta}^{2} := \text{MSE}( \widehat{P}_{\eta}(\bm{Z}, \bm s_{0}) )$ is obtained from 
the kriging equations. This variance is used as a measure of statistical performance relative 
to the global kriging variance $\sigma^2_{\text{gk}} := \sigma^2_{\text{gk}}(\bm s_{0})$. 
The prediction variance $\sigma_{\eta}^{2}$ increases monotonically with $\eta$ as fewer 
neighbors contribute to prediction. We have $\sigma^2_{\eta} \rightarrow \sigma^2_{\text{gk}}$ 
when $\eta \rightarrow 0$ (global kriging achieves minimum variance), and 
$\sigma^2_{\eta} \rightarrow \sigma^2_{\infty}$ when $\eta \rightarrow \infty$, where 
$\sigma^2_{\infty}$ corresponds to the variance using only the $p$ unpenalized nearest neighbors. 
To make the variance term comparable to the sparsity measure, we normalize it as
$$v_{\eta} = \frac{\sigma_{\eta}^{2} - \sigma^2_{\text{gk}}}{\sigma^2_{\infty} - \sigma^2_{\text{gk}}},$$
\noindent which ranges from $v_\eta = 0$ (global kriging, minimum variance) to $v_\eta = 1$ (maximum sparsity, maximum variance).

Our goal is to select $\eta$ that achieves an optimal balance between two competing objectives:
\begin{itemize}
    \item \textbf{Maximize sparsity} (large $s_\eta$): Remove redundant neighbors to reduce 
    computational cost and information redundancy
    \item \textbf{Minimize variance increase} (small $v_\eta$): Maintain prediction accuracy 
    close to global kriging
\end{itemize}

A simple average of $s_\eta$ and $v_\eta$ would be inappropriate because it treats both 
objectives equally, particularly on the boundaries: a solution with high sparsity but poor 
accuracy (e.g., $s_\eta = 0.9$, $v_\eta = 0.1$) would score the same as one with low sparsity 
but excellent accuracy (e.g., $s_\eta = 0.1$, $v_\eta = 0.9$). Instead, we use the harmonic 
mean, which heavily penalizes imbalance. The harmonic mean achieves its maximum only when 
both terms are relatively large and balanced.

To see this, note that for fixed sum $s_\eta + v_\eta = c$, the harmonic mean 
$\frac{2s_\eta v_\eta}{s_\eta + v_\eta} = \frac{2s_\eta v_\eta}{c}$ is maximized when 
$s_\eta = v_\eta = c/2$. When one term is much smaller than the other (e.g., $s_\eta \approx 0$ 
or $v_\eta \approx 0$), the harmonic mean is close to zero regardless of the other term's value. 
This makes the harmonic mean particularly appropriate for our multi-objective optimization 
problem. We therefore propose selecting $\eta$ by maximizing the harmonic mean:

\begin{equation}\label{eq:eta_selection}
    \eta^{\ast}=\argmax_{\eta \geq 0}\left\{  \frac{2 s_{\eta} v_{\eta}}{s_{\eta}  + v_{\eta}}\right\}.
\end{equation}

The objective function $\frac{2 s_{\eta} v_{\eta}}{s_{\eta}  + v_{\eta}}$ ranges from 0 to 1, 
achieving its maximum when sparsity and variance loss are balanced. This criterion favors 
solutions that remove substantial redundancy ($s_\eta$ large) without excessive variance 
penalty ($v_\eta$ not too large).

The criterion \eqref{eq:eta_selection} is computationally efficient, requiring only:
\begin{enumerate}
    \item Evaluation of the penalized kriging objective function at candidate $\eta$ values
    \item Computation of $n_\eta^{\text{ESS}} = \|\widehat{\bm\lambda}(\eta)\|_0^2 / (\bm e_{N,\eta}^{\top}\bm R\bm e_{N,\eta})$, 
    which involves summing elements of the correlation matrix $\bm R$ for the active set
    \item Calculation of kriging variance $\sigma_\eta^2$, obtained from the kriging equations
\end{enumerate}

All required quantities are by-products of solving the penalized kriging problem, so no additional matrix inversions or model fits are needed. This contrasts sharply with K-fold cross-validation, which requires $O(GK)$ model refits for $G$ candidate $\eta$ values and $K$ folds.

In practice, we avoid computing $\sigma_{\text{gk}}^2$ (which requires inverting the full 
$N \times N$ matrix) by restricting the search to a bounded interval $[\eta_{\min}, \eta_{\max}]$ where $\eta_{\min}$ is small enough that $\sigma_{\eta_{\min}}^2 \approx \sigma_{\text{gk}}^2$. Typically, we set $\eta_{\max}$ to the maximum absolute value of the ordinary kriging weights to ensure at least $p$ neighbors remain, and $\eta_{\min} = 10^{-9} \times \eta_{\max}$. 

The optimal value $\eta^*$ is found through an adaptive grid search algorithm that combines coarse initial exploration with iterative refinement. The algorithm is presented in Algorithm \ref{alg:eta_search}.
\begin{algorithm}[ht]
\caption{Two–stage search procedure for selecting $\eta$}
\label{alg:eta_search}
\begin{algorithmic}[1]
    \State Evaluate the criterion on a coarse logarithmic grid of $\kappa_{0}$ points spanning $[\eta_{\min}, \eta_{\max}]$.
    \State Identify the top $5$ candidate values with the highest criterion values.
    \For{each candidate $\eta_k$}
        \State Construct a local grid of $\kappa_{1}$ points in the interval $[\eta_k/2, 2\eta_k]$.
        \State Evaluate the criterion on this local grid.
    \EndFor
    \State \Return $\eta^{*}$ that maximizes the criterion.
\end{algorithmic}
\end{algorithm}
This procedure typically requires 200-300 evaluations of the criterion, which is computationally negligible compared to the matrix inversions required for cross-validation. The logarithmic spacing in Step 1 ensures good coverage across orders of magnitude, while the local refinement in Steps 3-4 provides precision near the optimum.

\begin{figure}
    \centering
    \includegraphics[scale = 0.23]{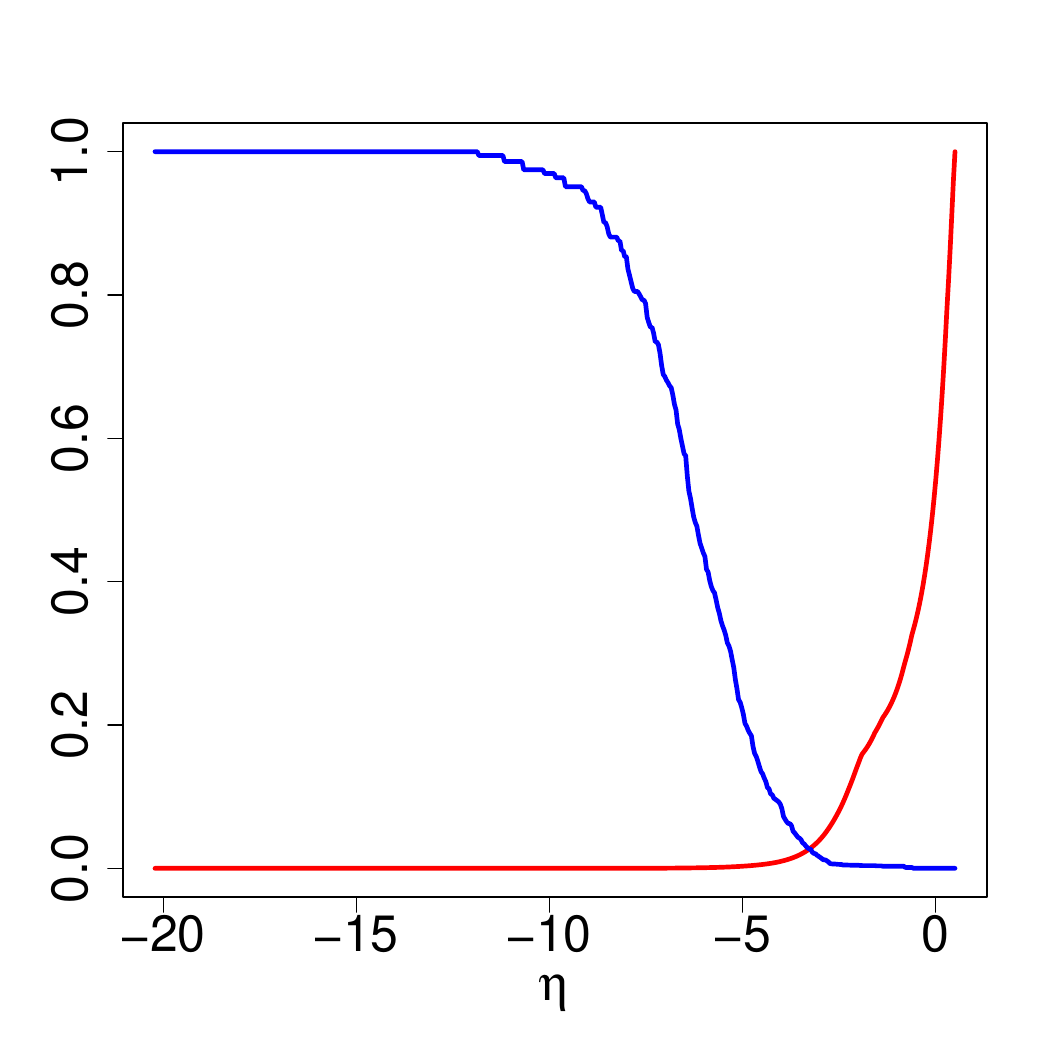}  \includegraphics[scale = 0.23]{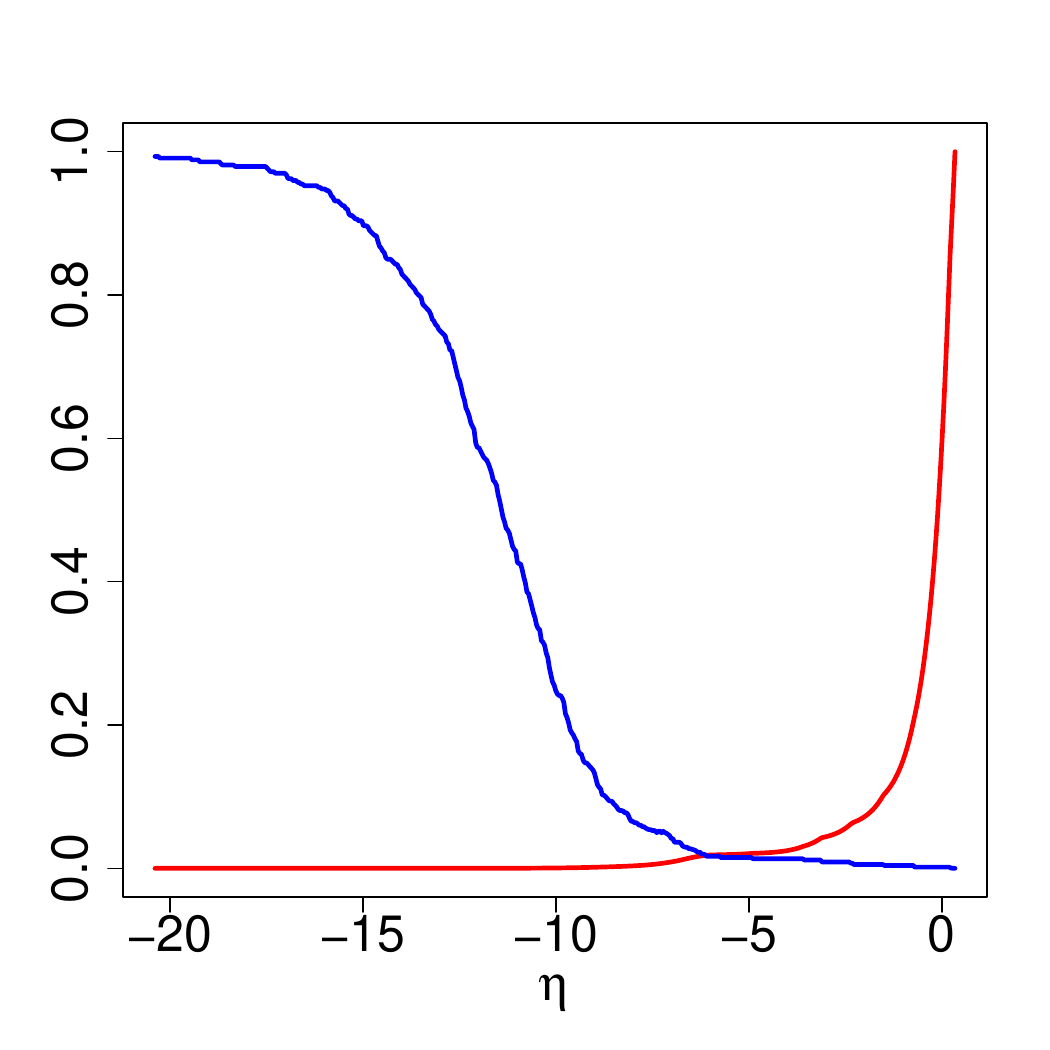}  \includegraphics[scale = 0.23]{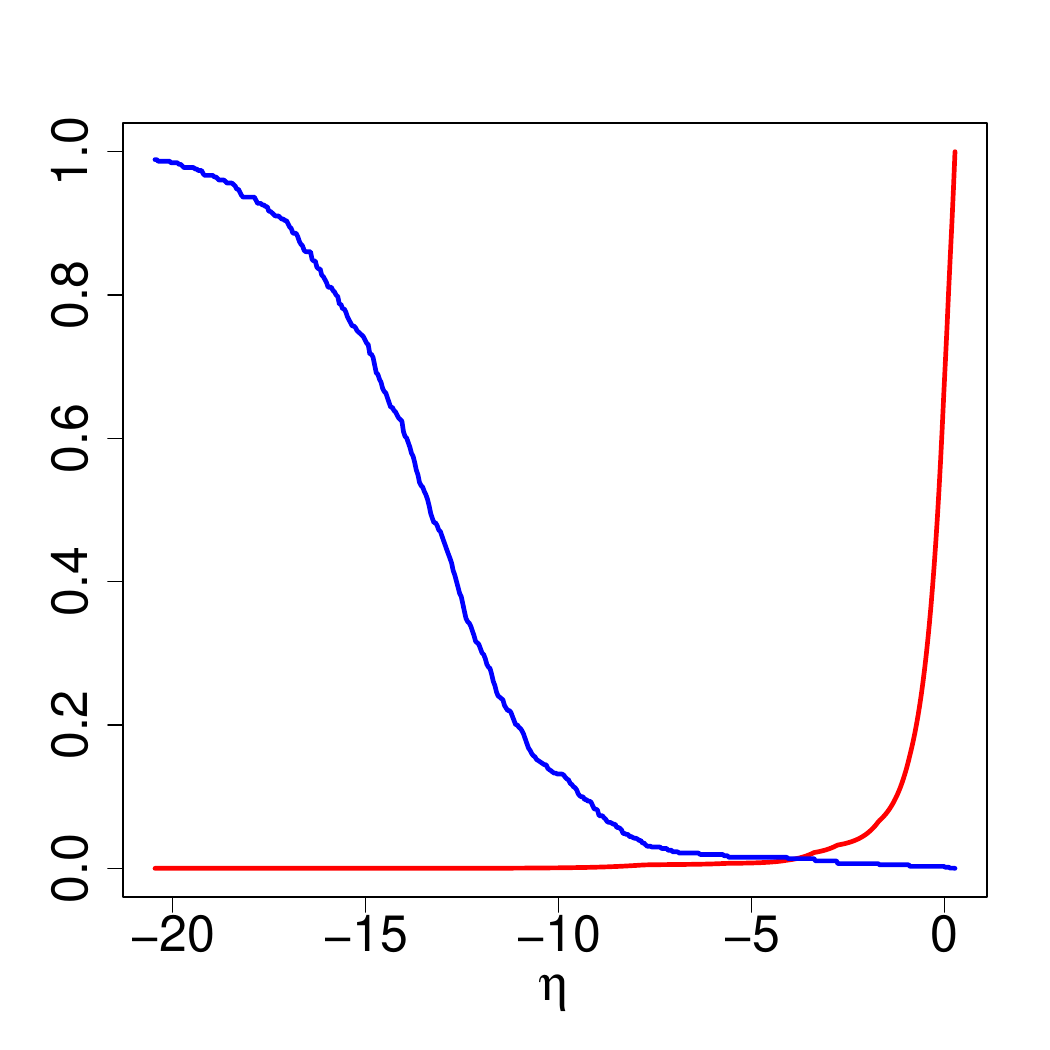} \\
    \includegraphics[scale = 0.23]{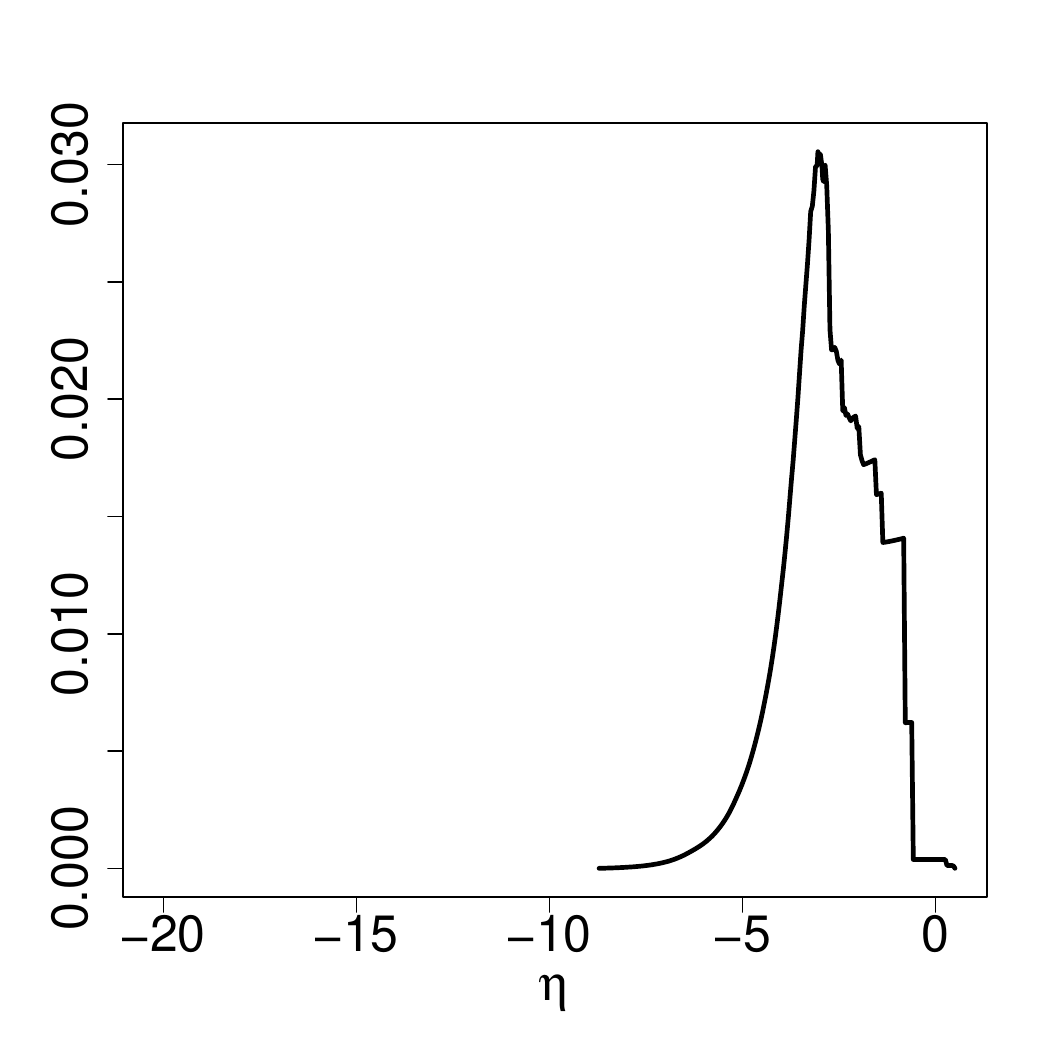} \includegraphics[scale = 0.23]{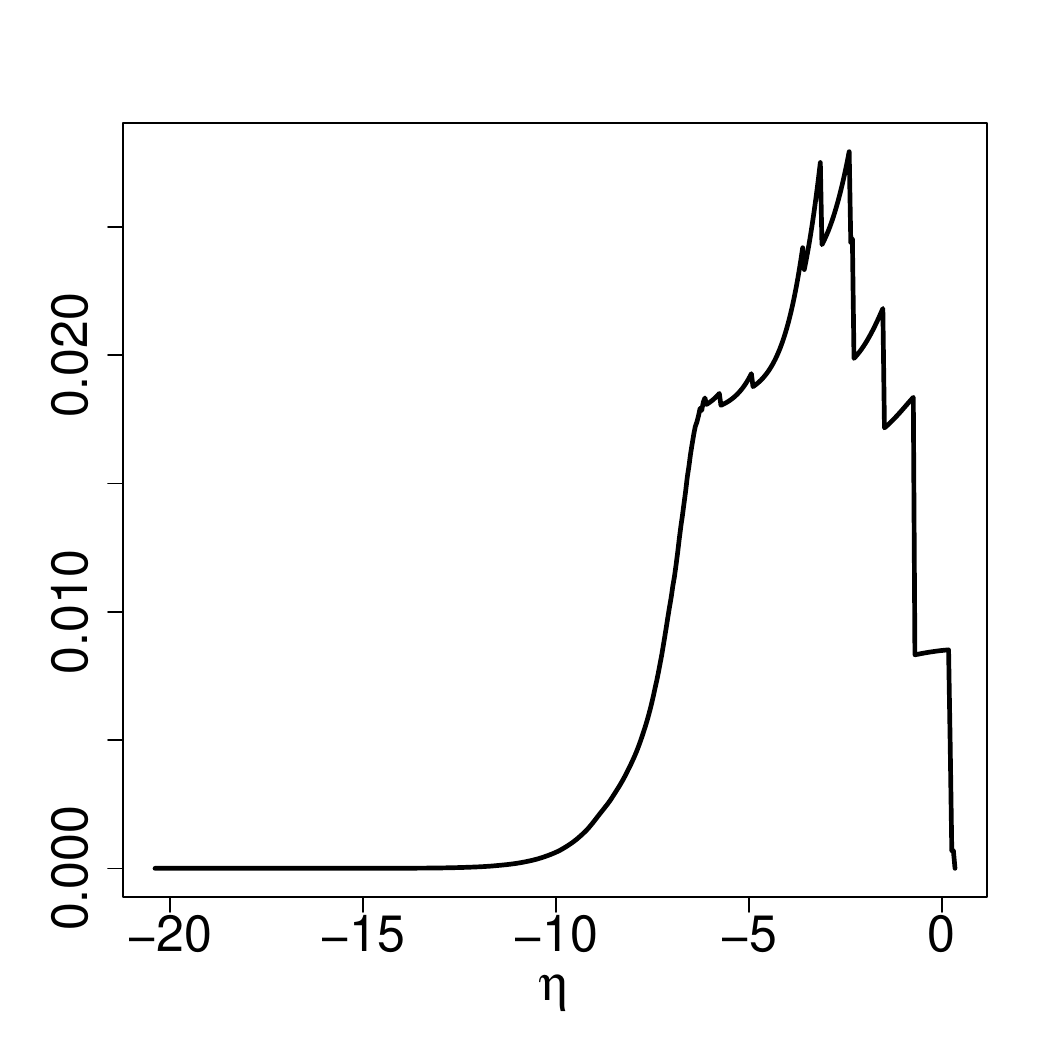} \includegraphics[scale = 0.23]{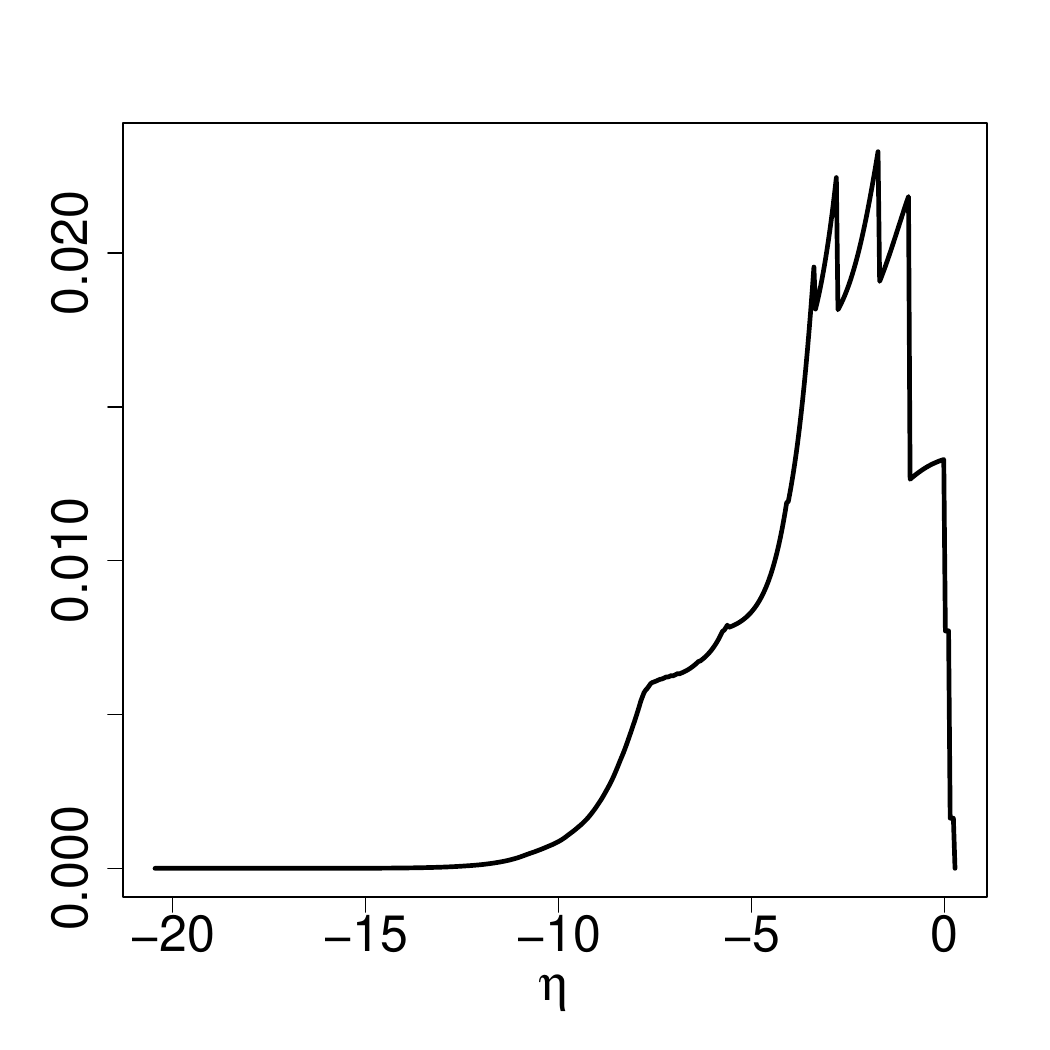}
    \caption{Behavior of the tuning parameter selection criterion for exponential covariance models with practical range (PR) of $0.05$ (left), $0.15$ (center), and $0.2$ (right). The variance ratio $v_{\eta}$ (red line) increases monotonically with $\eta$, while the sparsity measure $s_{\eta}$ (black line) shows non-monotonic behavior. The harmonic mean criterion (blue line) achieves its maximum at intermediate values, with optimal $\eta^{\star}$ increasing for smoother processes (larger PR).}
    \label{fig:harmonic_mean_example}
\end{figure}

As an example, Figure~\ref{fig:harmonic_mean_example} illustrates the behavior of the criterion for exponential covariance models with three practical ranges (PR), defined as the distance at which the correlation decays to $0.05$. Several patterns are apparent: (1) The variance ratio $v_\eta$ (red) increases monotonically with $\eta$, as expected when fewer neighbors are retained, while the sparsity measure $s_\eta$ exhibiths an opposite behaviour. (2) The harmonic mean (black) displays non-monotonic behavior, with discontinuous jumps whenever neighbors are removed from the active set. (3) The harmonic mean attains its maximum at intermediate $\eta$ values, avoiding both extremes: dense solutions with negligible variance reduction and overly sparse solutions with substantial efficiency loss. (4) For processes with larger PRs (Figure~\ref{fig:harmonic_mean_example}, right panel), the optimal $\eta^{\star}$ is higher, leading to sparser representations due to increased spatial correlation. (5) For rougher processes (smaller PRs, Figure~\ref{fig:harmonic_mean_example}, left panel), the optimal $\eta^{\star}$ is smaller, retaining more neighbors to adequately capture local variability. Overall, the ESS-based criterion adapts the neighborhood size to the underlying correlation structure without requiring user tuning or cross-validation.

\section{Numerical Experiments}\label{section:simulations}

\begin{figure}[h!]
    \centering
    \includegraphics[scale = 0.6]{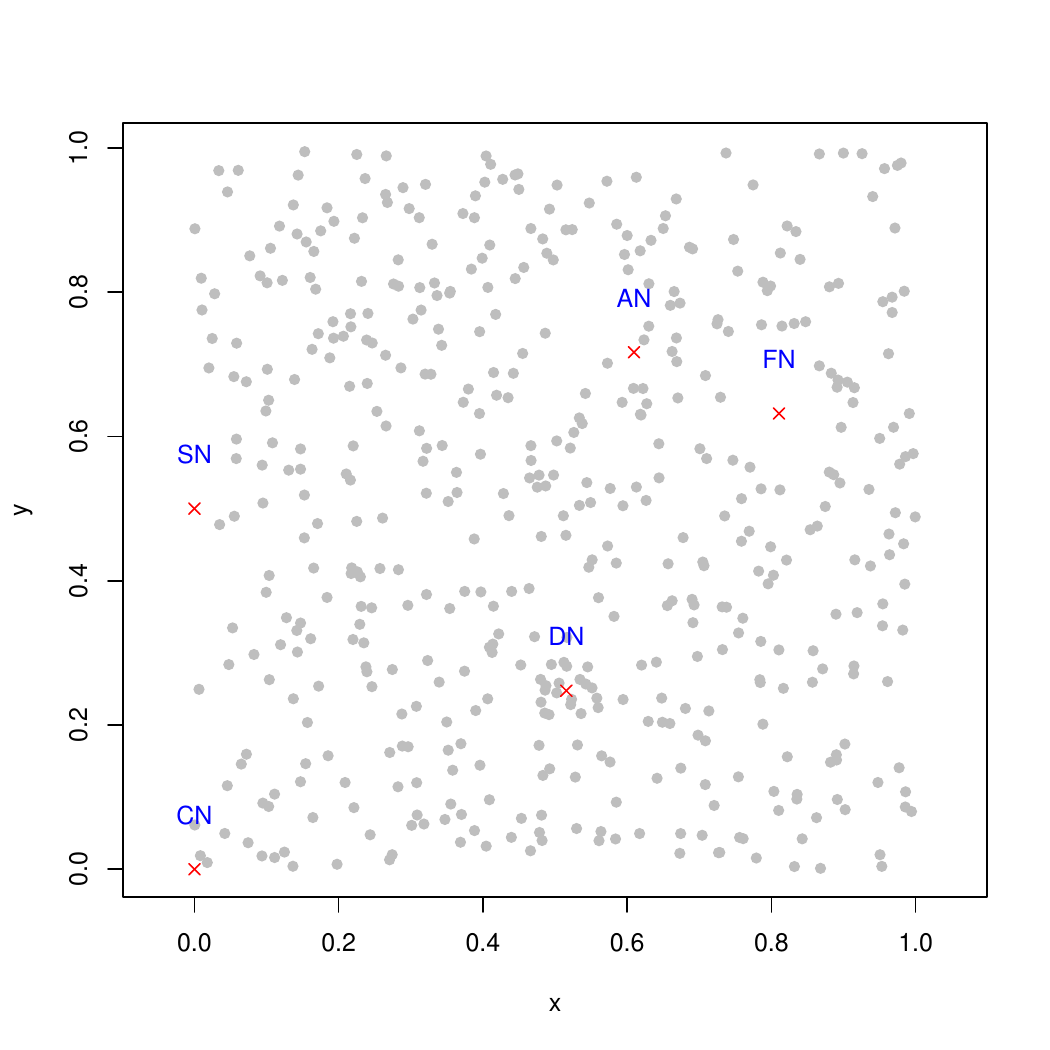} 
    \caption{Locations used in the simulation experiment. The gray points are used as the 
    training set, while the red points are the prediction locations used to explore the 
    neighborhood structure selected with penalized kriging. The labels stand for farthest 
    neighbor (FN), average neighbor (AN), densest neighbor (DN), side neighbor (SN), and 
    corner neighbor (CN).}
    \label{fig:coordinates_to_simulate}
\end{figure}

In all numerical experiments, we consider $500$ fixed locations uniformly distributed over $[0,1]\times[0,1]$. Prediction is performed at five representative locations characterized by distinct neighborhood configurations: (FN) the farthest neighbor, defined as the location whose nearest-neighbor distance is the largest among all points; (AN) the average neighbor, whose nearest-neighbor distance is closest to the global average; (DN) the densest neighbor, located in the region of highest point density; (SN) a side neighbor at $(0,0.5)$; and (CN) a corner neighbor at $(0,0)$. These configurations are illustrated in Figure~\ref{fig:coordinates_to_simulate}.

The underlying Gaussian random fields are generated using three covariance models: exponential, spherical, and Matérn with smoothness parameter $\nu=1.5$. To ensure comparability across models, all covariance functions are parameterized to share the same practical range (PR), defined as the distance at which the correlation drops to $0.05$. For each prediction point, we evaluate the proposed penalized kriging method, introduced in Section~\ref{section::penalized_kriging}, together with the tuning parameter selection strategy described in Section~\ref{section:tuning_parameter}, by analyzing some key features.

\begin{figure}[h!]
    \centering
    \subfigure[Exponential]{\includegraphics[width=0.32\linewidth]{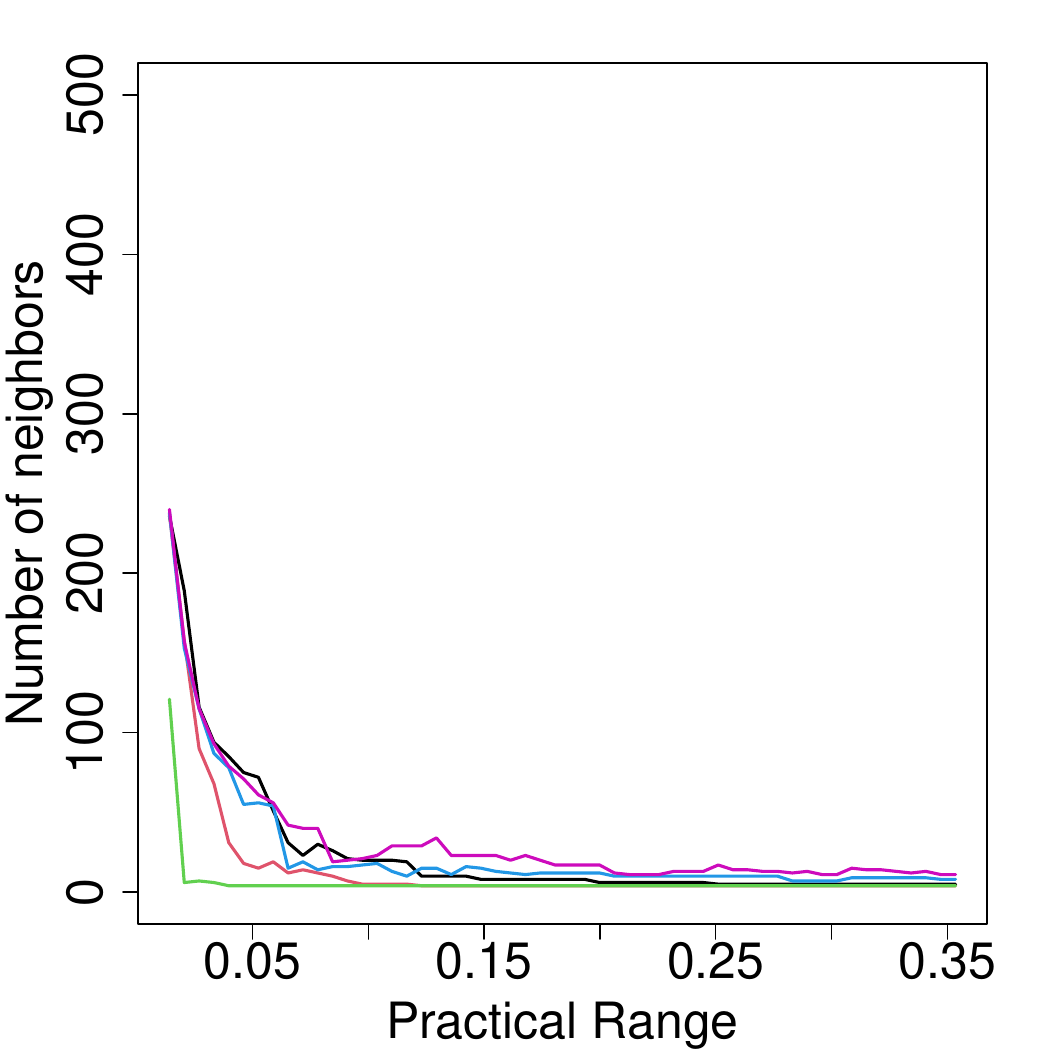}}
    \subfigure[Matérn ($\nu=1.5$)]{\includegraphics[width=0.32\linewidth]{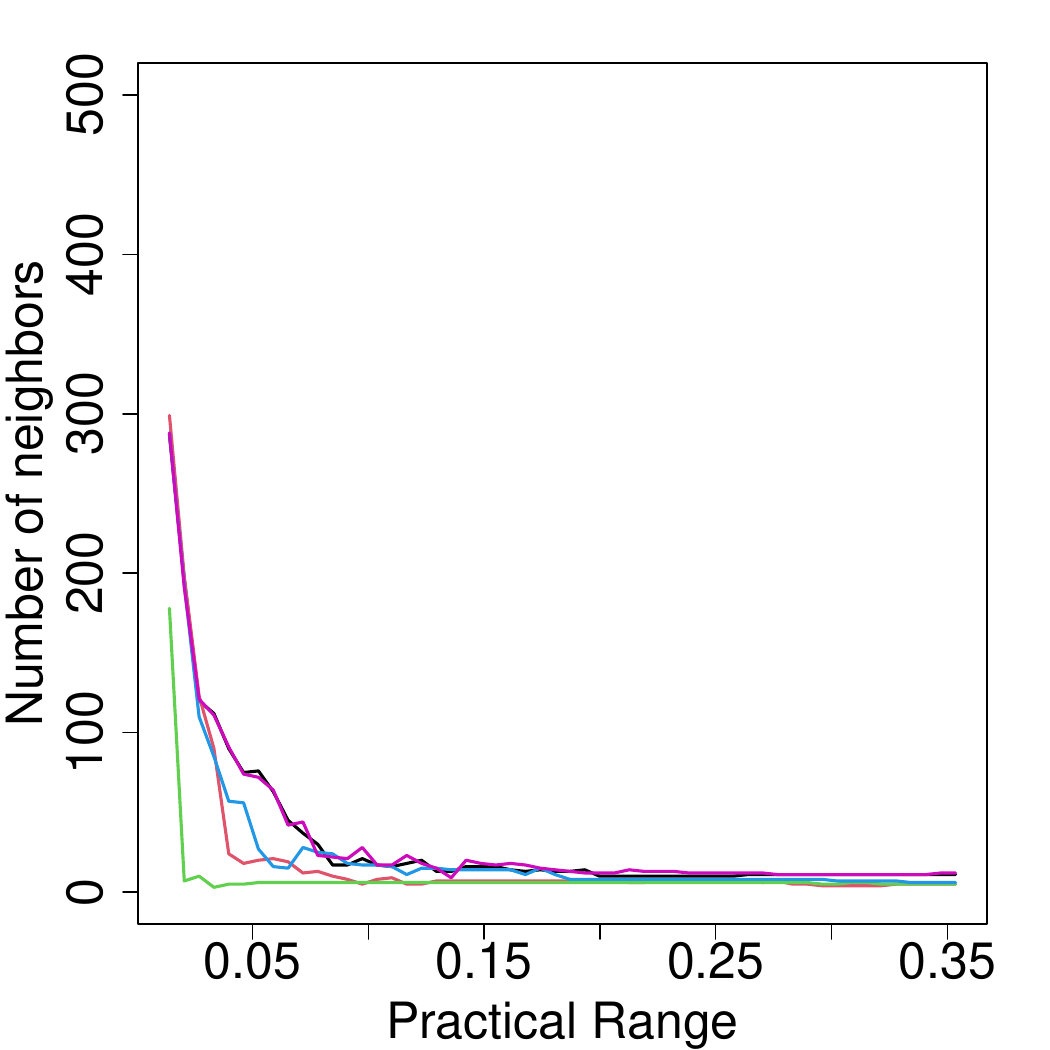}}
    \subfigure[Spherical]{\includegraphics[width=0.32\linewidth]{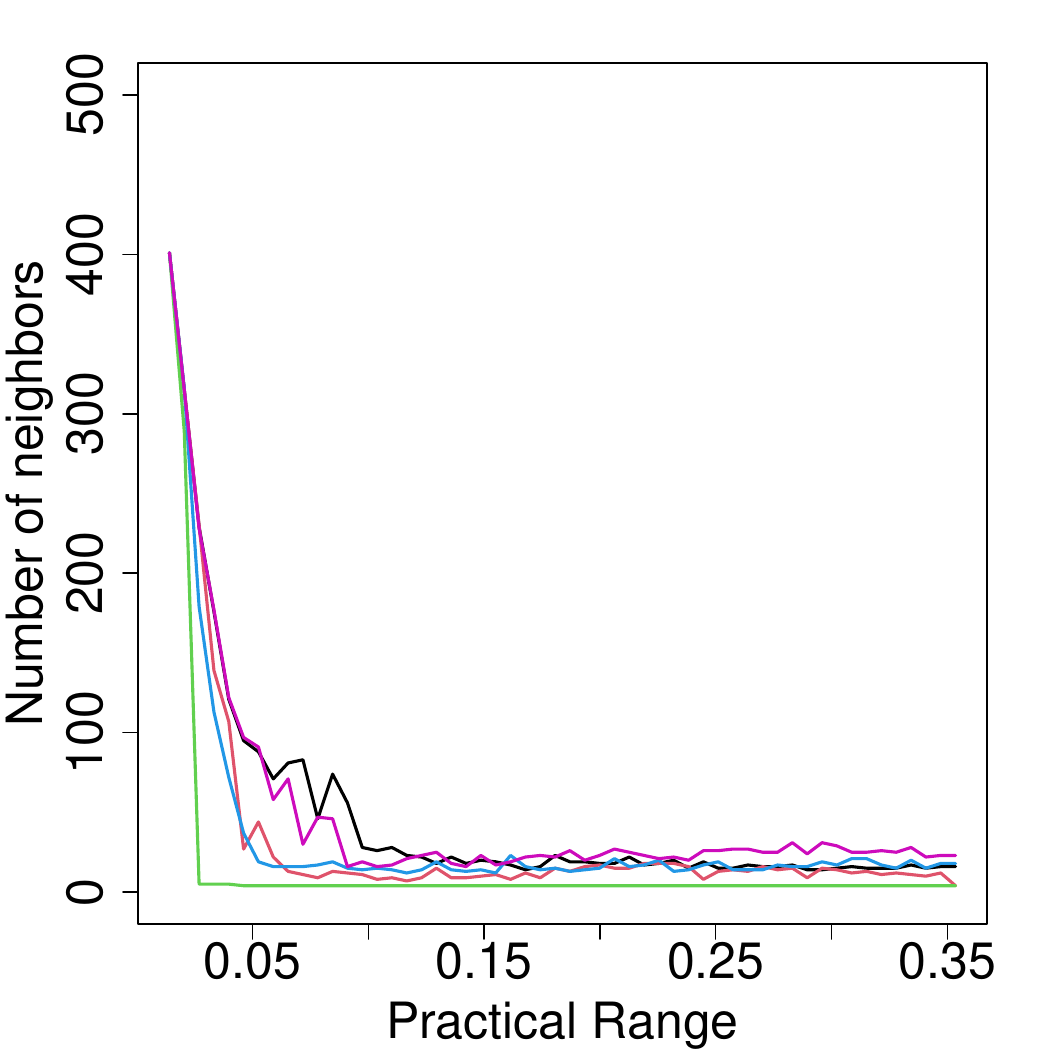}}
    \caption{Number of neighbors selected by penalized kriging as a function of the practical range for three covariance models. Each curve represents one of the five prediction locations (FN: black line; AN: red lines; DN: green line; CN: blue line; SN: magenta line). The number of selected neighbors decreases as the practical range increases for all models, with the rate of decrease depending on the process smoothness and the geometry of the prediction locations.}
    \label{fig:n_vs_pr}
\end{figure}

Figure~\ref{fig:n_vs_pr} summarizes how the selected number of spatial locations changes with the practical range across covariance structures and prediction locations. We observe that as spatial correlation increases, the algorithm systematically reduces the number of neighbors. For instance, under the exponential model, the average neighbor (AN) decreases from $237$ at $PR=0.014$ to 15 at $PR=0.053$, and stabilizes at $4$ neighbors for $PR\ge 0.117$. This corresponds to a reduction from $47\%$ to less than $1\%$ of all available observations, while preserving prediction accuracy. 
The magnitude of this reduction depends strongly on the local configuration around the prediction point. For example, DN reaches minimal neighborhood sizes at smaller $PR$ values because nearby observations provide sufficient information, making additional neighbors redundant. In contrast, SN and CN require larger neighborhoods at comparable $PR$ levels. Overall, we observe no substantial differences between covariance models, indicating that the reduction in neighborhood size is driven primarily by spatial configuration and correlation strength rather than by the specific covariance family.

\begin{figure}[h!]
    \centering
    \subfigure[AN - Exponential]{\includegraphics[width=0.30\linewidth]{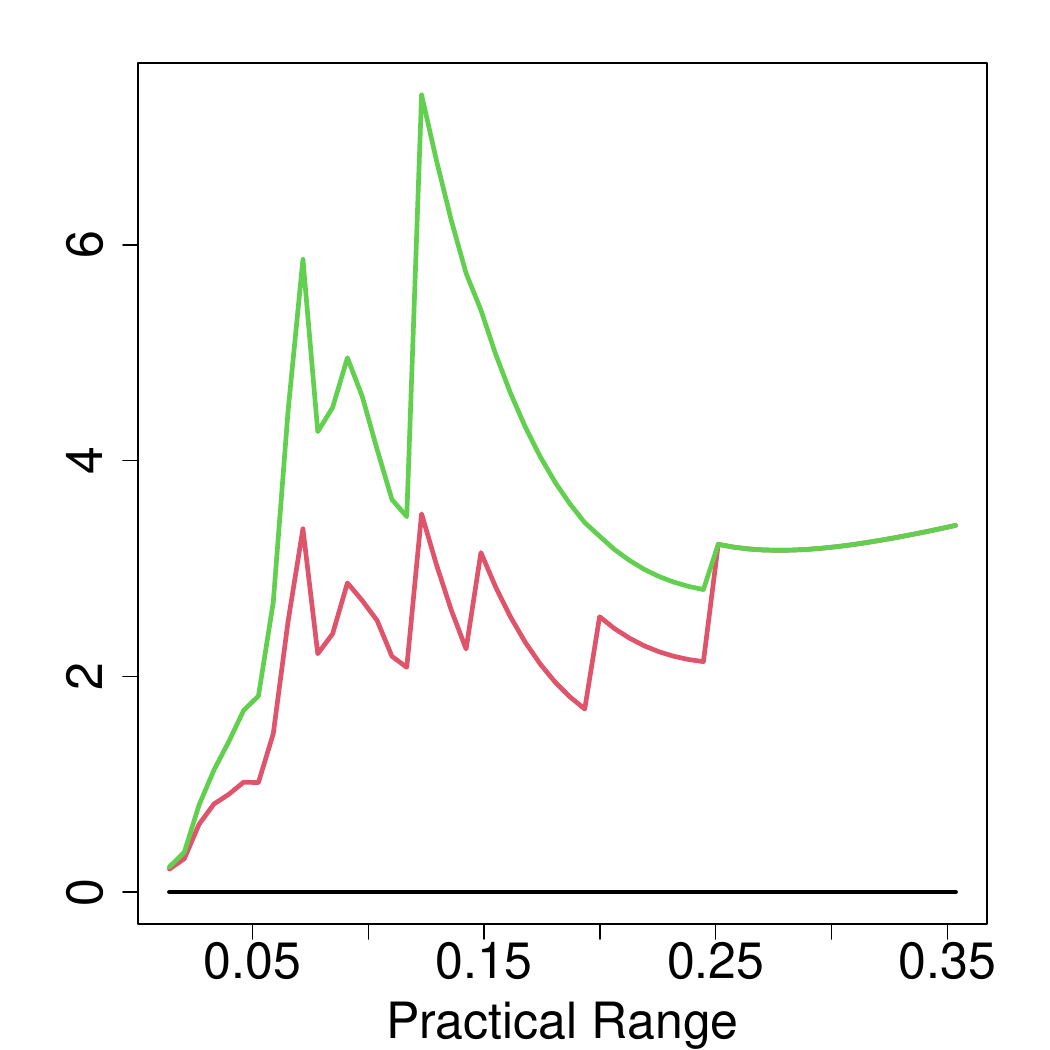}}
    \subfigure[AN - Matérn]{\includegraphics[width=0.30\linewidth]{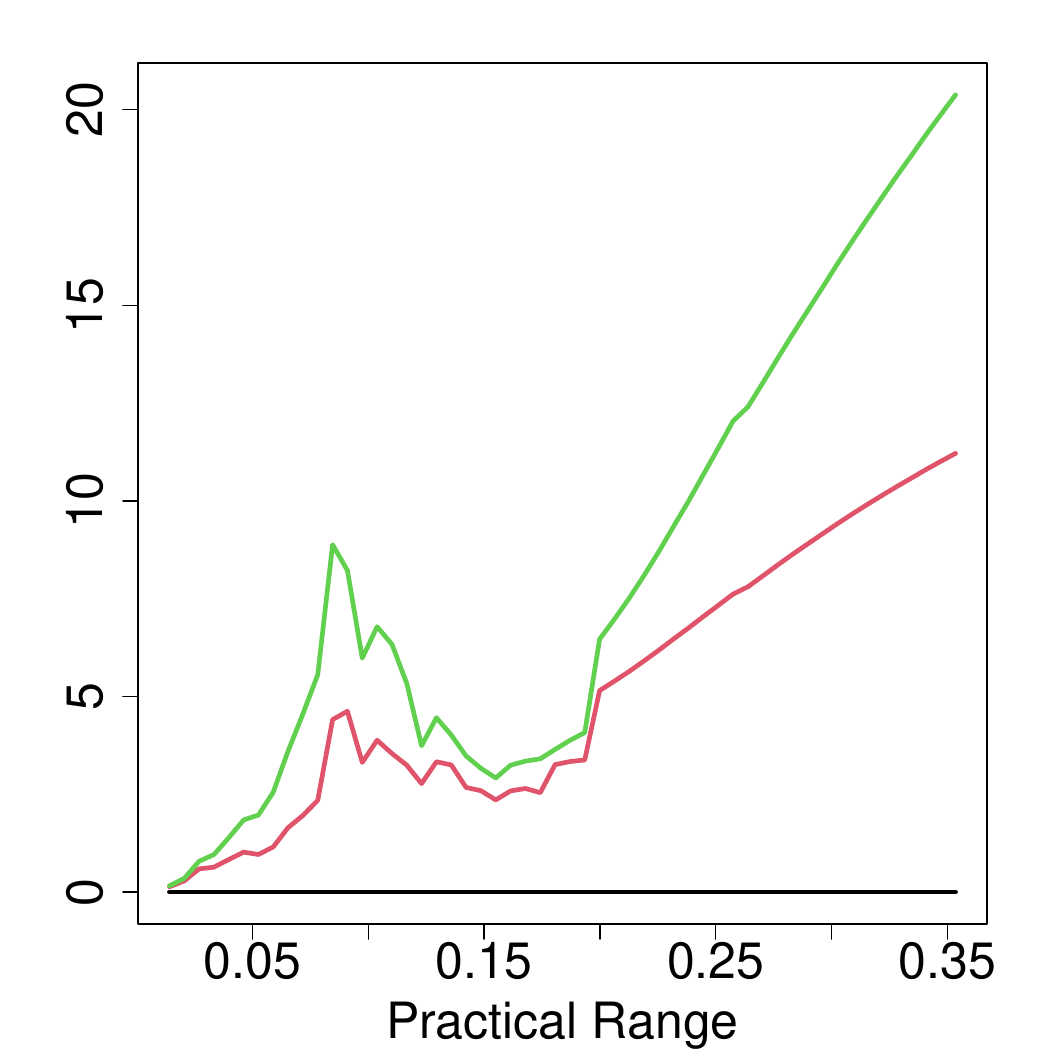}}
    \subfigure[AN - Spherical]{\includegraphics[width=0.30\linewidth]{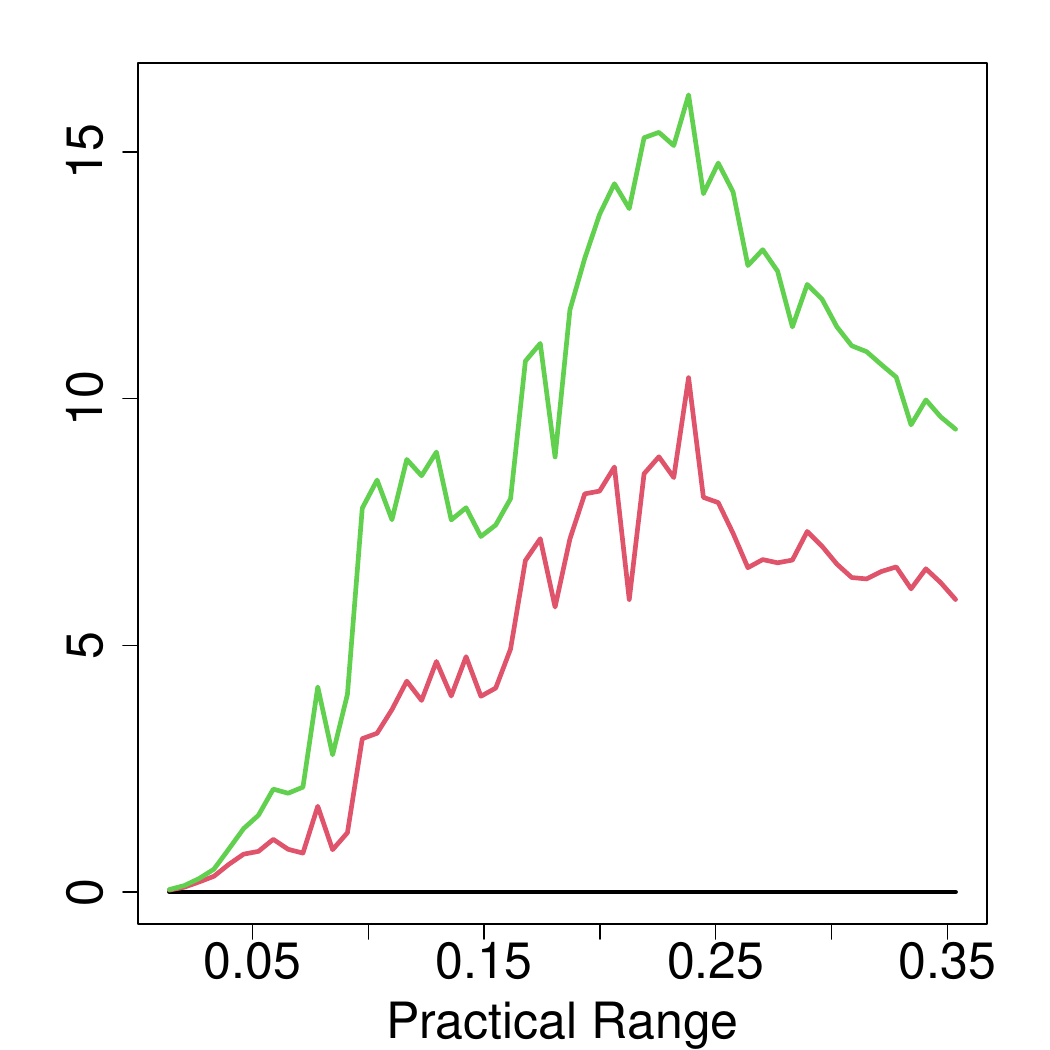}}  
    \subfigure[DN - Exponential]{\includegraphics[width=0.30\linewidth]{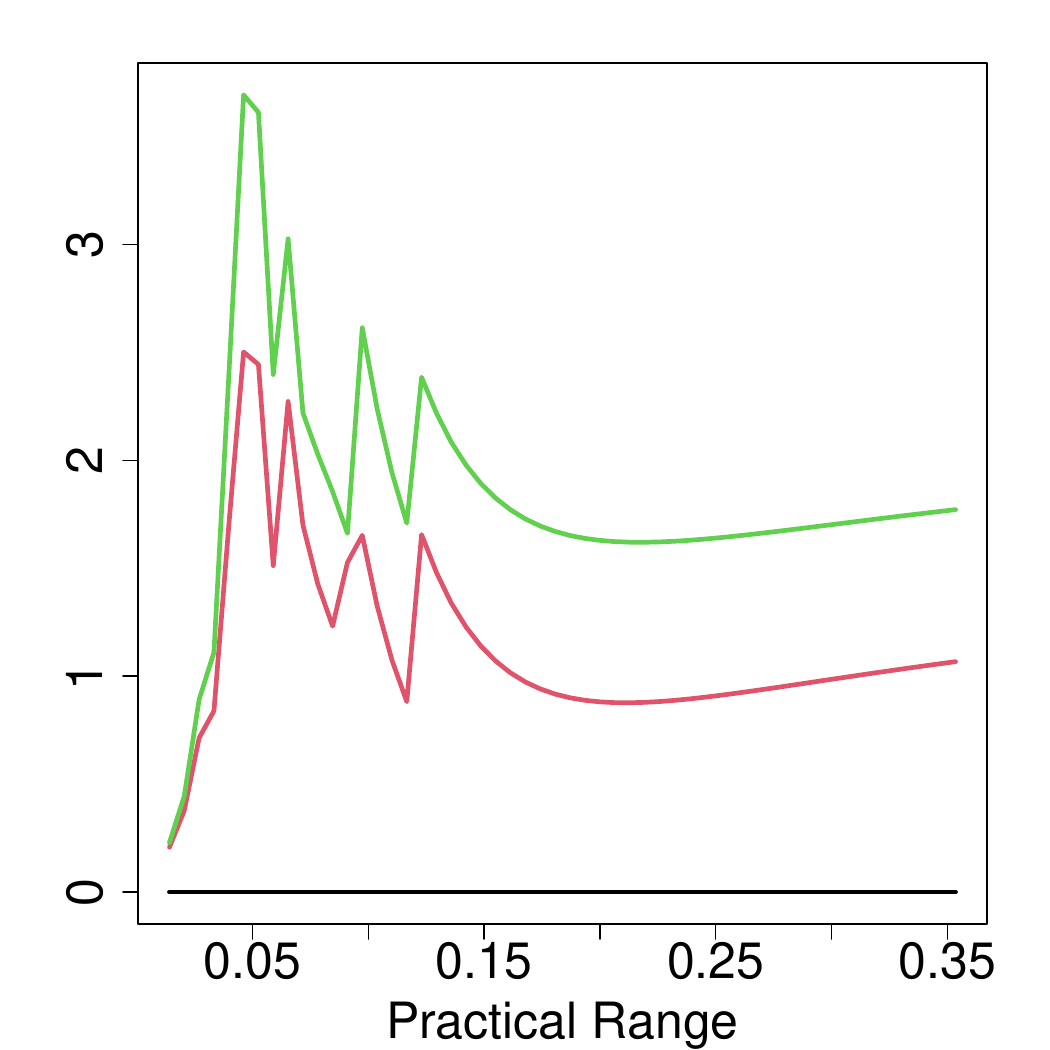}}
    \subfigure[SN - Exponential]{\includegraphics[width=0.30\linewidth]{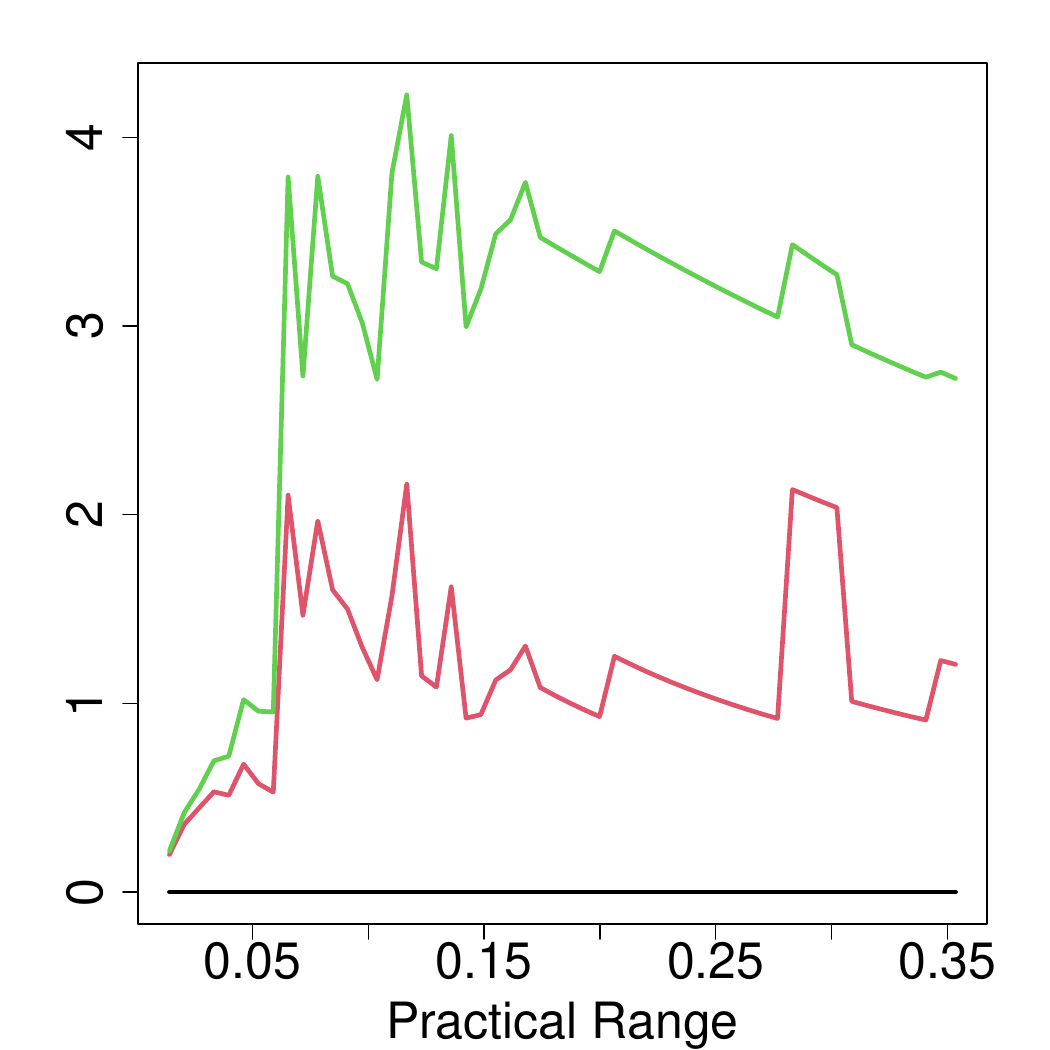}}
    \subfigure[CN - Exponential]{\includegraphics[width=0.30\linewidth]{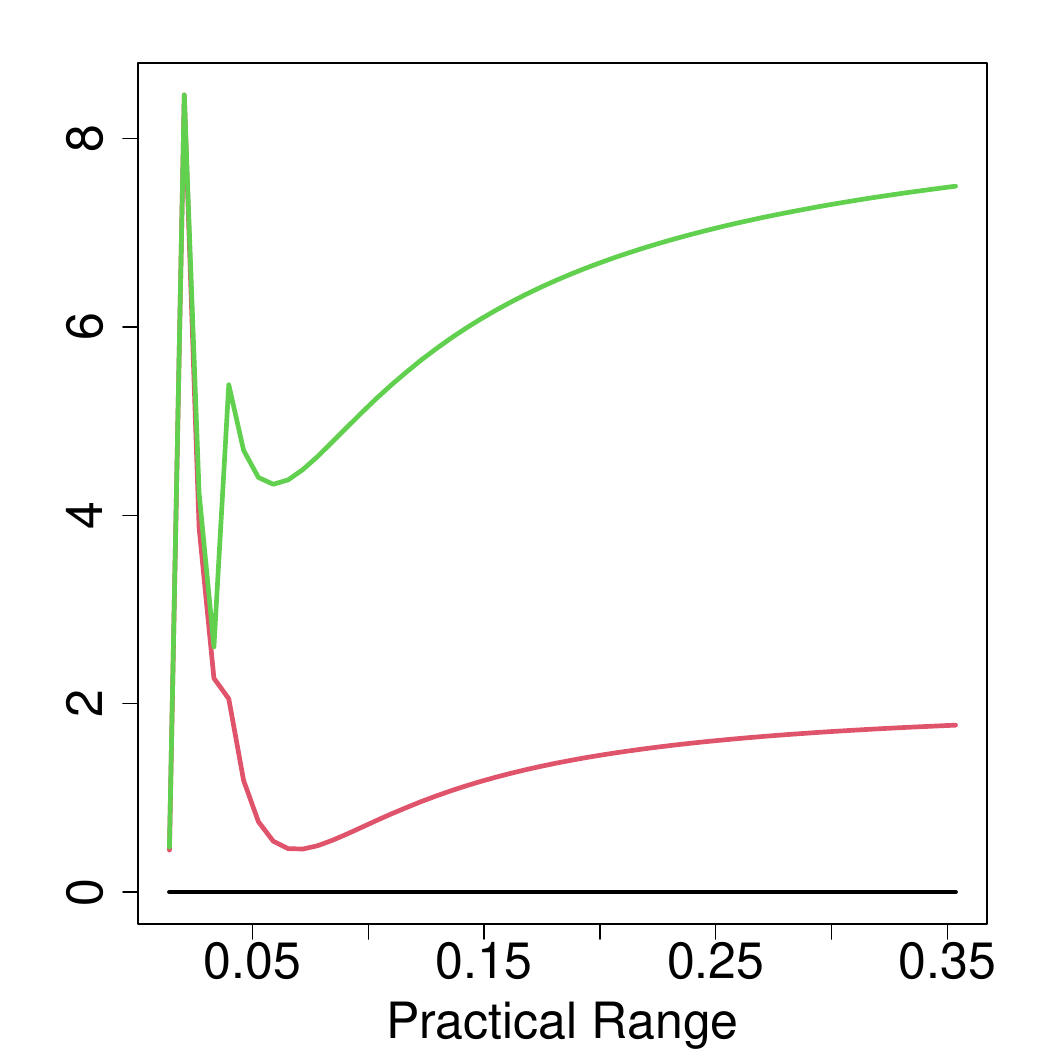}}
    \caption{Relative variance increase (\%) compared to global kriging (GK) using all 500 observations. Red lines correspond to penalized kriging (PK), green lines to local kriging (LK) using the same number of $K$-nearest neighbors as PK, and the black horizontal line indicates the global kriging baseline (0\%). Top row shows the effect of the covariance model with prediction location fixed at AN, while the bottom row shows the effect of prediction location under exponential covariance.}
    \label{fig:variance_comparison}
\end{figure}

Figure~\ref{fig:variance_comparison} compares the relative variance increase for penalized kriging (PK) and local kriging (LK) across practical ranges. For a fair comparison, local kriging is implemented using the same neighborhood size as penalized kriging, selecting the corresponding $K$ nearest neighbors. Penalized kriging performs as well as or better than local kriging across all considered settings.

The top row shows how different covariance models affect the variance comparison, holding the prediction location fixed at AN. The bottom row shows how prediction location affects the comparison, holding the covariance model fixed at exponential. Complete results for all $15$ combinations of covariance model and prediction location are provided in Appendix \ref{appendix:variance_full}.

The results in Figure~\ref{fig:variance_comparison} show that penalized kriging achieves remarkably low variance increases despite substantial reductions in neighborhood size, remaining close to the global kriging (GK) benchmark that uses all available observations. For the exponential covariance model at $PR = 0.053$, the average neighbor (AN) requires only $15$ neighbors (approximately $3\%$ of the data), with a variance increase of just $2.44\%$ relative to GK. At $PR = 0.354$, AN relies on only $4$ neighbors, yielding a variance increase of $1.07\%$. Similar behavior is observed for the Matérn model with smoothness parameter $\nu = 1.5$: at $PR = 0.117$, AN uses $5$ neighbors with a variance increase of only $0.88\%$. These results indicate that the proposed method retains most of the effective information used by global kriging while dramatically reducing the number of observations involved in prediction.

When compared to local kriging (LK) using the same number of $K$-nearest neighbors, penalized kriging consistently delivers superior predictive efficiency across all configurations. This advantage is clearly visible in Figure~\ref{fig:variance_comparison}, where PK curves systematically lie below those of LK. For instance, under exponential covariance at $PR = 0.053$, penalized kriging with $15$ neighbors yields a $2.44\%$ variance increase, whereas LK with the same neighborhood size results in a $3.61\%$ increase, corresponding to a relative improvement of $32\%$. The contrast is even more pronounced for the densest neighbor (DN): penalized kriging achieves a variance increase of $0.75\%$ using only $4$ neighbors, while LK incurs a $4.40\%$ increase with the same number of neighbors. This demonstrates that selecting neighbors based solely on proximity leads to substantial redundancy, whereas the ESS-based criterion identifies the most informative observations.

The superiority of penalized kriging over local kriging is consistently observed across all covariance models, practical ranges, and prediction locations. While the exponential model is used as a representative reference in several panels, similar patterns are observed for the Matérn ($\nu = 1.5$) and spherical covariance models. In particular, for the Matérn model, penalized kriging often yields comparable or larger relative improvements at moderate and large practical ranges. The spherical covariance highlights the effect of compact support: although both methods perform well at small practical ranges, penalized kriging maintains its advantage as the range increases and neighbor inclusion becomes more discrete. Across prediction locations, the densest neighbor (DN) benefits most from the proposed approach, as the ESS-based selection effectively filters redundant information in highly clustered regions, while boundary locations (SN and CN) show more modest but systematic gains.

Overall, these results confirm that penalized kriging achieves an effective compromise between sparsity and prediction accuracy. The method substantially outperforms local kriging by exploiting the spatial dependence structure rather than relying solely on proximity, while closely approximating the predictive performance of global kriging at a fraction of its computational cost. By automatically adapting to the correlation structure, process smoothness, and local geometry, the proposed approach provides a principled and computationally efficient solution to the neighbor selection problem in spatial prediction.

\begin{figure}[h!]
    \centering
        \subfigure[]{\includegraphics[width=0.08\textwidth]{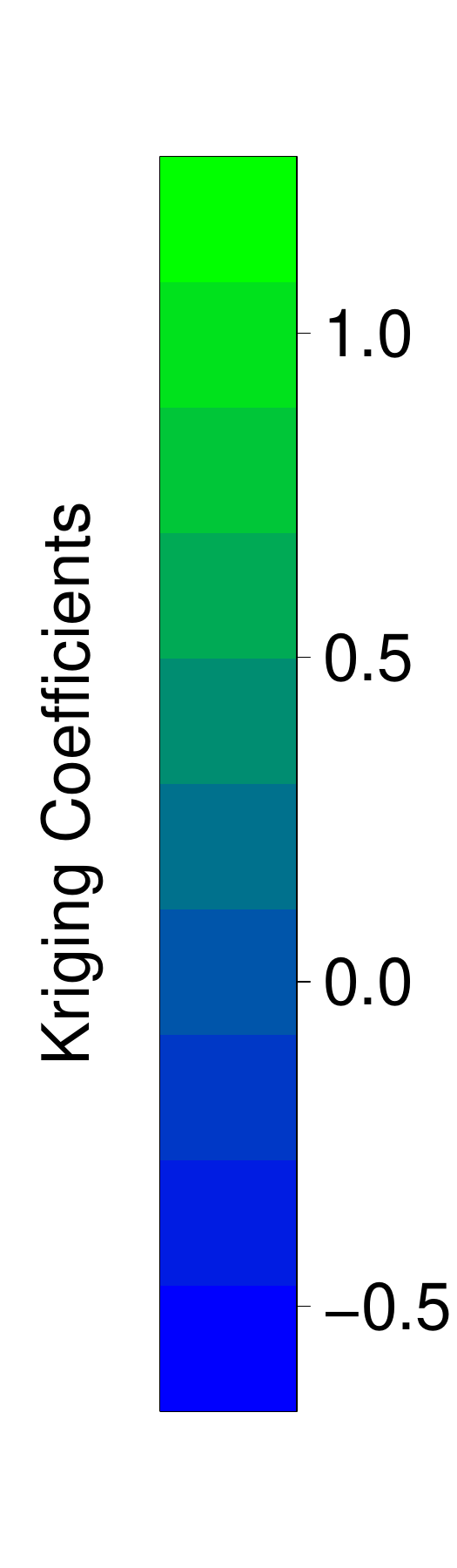}}
        \subfigure[Low PR (0.014)]{\includegraphics[width=0.28\textwidth]{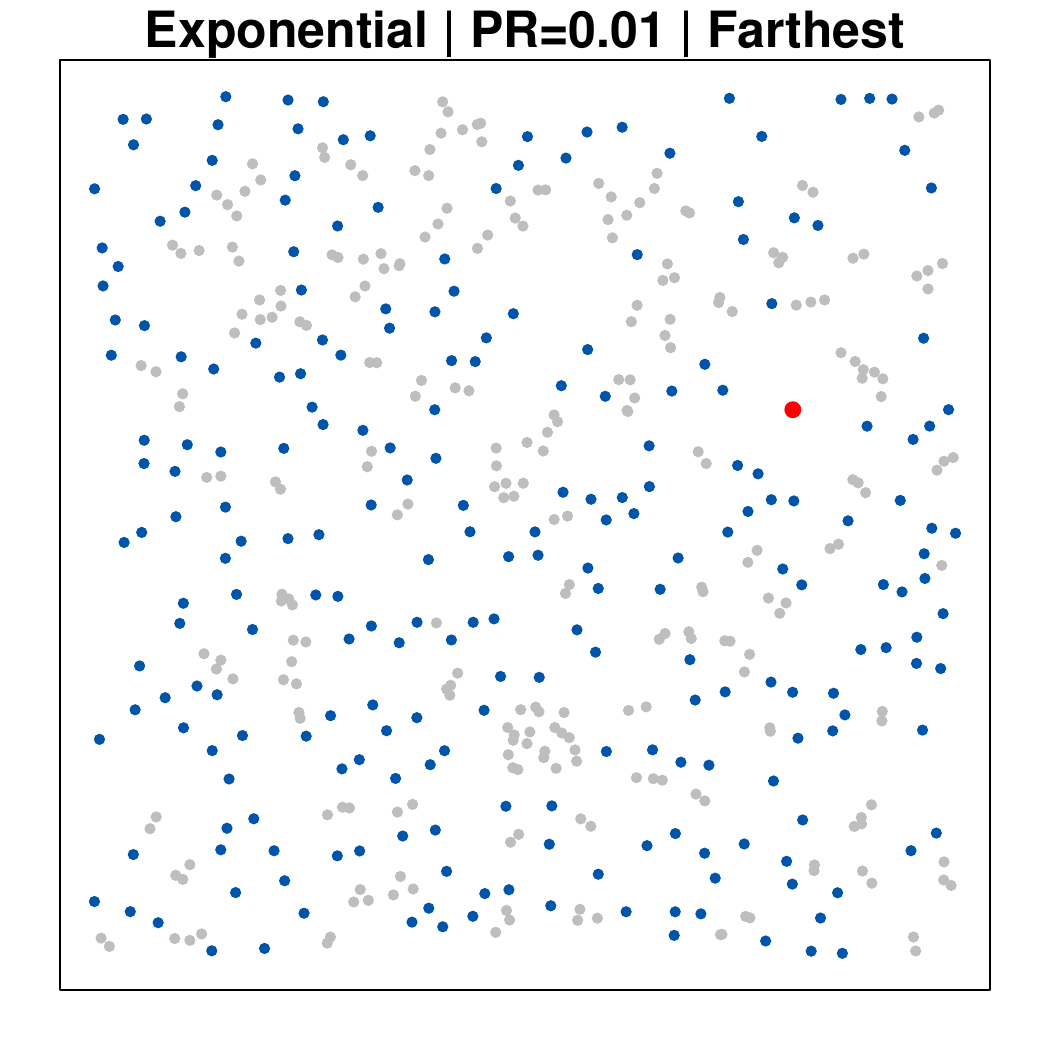}}
        \subfigure[Med PR (0.104)]{\includegraphics[width=0.28\textwidth]{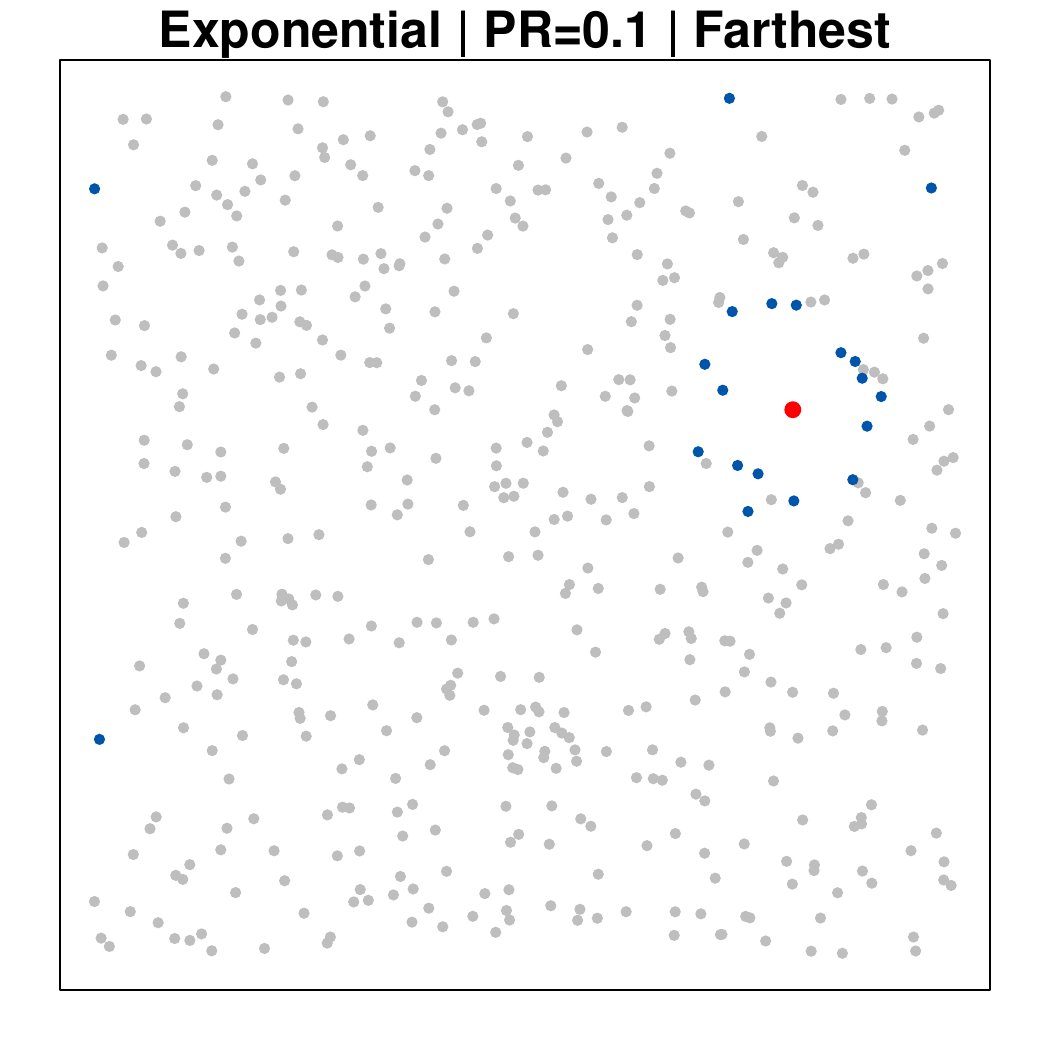}}
        \subfigure[Lar PR (0.251)]{\includegraphics[width=0.28\textwidth]{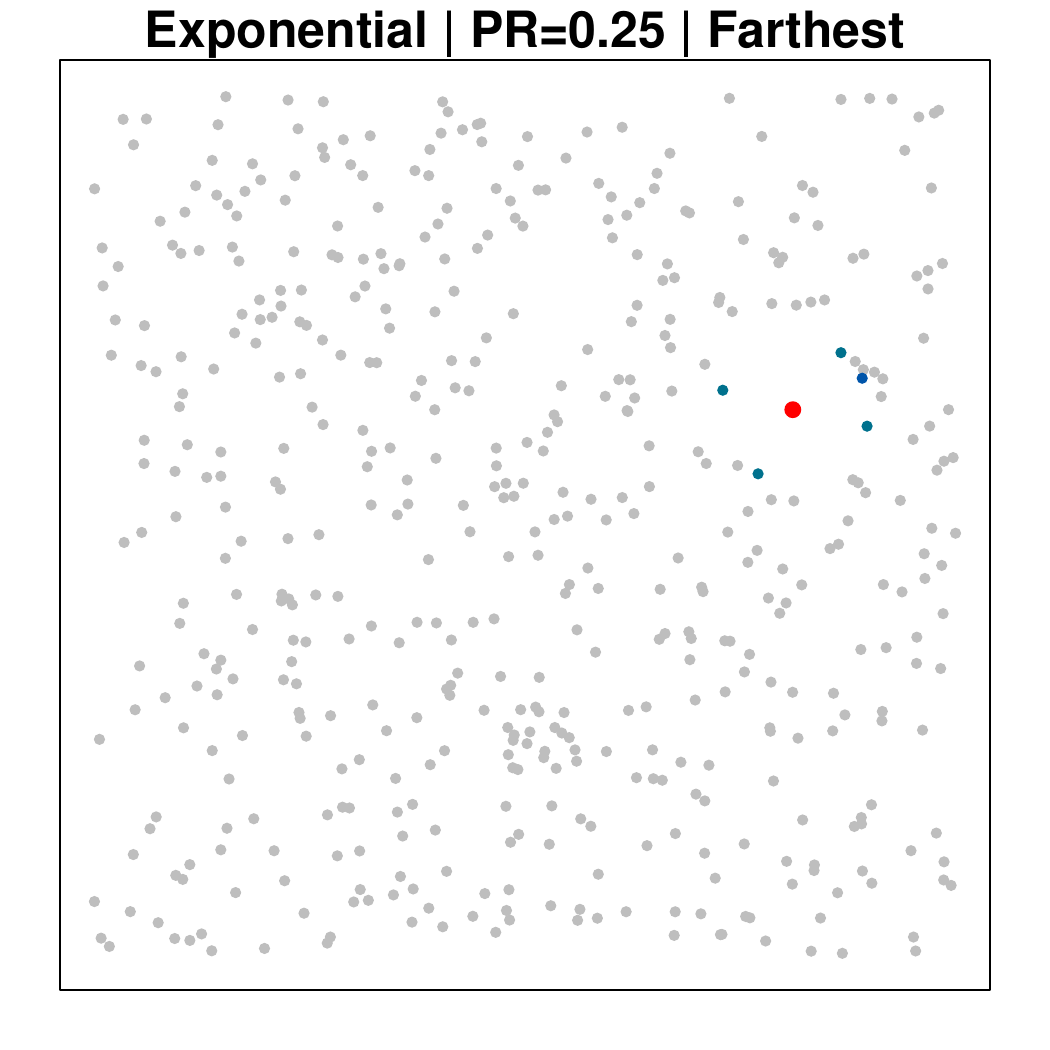}}
    \caption{Spatial neighbor selection for the farthest neighbor (FN) location under exponential covariance across three practical range levels. Grey points indicate observations with zero kriging coefficients (not selected), colored points show selected neighbors with color intensity indicating coefficient magnitude (see legend, blue = small, green = large), and the red point marks the prediction location. As the practical range increases from left to right, the selected neighborhood becomes progressively sparser. }
    \label{fig:spatial_selection_example}
\end{figure}

To illustrate how the method adapts spatially, Figure~\ref{fig:spatial_selection_example} visualizes the selected neighbors and their coefficient magnitudes for the farthest neighbor (FN) location under exponential covariance at three practical range levels. The spatial patterns reveal several key features of the adaptive selection mechanism.
First, the selected neighbors form a spatially coherent set surrounding the prediction location, rather than being scattered randomly across the domain. Second, the coefficient magnitudes (indicated by color intensity) decay smoothly with distance from the prediction location, with nearby observations receiving larger weights (green) and more distant observations receiving smaller weights (blue) before being eliminated entirely. This spatial coherence indicates that the LASSO penalty, combined with the ESS-based criterion, effectively identifies the most informative subset of neighbors while respecting the underlying spatial correlation structure.

However, this analysis also highlights why the proposed selection method differs from nearest-neighbor approaches. Although there is a clear tendency to select observations surrounding the prediction location, the method may also retain informative points located farther away, when they contribute non-redundant information. This behavior is consistently observed across all covariance models, practical ranges, and prediction locations; complete spatial coefficient visualizations are provided in Appendix~\ref{appendix:spatial_coeffs}.

\section{Data Analysis}\label{section:datanalysis}

We illustrate the practical utility of our penalized kriging methodology through two real-data applications that span different spatial scales. The Jura dataset (359 locations) represents a moderate-scale problem where global kriging remains computationally feasible, allowing us to assess our method's accuracy against the theoretical optimum. The COBE Sea Surface Temperature dataset (43,799 locations) represents a large-scale problem where global kriging becomes computationally prohibitive, demonstrating the method's scalability and practical necessity. Together, these applications illustrate how penalized kriging adapts automatically to different spatial scales, correlation structures, and prediction scenarios.

\subsection{Jura Dataset}

\begin{figure}
    \centering
    \subfigure[Chromium observation from the Jura region]{ \includegraphics[scale = 0.35]{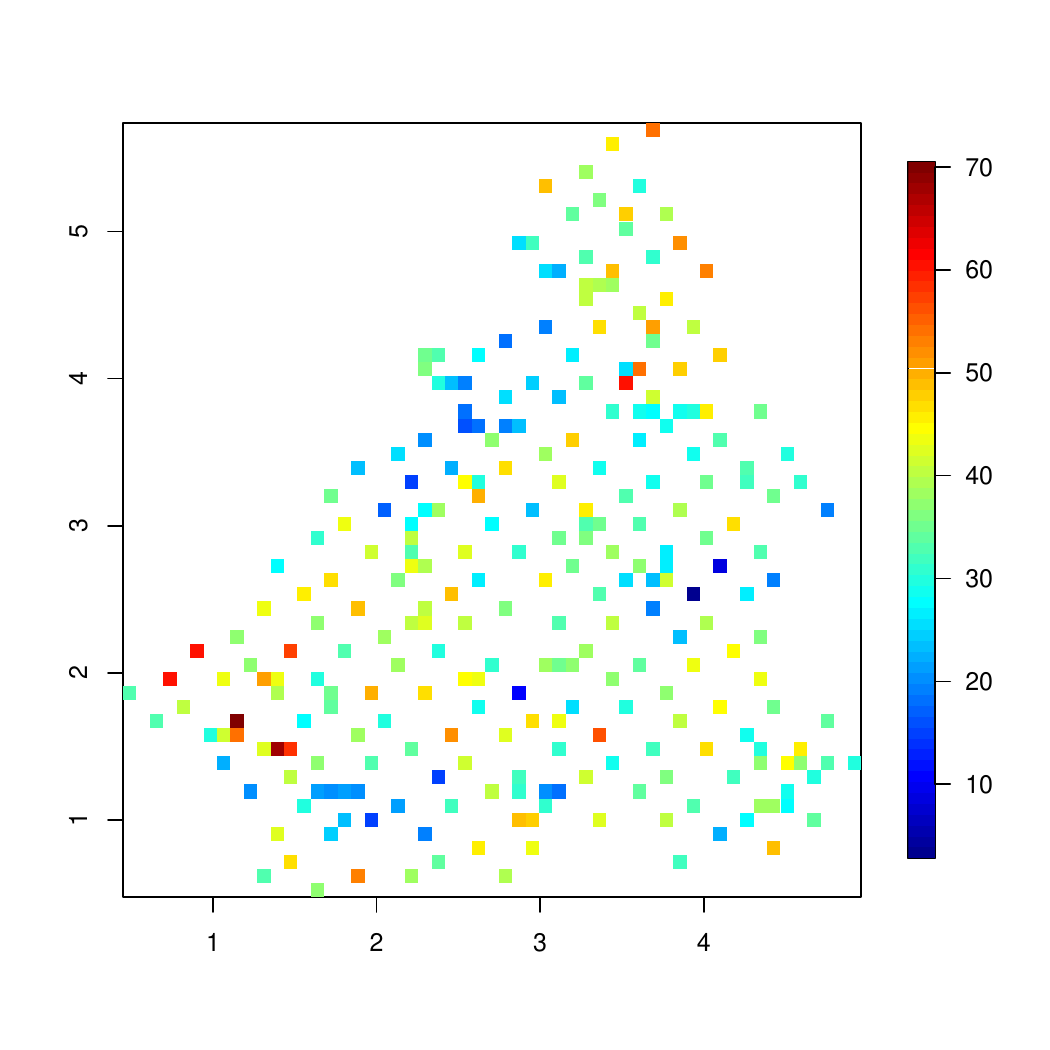} \label{fig:dataset1:a}}  
    \subfigure[Empirical variogram and fitted variogram]{ \includegraphics[scale = 0.35]{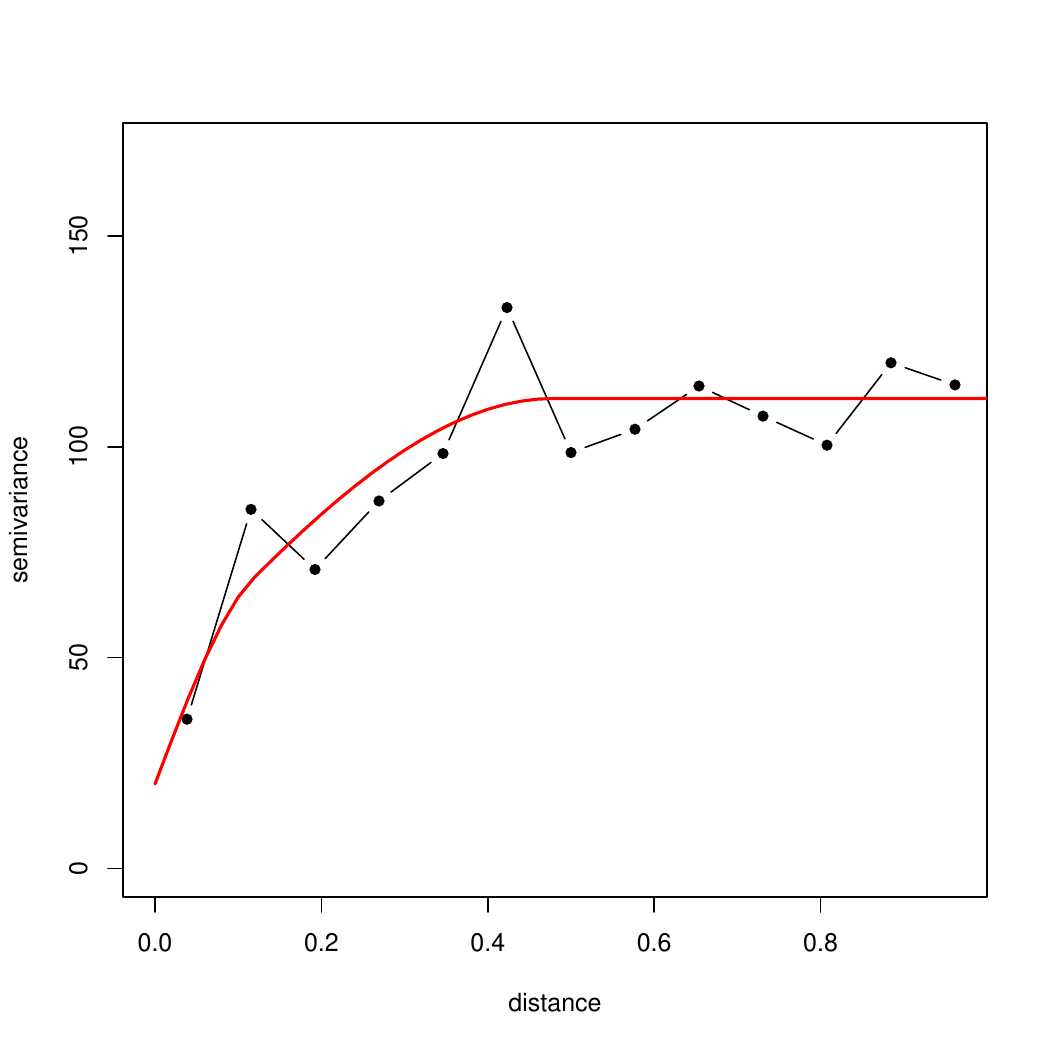} \label{fig:dataset1:b}}
    \caption{Left: Chromium observation from the Jura region, right: estimated (black dots and fited variogram (red line)}
    \label{fig:dataset1}    
\end{figure}

We evaluate the performance of our penalized kriging methodology using the Jura dataset \citep{atteia1994, webster1994, goovaerts1997}, a well-studied benchmark from the \texttt{compositions} package in R \citep{compositions} containing measurements of heavy metal concentrations at 359 spatial locations in the Swiss Jura region. We focus on chromium (Cr) concentrations, which exhibit substantial spatial variation requiring careful neighbor selection for accurate prediction, as discussed by \citet{emery2009kriging} in the context of optimal neighborhood selection. This dataset is small enough that global kriging is computationally feasible, allowing direct comparison of our method's accuracy and efficiency against the theoretical optimum. The data is depicted in Figure \ref{fig:dataset1:a}.

The empirical variogram exhibits clear spatial structure at two distinct scales, motivating a nested spherical covariance model
\begin{equation}
C(h) = C_1(h; \theta_1) + C_2(h; \theta_2) + \tau^2 \mathbb{I}_{h=0},
\end{equation}
where $C_i(h; \theta_i)$ denotes the spherical covariance function with partial sill $\sigma_i^2$ and range $\phi_i$, and $\tau^2$ represents measurement error (nugget effect). The first component captures short-range variation (local geological features) while the second captures regional trends. The model was fitted by weighted least squares \citep{cressie1993statistics}, minimizing
$$\sum_{j=1}^{n_b} w_j \left(\hat{\gamma}(h_j) - \gamma(h_j; \bm{\theta})\right)^2,$$
where weights $w_j = n_j$ are based on the number of pairs in each distance bin. The estimated parameters are $\hat{\sigma}_1^2 = 50.0$, $\hat{\phi}_1 = 0.15$, $\hat{\sigma}_2^2 = 50.0$, $\hat{\phi}_2 = 0.65$, and $\hat{\tau}^2 = 20.0$, yielding a total sill of approximately 120 units. The estimated variogram is depicted in Figure \ref{fig:dataset1:b}.

To assess spatial prediction performance, we constructed a regular $50 \times 50$ grid over the bounding box of the observation locations. After removing 1,023 grid points falling outside the convex hull of the observations (to avoid excessive extrapolation), we obtain 1,477 prediction locations where we compare three kriging approaches. First, Ordinary Kriging (GK) uses all 359 observations, representing the theoretical optimum. Second, penalized kriging (PK) employs sparse kriging with automatic neighbor selection via adaptive LASSO penalization. Third, local kriging (LK) uses the same amount of nearest neighbors selected by PK.

For the PK, the adaptive LASSO method employed penalty weights $w_j = 1/|\hat{\lambda}_j|$ for $j = 2, \ldots, n$, where $\hat{\lambda}_j$ are the ordinary kriging weights. The optimal penalty parameter $\eta$ was selected using the adaptive grid search algorithm described in Section \ref{section:tuning_parameter} over the range $[\eta_{\min}, \eta_{\max}]$, where $\eta_{\max} = \max_j |\hat{\lambda}_j|$ and $\eta_{\min} = \eta_{\max} \times 10^{-9}$. The grid search employed an initial coarse grid of 150 points, followed by iterative refinement with 50 additional points in promising regions.

\begin{figure}[htbp]
    \centering
    \begin{minipage}[b]{0.24\textwidth}
        \centering
        \includegraphics[width=\textwidth]{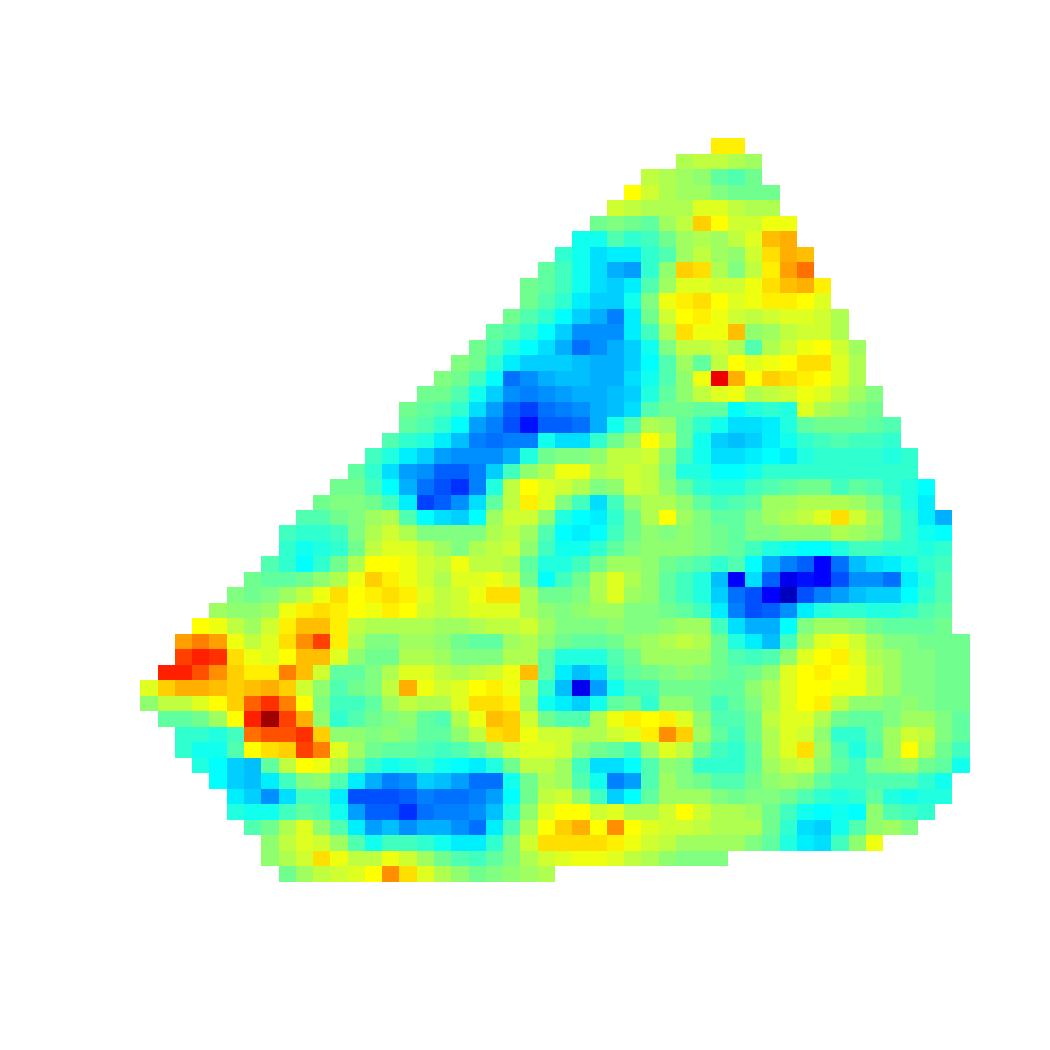}
    \end{minipage}
    \hfill
    \begin{minipage}[b]{0.24\textwidth}
        \centering
        \includegraphics[width=\textwidth]{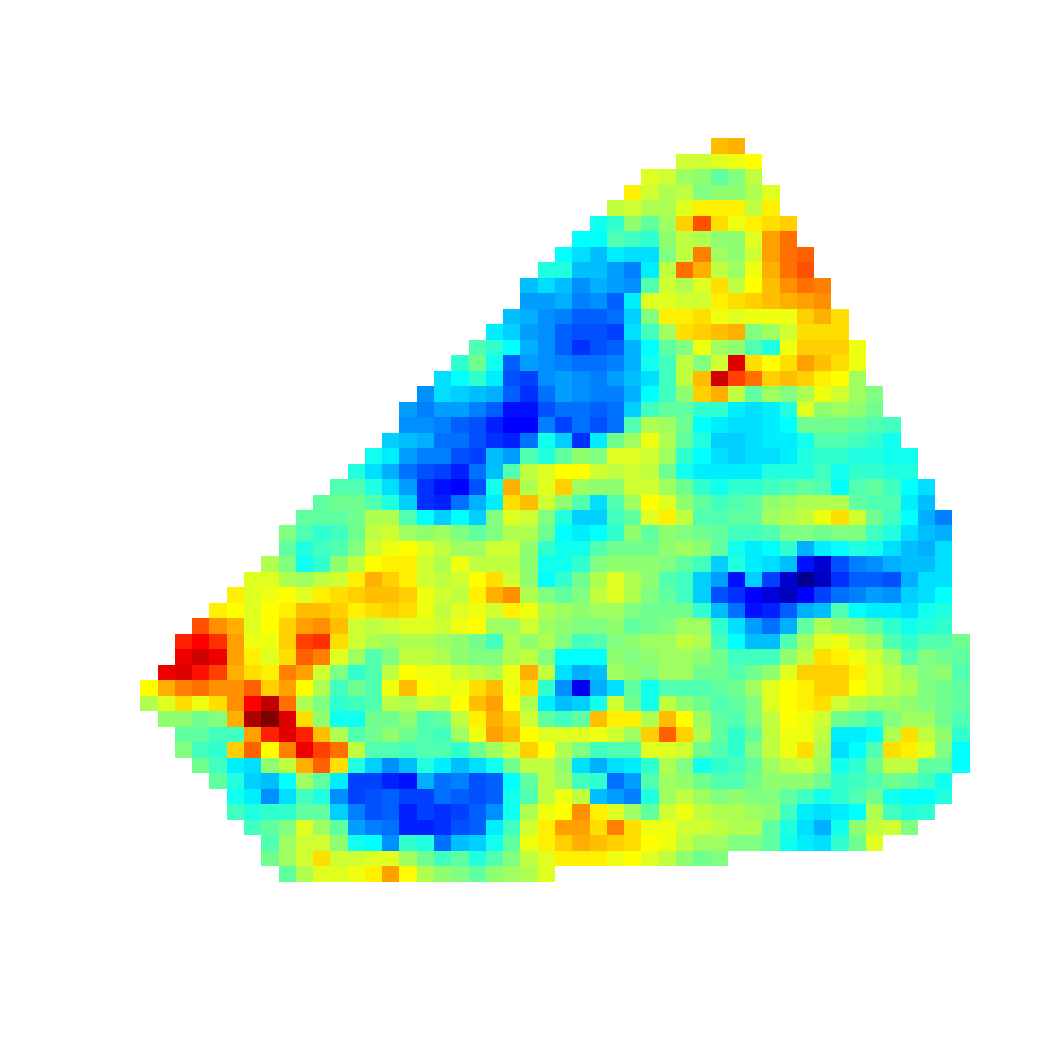}
    \end{minipage}
    \hfill
    \begin{minipage}[b]{0.24\textwidth}
        \centering
        \includegraphics[width=\textwidth]{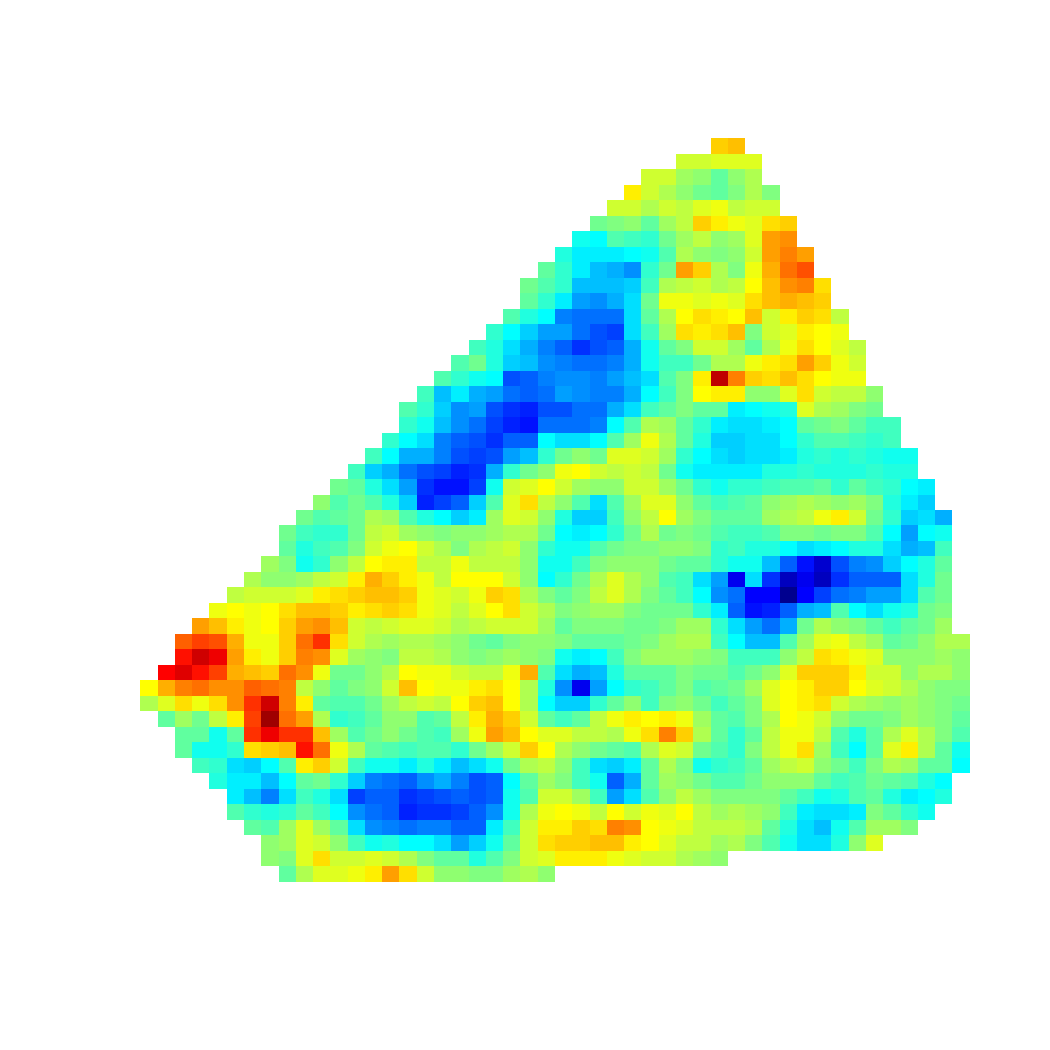}
    \end{minipage}
    \hfill
    \begin{minipage}[b]{0.075\textwidth}
        \centering
        \includegraphics[width=\textwidth,height=3cm]{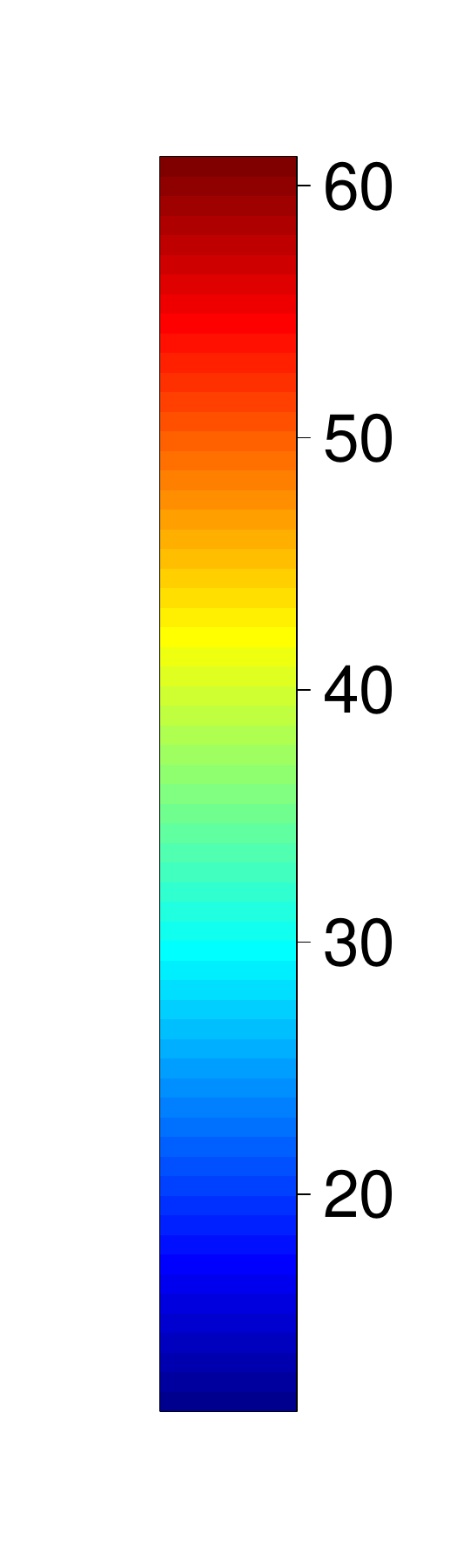}
    \end{minipage}
    
    \vspace{0.5cm}
    
    \begin{minipage}[b]{0.24\textwidth}
        \centering
        \includegraphics[width=\textwidth]{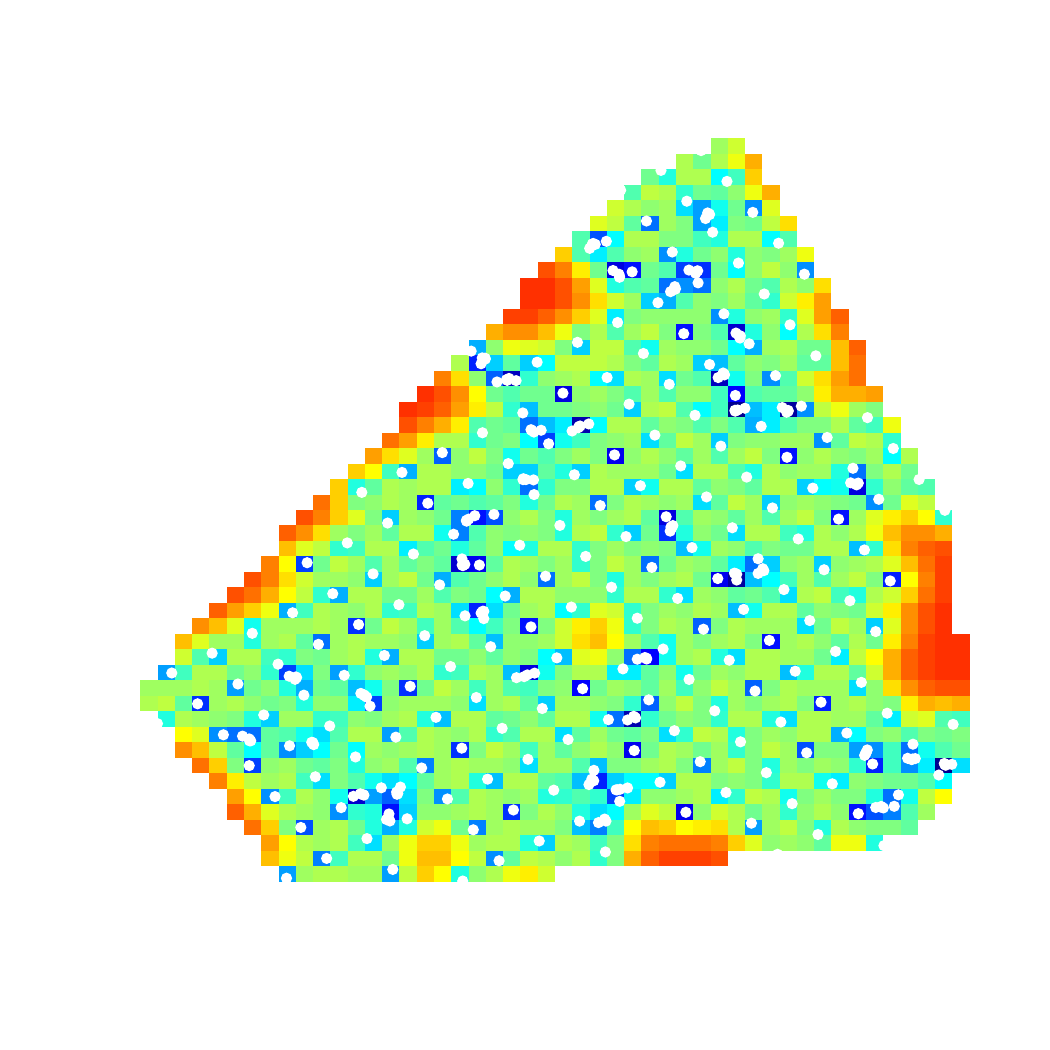}
    \end{minipage}
    \hfill
    \begin{minipage}[b]{0.24\textwidth}
        \centering
        \includegraphics[width=\textwidth]{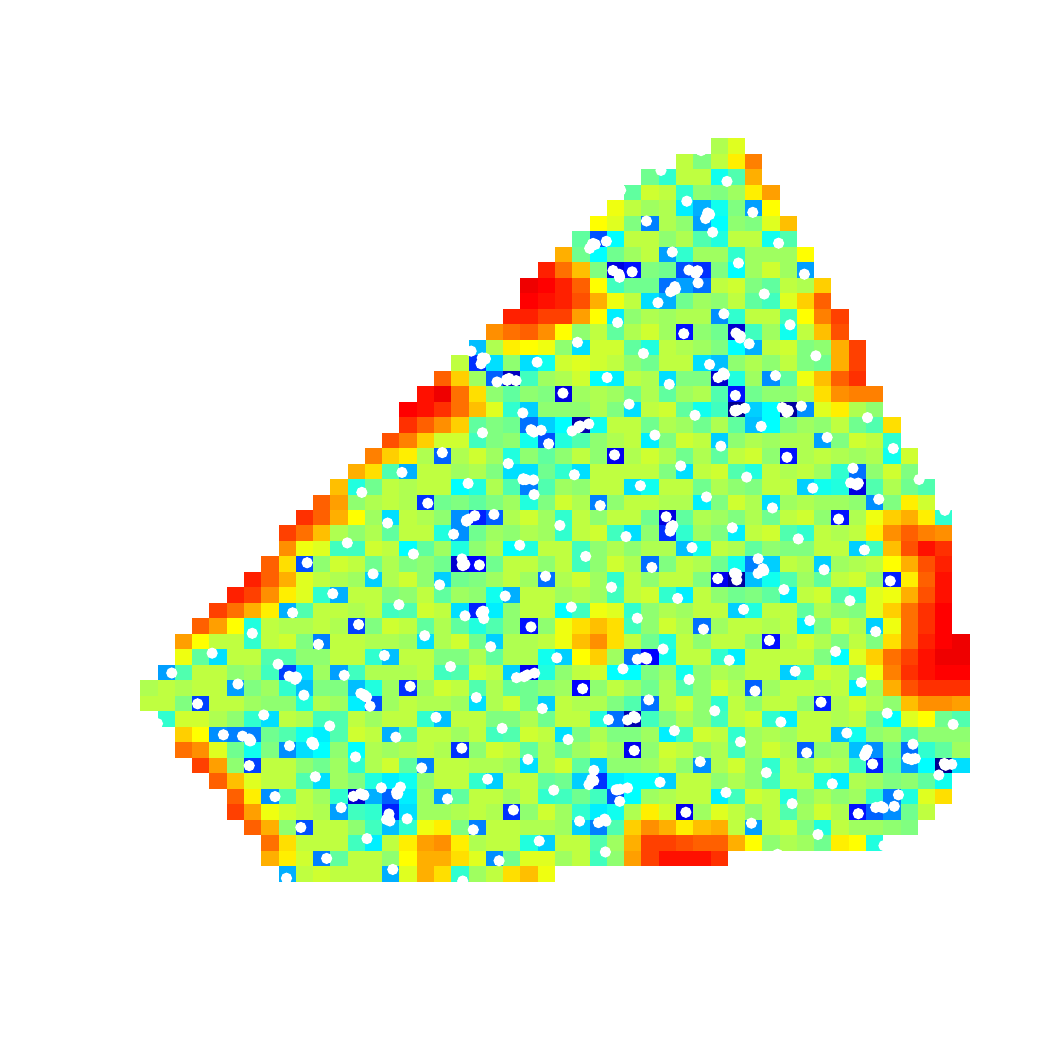}
    \end{minipage}
    \hfill
    \begin{minipage}[b]{0.24\textwidth}
        \centering
        \includegraphics[width=\textwidth]{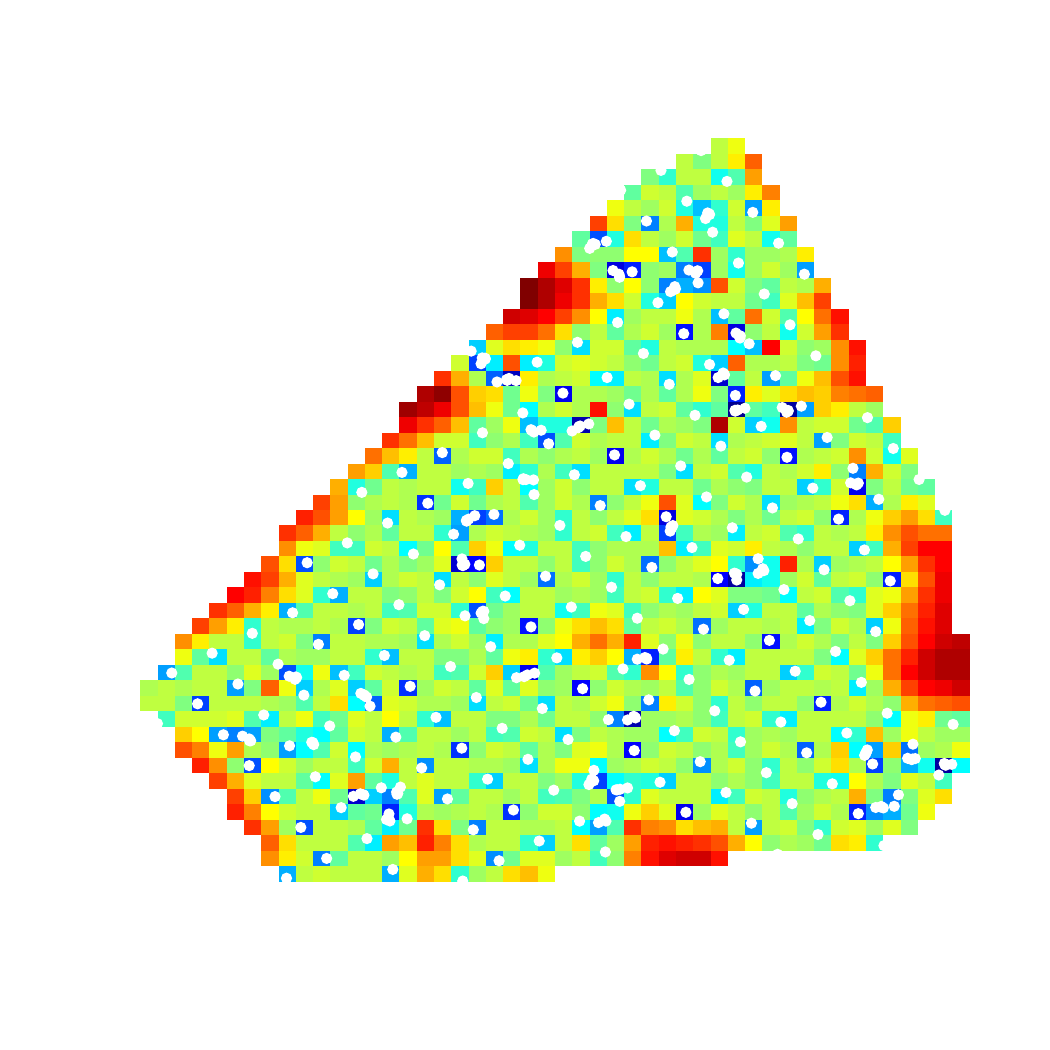}
    \end{minipage}
    \hfill
    \begin{minipage}[b]{0.075\textwidth}
        \centering
        \includegraphics[width=\textwidth,height=3cm]{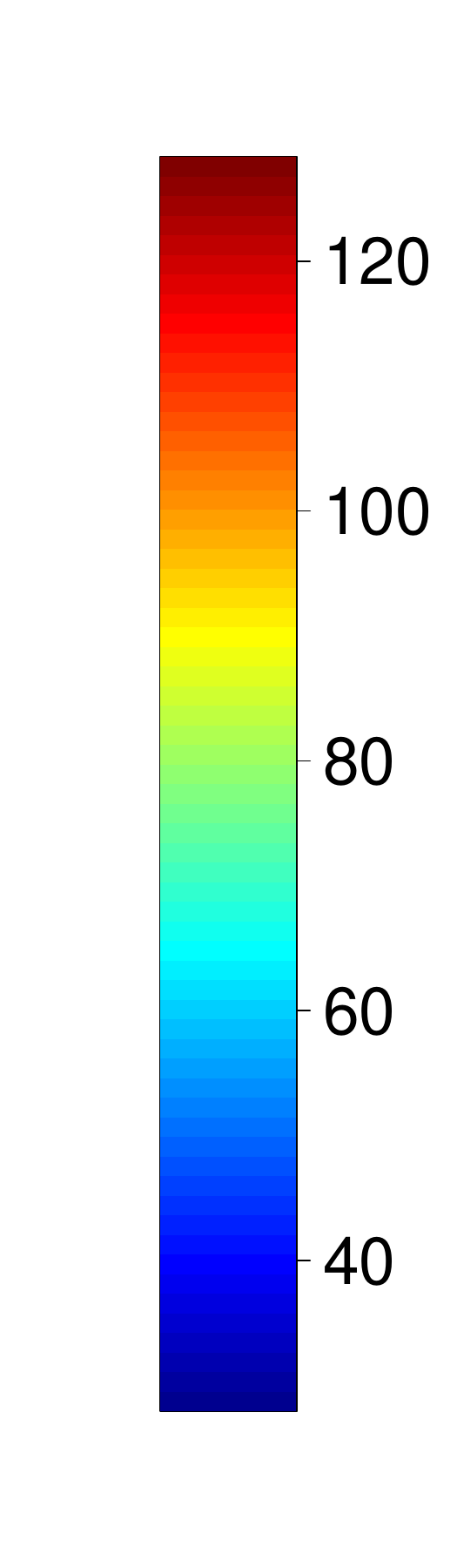}
    \end{minipage}
    
    \caption{Prediction maps (top row) and kriging variance maps (bottom row) for the Jura dataset using Ordinary Kriging (left column), penalized kriging (center column), and Local Kriging (right column).
    Colors represent relative magnitude (blue = low, red = high). Prediction and variance maps use separate continuous color scales, and scales are not comparable across rows. }
    \label{fig:jura_dataset_results}    
\end{figure}

Figure \ref{fig:jura_dataset_results} presents prediction maps (top row) and kriging variance maps (bottom row) for the three methods. Visual inspection reveals strong agreement in the spatial features of predicted chromium concentrations, with all three methods successfully capturing the main features: elevated concentrations in the northeastern portion of the study region and lower values in the southwest, large variance where there are no points, low variance close to an observed location.

\begin{figure}[htbp]
    \centering
    \begin{minipage}[c]{0.6\textwidth}
        \raggedleft
        \includegraphics[width=1.2\textwidth]{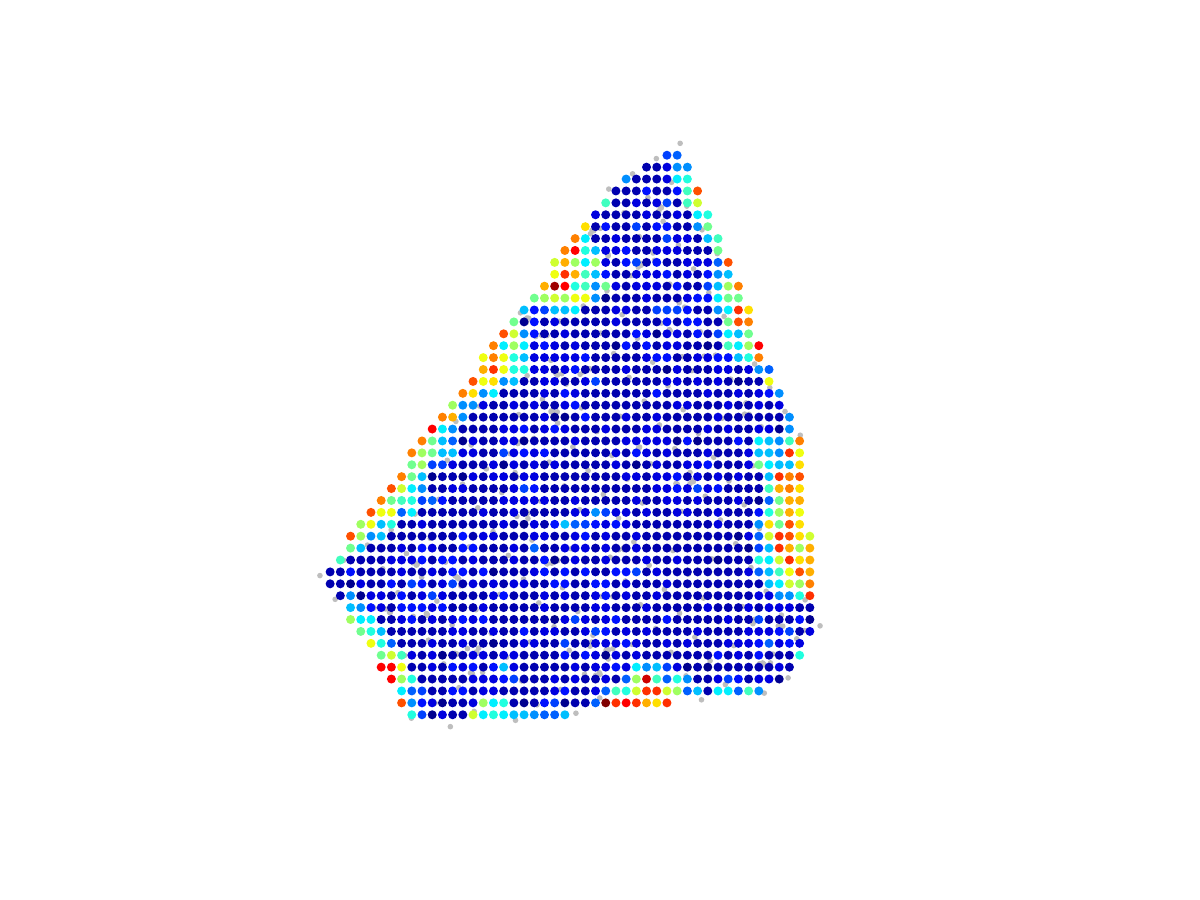}
    \end{minipage}
    \hfill
    \begin{minipage}[c]{0.39\textwidth}
        \raggedright
        \includegraphics[width=0.25\textwidth]{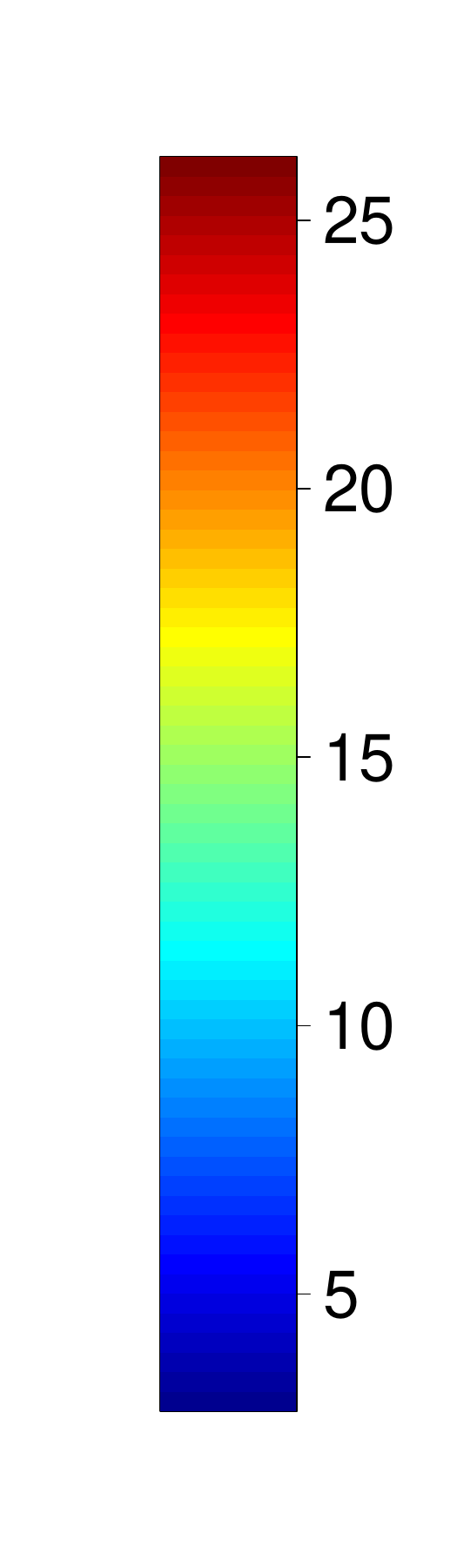} 
    \end{minipage}
    \hfill\vspace{-10mm}
    
    \caption{Spatial distribution of the number of neighbors selected by penalized kriging for the Jura dataset.
    The selected neighborhood size ranges from as few as 3 neighbors to 25 neighbors. In interior regions, the method typically selects around 5 neighbors, while near domain boundaries the number increases and varies between 10 and 25 neighbors, illustrating the method’s automatic adaptation to the local spatial configuration. }
    \label{fig:jura_selected_nn}
\end{figure}

The number of neighbors selected by penalized kriging varies spatially (Figure \ref{fig:jura_selected_nn}), ranging from approximately 3-5 neighbors in well-connected interior regions to 10-25 neighbors near boundaries and in areas with complex local variation. This spatial adaptation demonstrates the advantage of  to automatically adjust neighborhood size based on local geometry and correlation structure, unlike the nearest neighbor approach.

In the variance maps we can see some small differences. First, GK (Figure \ref{fig:jura_dataset_results}, bottom left) achieves the lowest prediction variance in all the locations (see Figure \ref{fig:jura_selected_nn}). Then, PK (Figure \ref{fig:jura_dataset_results}, bottom center) reproduces the variance behaviour of GK with less spatial locations and has small variance than the LK (Figure \ref{fig:jura_dataset_results}, bottom right). The largest differences appears when the kriging is trying to predict outside In the interior of the observed spatial locations. As is depicted in the simulation study, PK tries to compensate this effect by selecting more spatial locations (see Figure \ref{fig:jura_selected_nn}). Besides, the three methods are in total agreement. 

\begin{figure}
    \centering
    \includegraphics[width=0.5\linewidth]{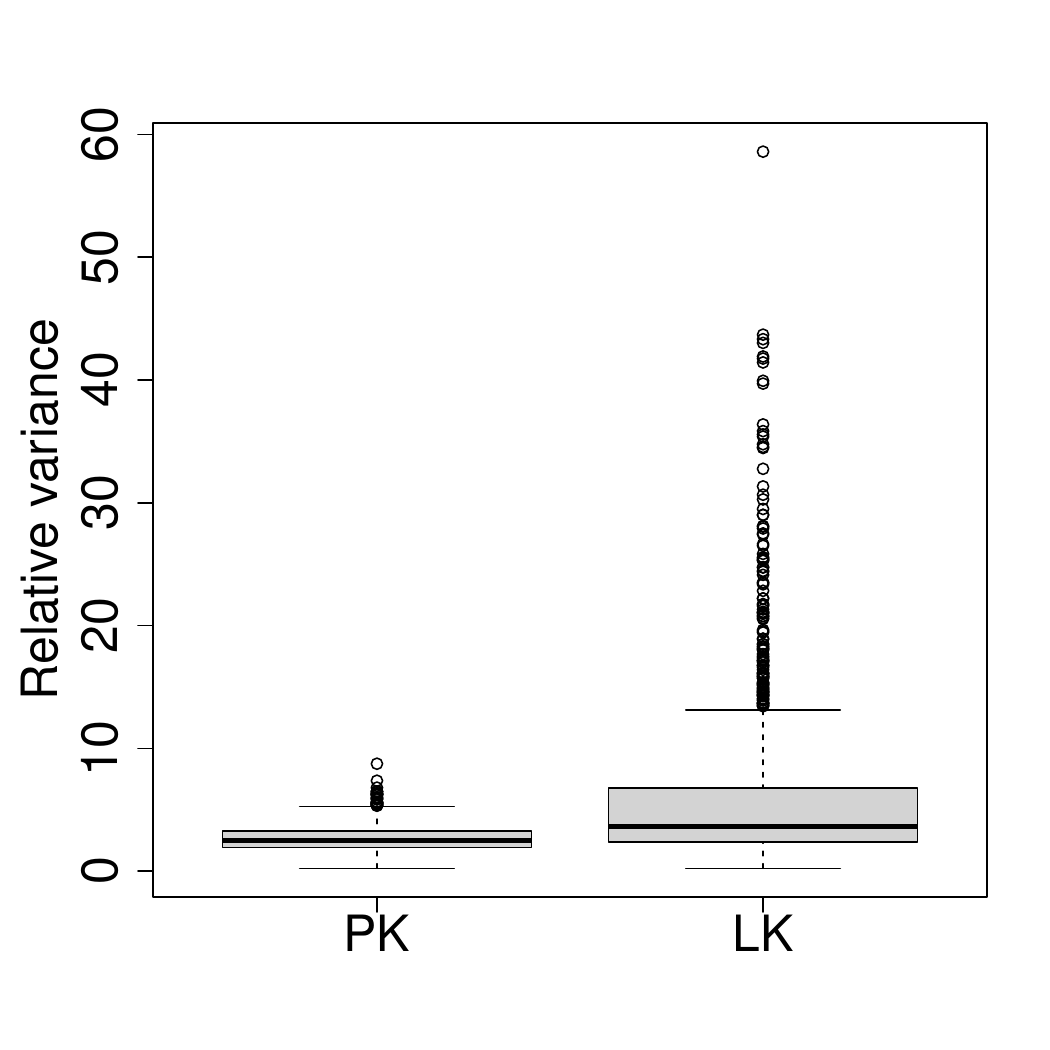}
    \caption{Boxplots of relative kriging variance increase (in \%) of penalized kriging (PK) and local kriging (LK) compared to ordinary kriging (GK) for the Jura dataset.}
    \label{fig:boxplot_rel_variance}
\end{figure}

Figure \ref{fig:boxplot_rel_variance} quantifies the variance comparison through boxplots of relative variance increase via the formula $(\sigma^2_{\text{method}} - \sigma^2_{\text{GK}})/\sigma^2_{\text{GK}}$. 
The adaptive LASSO method achieves a much better performance than the local kriging in all the locations. Indeed, adaptive LASSO variance is smaller or equal than local kriging variance in 98.5\% of the locations.

Notice than the PK achieves this performance by using an average of 6 neighbors. At some interior locations with strong local correlation, PK uses as few as 3 neighbors while maintaining variance within 5\% of GK, demonstrating that we can achieve accurate sparse solutions via intelligent neighbor selection.

\subsection{COBE Sea Surface Temperature Dataset: Large-Scale Application}

Having demonstrated the method's performance on a moderate-scale dataset where global kriging is feasible, we now examine a large-scale application where global kriging becomes computationally prohibitive. We analyze the COBE Sea Surface Temperature dataset from NOAA PSL, Boulder, Colorado \citep{ishii2005objective}, which provides monthly gridded SST analyses at $1^\circ \times 1^\circ$ resolution globally. With 43,799 spatial locations covering the world's oceans, matrix inversion for global kriging ($O(N^3) \approx 10^{14}$ operations) is infeasible on standard computing resources. This dataset provides a realistic test of our method's scalability and ability to handle large-scale spatial prediction problems. 

\begin{figure}[h]
    \centering
    \subfigure[Sea Surface Temperature mean anomalies for March 2012]{ \includegraphics[scale = 0.35]{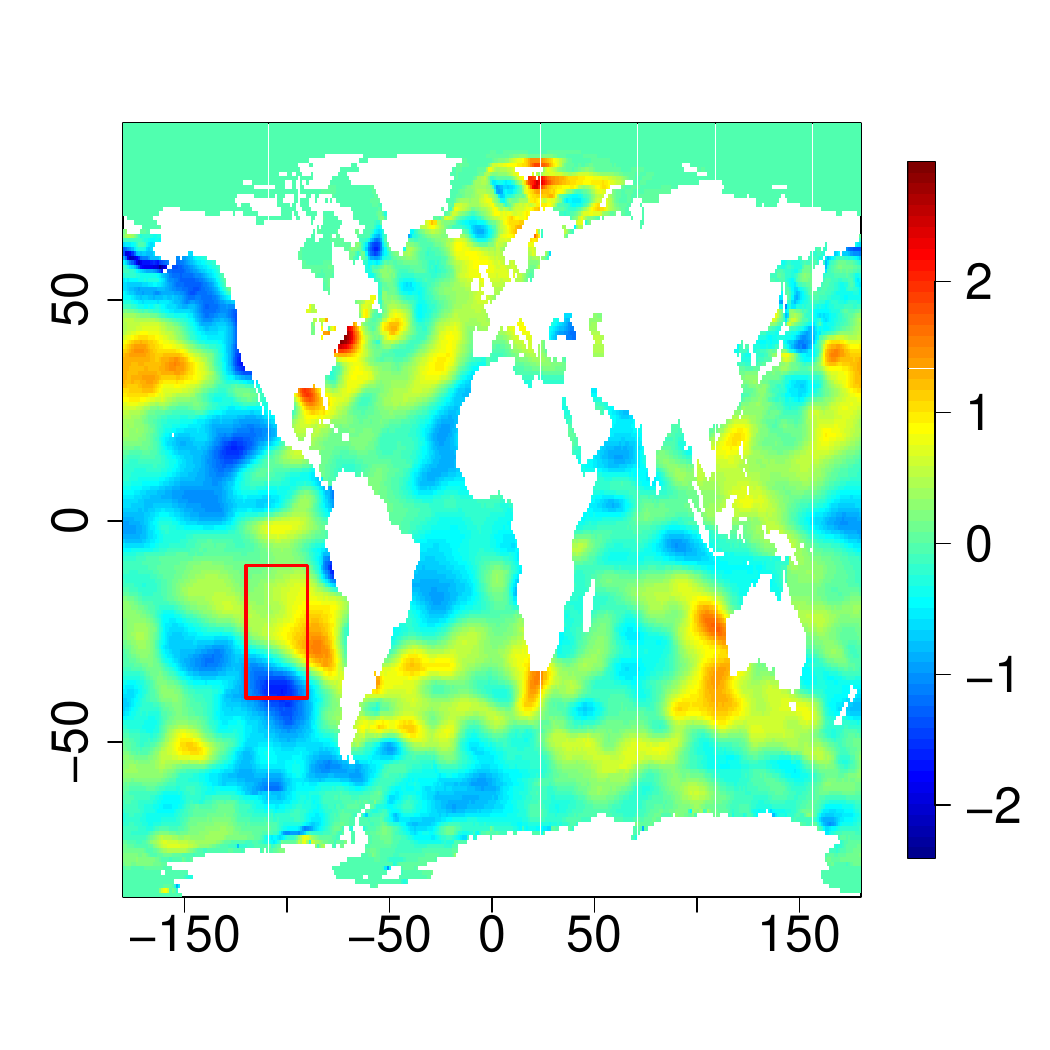} \label{fig:dataset2:a}}  
    \subfigure[Empirical variogram and fitted variogram]{ \includegraphics[scale = 0.35]{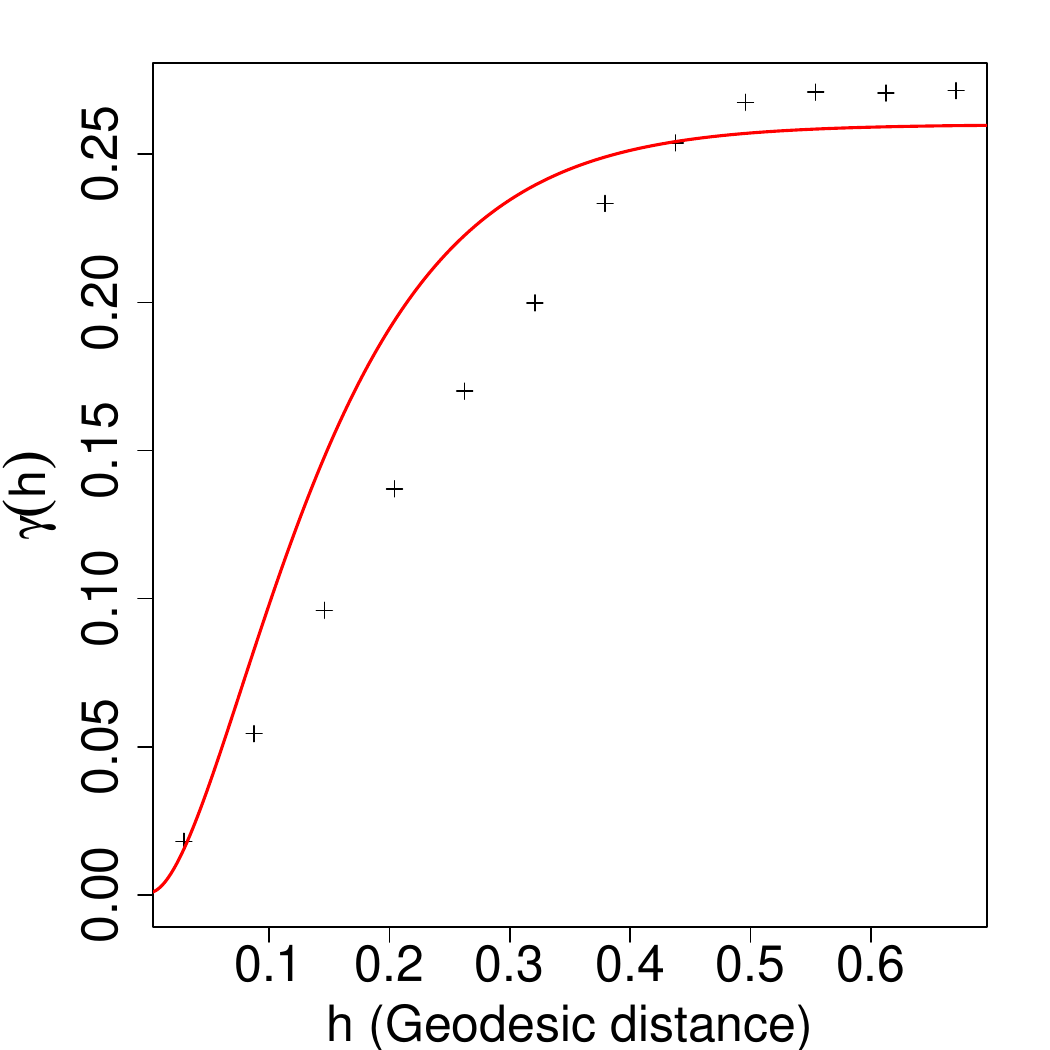} \label{fig:dataset2:b}}
    \caption{Left: Sea Surface Temperature mean anomalies for March 2012 from the COBE dataset ($N = 43,799$ locations). The red box indicates the downscaling prediction region spanning longitudes $-120°$ to $-90°$ and latitudes $-40°$ to $-10°$ in the southeastern Pacific. Right: Empirical variogram (crosses) together with the fitted Matérn model with $\nu = 1.5$ (solid line), estimated via composite likelihood.}
    \label{fig:dataset2}    
\end{figure}

We focus on mean temperature anomalies for March 2012 (Figure \ref{fig:dataset2}a), which exhibit complex spatial patterns including the remnants of a La Niña event in the tropical Pacific. We perform a downscaling analysis over a rectangular region in the southeastern Pacific (Figure \ref{fig:dataset2}a, red box), spanning longitudes from $-120^\circ$ to $-90^\circ$ and latitudes from $-40^\circ$ to $-10^\circ$. This region was chosen for several reasons: (1) it exhibits substantial spatial variation in temperature anomalies, (2) it includes both near-coastal and open-ocean areas with different correlation structures, and (3) it is large enough ($30^\circ \times 30^\circ$) to demonstrate scalability while allowing detailed visualization. A prediction grid with resolution $0.25^\circ \times 0.25^\circ$ was constructed within this region, yielding 14,650 prediction locations—a four times increase in spatial resolution compared to the original $1^\circ \times 1^\circ$ data. This represents a realistic downscaling scenario where coarse global model output is refined for regional applications.

We follow \citet{alegria2024assessing} and we fit a Gaussian random field with constant mean and Matérn covariance function with fixed smoothness parameter $\nu = 1.5$ and nugget effect, using chordal distance \citep{guinness2016isotropic} to account for the Earth's spherical geometry. The covariance model is $C(h) = \sigma^{2}\mathcal{M}(h; \phi, 1.5) + \tau^{2}\mathbb{I}(h = 0)$ where $\mathcal{M}$ denotes the Matérn correlation function and $h$ is the chordal distance between locations. 

Due to the massive scale of this dataset, we cannot use maximum likelihood to estimate covariance parameters. Instead, we employ composite likelihood estimation \citep{bevilacqua2012estimating, caamano2024nearest}. Specifically, we use a pairwise likelihood approach with three nearest neighbors on all the available data (implemented in \cite{GeoModels}). The estimated parameters are  $\hat{\mu}_{0} = 0.002$, $\hat{\sigma}^{2} = 0.2592$, $\hat{\phi} = 0.0767$, $\hat{\tau}^{2} = 0.0008$ with practical range of approximately $0.363$ radians (about 2,312 km on Earth's surface), and a nugget-to-sill ratio of $\hat{\tau}^2/(\hat{\sigma}^2+\hat{\tau}^2) \approx 0.003$, indicating that measurement error accounts for roughly $0.3\%$ of total variation. Figure \ref{fig:dataset2}b shows the empirical variogram together with the fitted model, demonstrating reasonable agreement. The estimated effective sample size for the full COBE dataset is $n^{\text{ESS}} = 67.15$ despite $N = 43,799$ total locations, revealing a high level of information redundancy ($99.85\%$).
This result highlights the severe information redundancy present in large spatial datasets and underscores the practical relevance of adaptive neighborhood selection strategies for scalable kriging.

The procedure for implementing penalized kriging on a very large dataset is as follows. For each prediction location $\mathbf{s}_0$, we first identify the $K = 300$ nearest neighbors from the full dataset of 43,799 locations, accounting for spherical geometry. The choice $K = 300$ was motivated by the estimated ESS and corresponds to a conservative and practical upper bound on the candidate neighborhood size, adopted to ensure that the candidate set contains all informative observations rather than as a strict selection rule.

 The adaptive weights used in penalized kriging (PK) were computed as $w_i= 1/|\hat{\lambda}_i|$, where $\hat{\lambda}_i$ denote the kriging weights obtained from ordinary kriging using the 300 nearest neighbors (GK-300). The penalty parameter $\eta$ was selected using the criterion defined in \eqref{eq:eta_selection}.

\begin{figure}[h!]
    \centering
       \subfigure[Observed sea surface temperature field]{\includegraphics[width=0.45\textwidth]{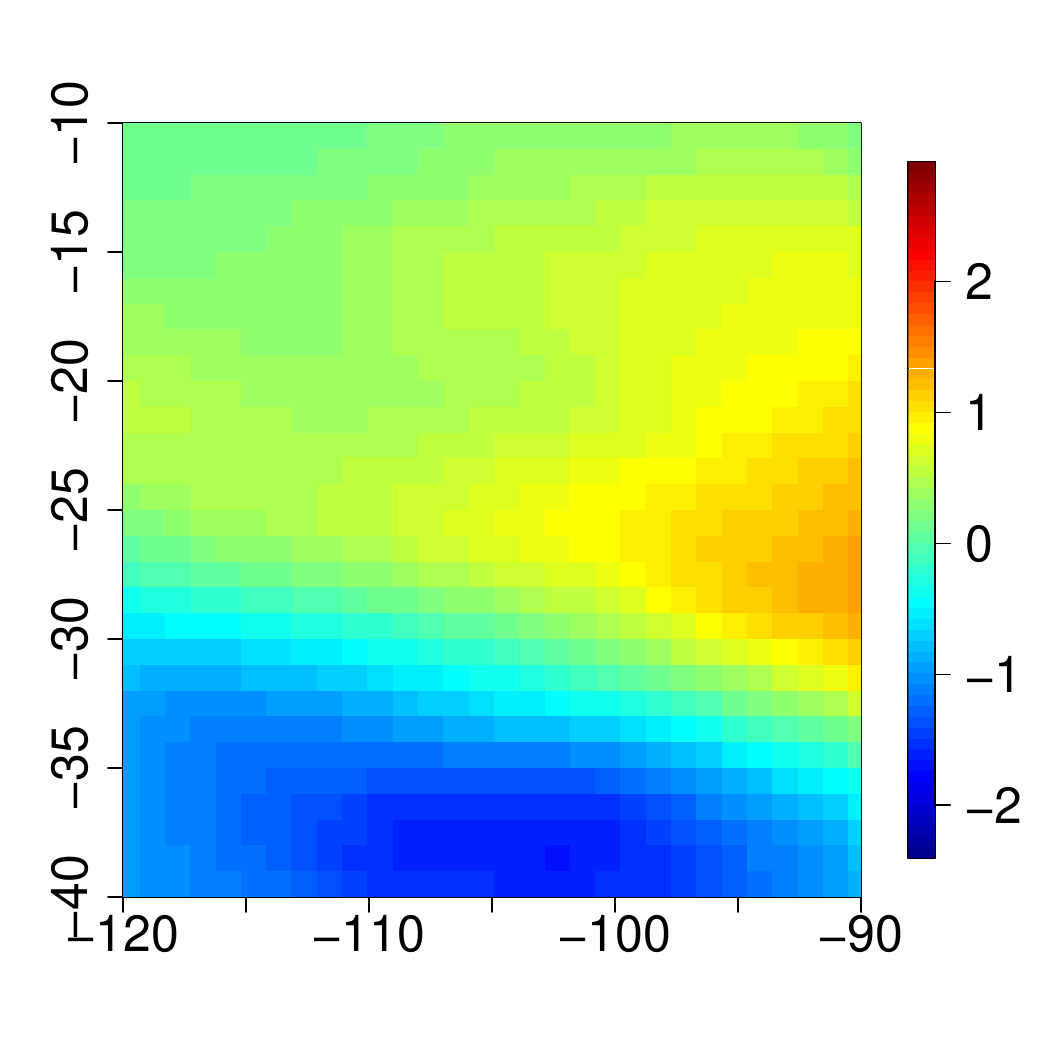}\label{fig:cobe_predictions:a}}
       \subfigure[downscaled field obtained using penalized kriging]{\includegraphics[width=0.45\textwidth]{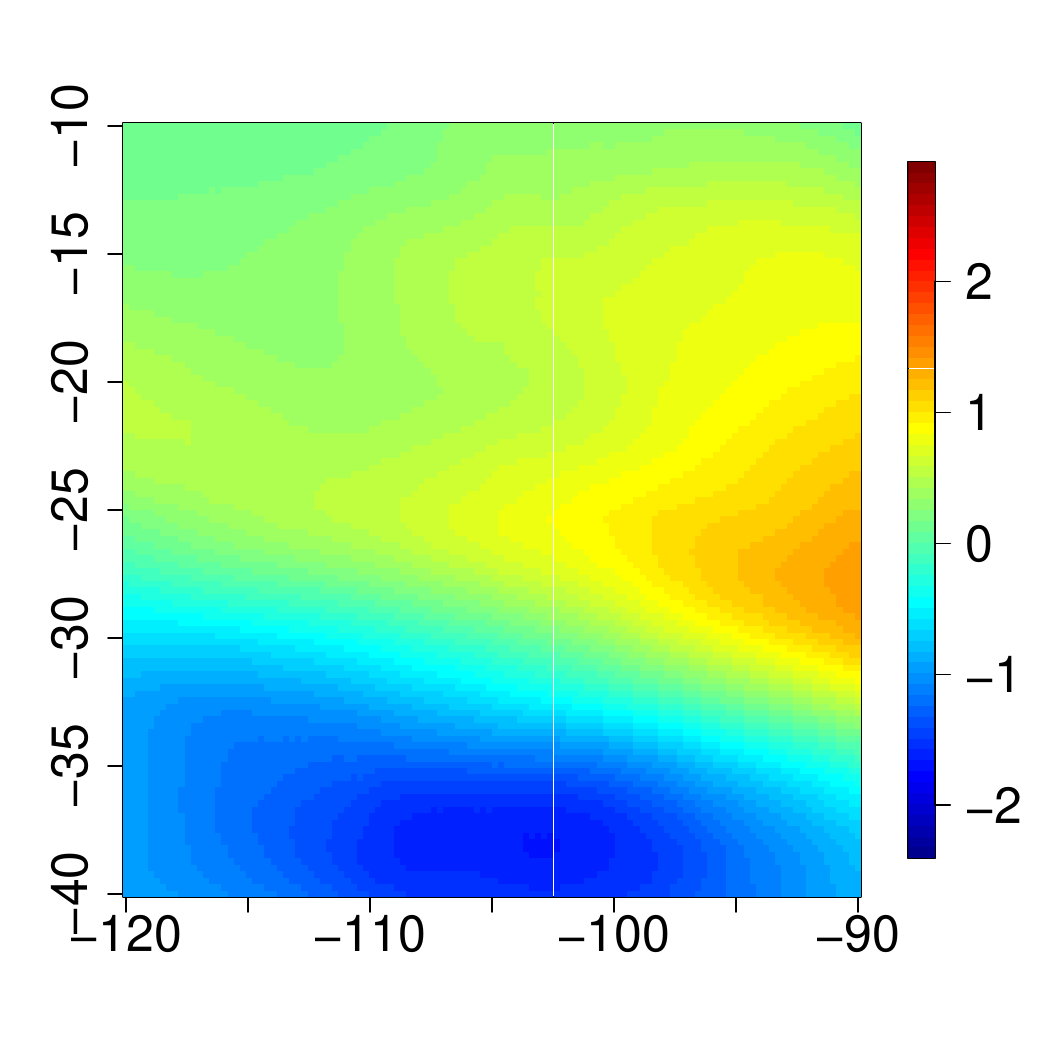}\label{fig:cobe_predictions:b}}
       
       \subfigure[Kriging variance $\times 10^{-3}$]{\includegraphics[width=0.475\textwidth]{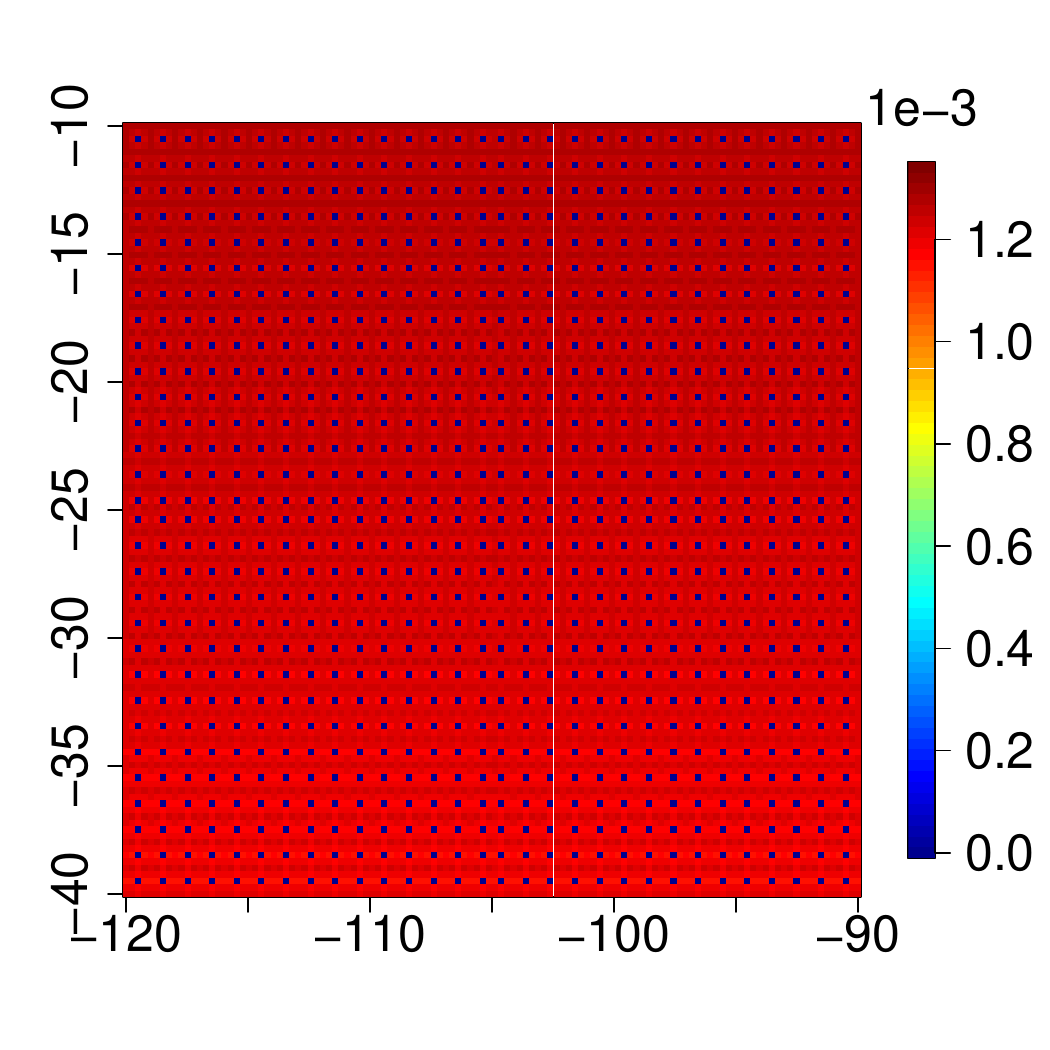}\label{fig:cobe_predictions:c}}
       \subfigure[Distribution of selected neighborhood sizes]{\includegraphics[width=0.45\linewidth]{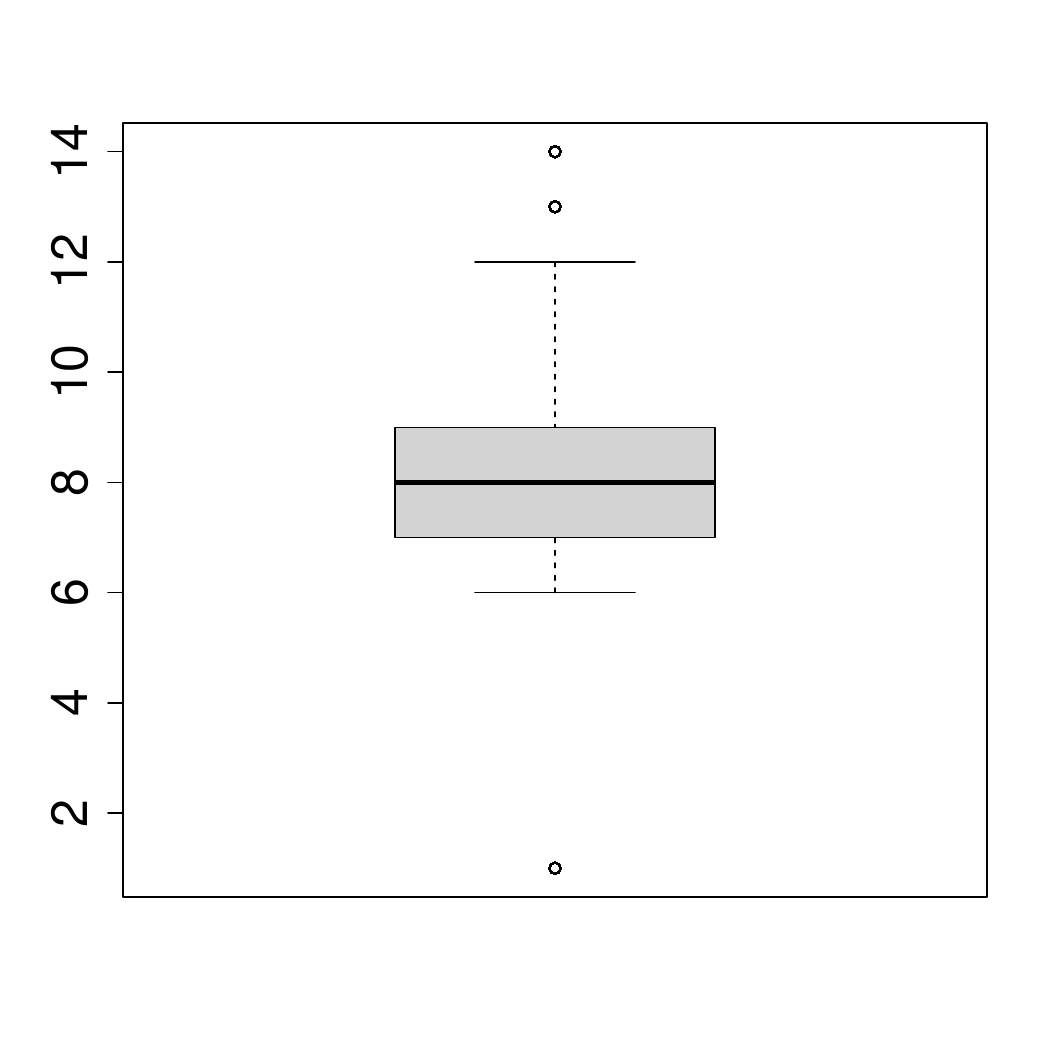}\label{fig:cobe_predictions:d}}
    \caption{Results for the COBE sea surface temperature dataset. Panels show \ref{fig:cobe_predictions:a} the observed sea surface temperature field, \ref{fig:cobe_predictions:b} the downscaled field obtained using penalized kriging, \ref{fig:cobe_predictions:c} the corresponding kriging variance ($\times 10^{-3}$), and \ref{fig:cobe_predictions:d} the boxplot of selected neighborhood sizes. The color scale is the same for the observed data and the predictions.}
    \label{fig:cobe_predictions}
\end{figure}

 Figure~\ref{fig:cobe_predictions} displays \ref{fig:cobe_predictions:a} the spatial distribution of the observed sea surface temperature data, \ref{fig:cobe_predictions:b} the downscaled spatial field obtained using penalized kriging, \ref{fig:cobe_predictions:c} the corresponding kriging variance, and \ref{fig:cobe_predictions:d} the boxplot of number of neighbors selected.

 The downscaled field exhibits smooth and physically plausible spatial patterns, with gradual transitions between warmer and cooler regions that are consistent with large-scale oceanic circulation. The resulting predictions closely resemble the original temperature field, despite relying on relatively small local neighborhoods. In particular, penalized kriging selects an average of approximately 7 neighbors per prediction location, with most values concentrated between 7 and 9, as shown in panel \ref{fig:cobe_predictions:d}. This highlights the method’s ability to achieve effective spatial refinement while substantially reducing the number of observations involved. The corresponding kriging variance map indicates low and spatially coherent prediction uncertainty across the domain, remaining comparable to that of ordinary kriging despite the strong reduction in neighborhood size.

Together, these two real-data applications illustrate the practical utility of penalized kriging across spatial scales. For moderate-size datasets (Jura, 359 locations), the method achieves computational efficiency while maintaining accuracy within 4\% of global kriging. For large datasets (COBE, 43,799 locations) where global kriging is infeasible, the method provides intelligent neighbor selection that substantially reduces the ammount of neighbors needed to perform accurate predictions.

\section{Conclusions}\label{section:conclusions}

We have developed a penalized kriging framework that incorporates $\ell_1$ regularization directly into the kriging equations, enabling automatic neighbor selection without requiring user-specified neighborhood sizes or cross-validation procedures. The proposed approach addresses the computational challenges of kriging for large spatial datasets while preserving predictive accuracy, providing a principled and scalable alternative to traditional neighborhood selection strategies.

This work makes three main contributions. First, we reformulate the kriging problem using $\ell_1$ penalties that respect the unbiasedness constraints of kriging, and extend this formulation to an adaptive LASSO version with data-driven penalty weights. Second, we introduce a criterion for automatic selection of the penalty parameter based on the effective sample size, which balances sparsity and prediction variance through a harmonic combination of normalized information measures, avoiding the computational burden of cross validation. Third, through simulations and real-data applications, we demonstrate that the method automatically adapts the neighborhood structure to spatial correlation range, process smoothness, and the geometry of the prediction location.

The simulation experiments illustrates the adaptive behavior of the method across a wide range of scenarios. As spatial correlation increases, the selected neighborhoods become progressively sparser, reflecting increased redundancy among nearby observations. Smoother processes require fewer neighbors than rough processes at comparable correlation ranges, and prediction locations near domain boundaries require larger neighborhoods than interior locations. These patterns emerge naturally from the penalized formulation and do not depend on ad hoc tuning or manual intervention.

These findings are closely related to the conclusions of \citet{emery2009kriging}, who showed through intensive combinatorial simulations that selecting the nearest neighbors does not necessarily minimize kriging variance. In that work, optimal neighborhoods are identified through iterative evaluation of multiple data configurations using kriging update equations. In contrast, the approach proposed here reaches the same substantive conclusion—that proximity alone is insufficient for optimal neighbor selection—through a fundamentally different mechanism: a single-step penalized formulation combined with an effective sample size criterion. This avoids exhaustive combinatorial searches while remaining computationally scalable to very large datasets.

The proposed framework has several limitations that suggest directions for future research. It relies on a valid covariance model estimated from the data, and misspecification may affect neighbor selection. The current methodology assumes Gaussian processes and stationarity; extensions to non-Gaussian settings, such as trans-Gaussian or indicator kriging, and to non-stationary covariance structures requires further investigation. Additional work could establish theoretical properties of the spatial adaptive LASSO, connect the ESS criterion to formal model selection frameworks, and extend the approach to multivariate and space–time processes. Overall, the ESS-based penalized kriging framework provides an effective compromise between sparsity and prediction accuracy, offering a principled and computationally efficient solution to the long-standing problem of neighbor selection in spatial prediction. Also, study the behaviour of conditional simulations under Lasso Kriging instead of Simple kriging.

\section*{Acknowledgments}
This work was funded by the National Agency for Research and Development of Chile, under grants ANID FONDECYT INICIACION 11230502 (J. Acosta), and ANID FONDECYT INICIACION 11240330 (F. Cuevas-Pacheco). 

\bibliography{bibliography}

\appendix

\section{Complete Variance Comparison Results}\label{appendix:variance_full}

This appendix presents the complete set of variance comparison results for all 15 combinations of covariance model (Exponential, Matérn $\nu=1.5$, Spherical) and prediction location (Farthest Neighbor, Average Neighbor, Densest Neighbor, Corner Neighbor, Side Neighbor). 
Each panel shows the relative variance increase (\%) compared to ordinary kriging using all 500 observations as a function of practical range. The red lines represent penalized kriging, green lines represent local kriging with the same number of neighbors selected as K-nearest, and black lines represent the ordinary kriging baseline (0\%). These figures complement the selected results presented in Figure \ref{fig:variance_comparison} in the main text.

\begin{figure}[h!]
    \centering
    \subfigure[Exponential -- FN]{\includegraphics[width=0.1925\linewidth]{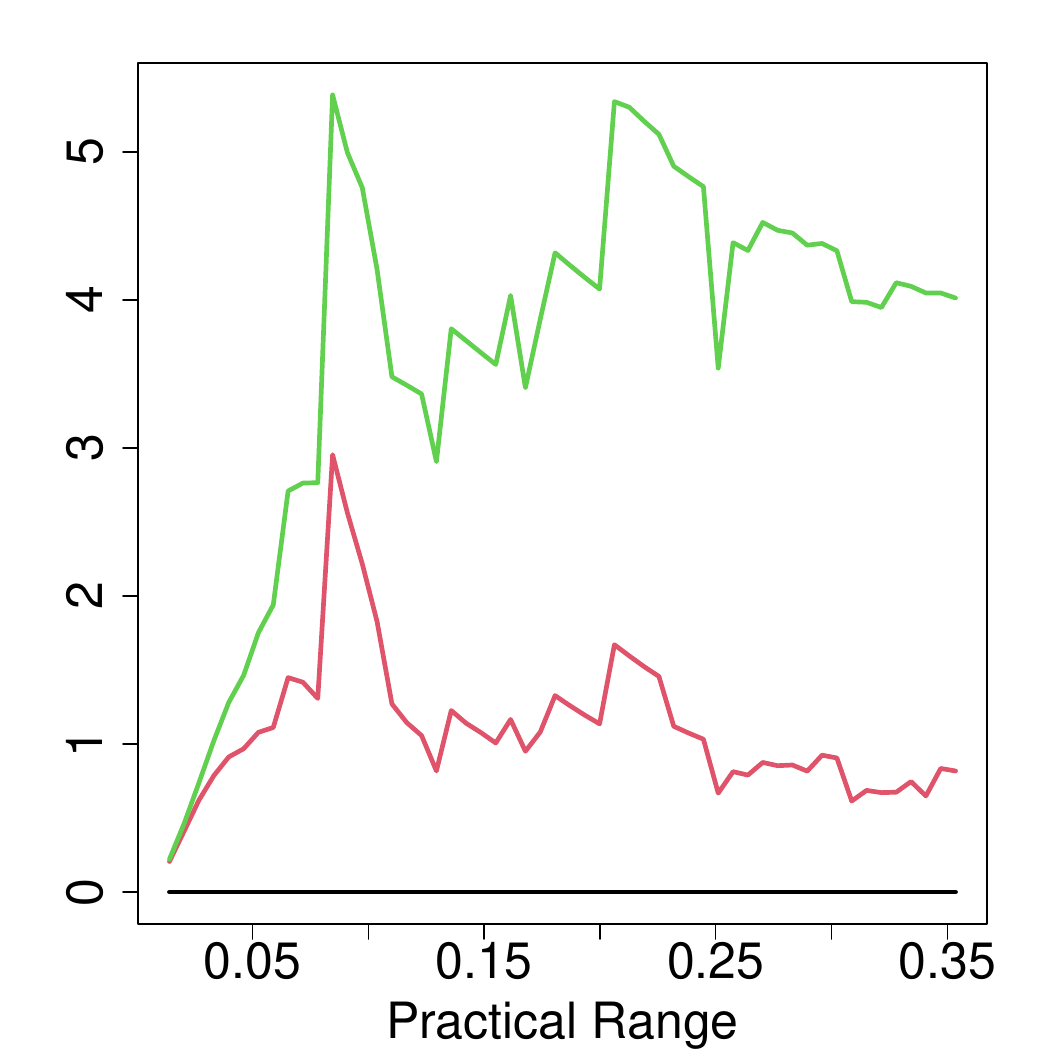}}
    \subfigure[Exponential -- AN]{\includegraphics[width=0.1925\linewidth]{figs/Variance_ratio_exponential_Average.pdf}}
    \subfigure[Exponential -- DN]{\includegraphics[width=0.1925\linewidth]{figs/Variance_ratio_exponential_Densest.pdf}}
    \subfigure[Exponential -- SN]{\includegraphics[width=0.1925\linewidth]{figs/Variance_ratio_exponential_Side.pdf}}
    \subfigure[Exponential -- CN]{\includegraphics[width=0.1925\linewidth]{figs/Variance_ratio_exponential_Corner.pdf}}

    \vspace{0.3em}

    \subfigure[Matérn -- FN]{\includegraphics[width=0.1925\linewidth]{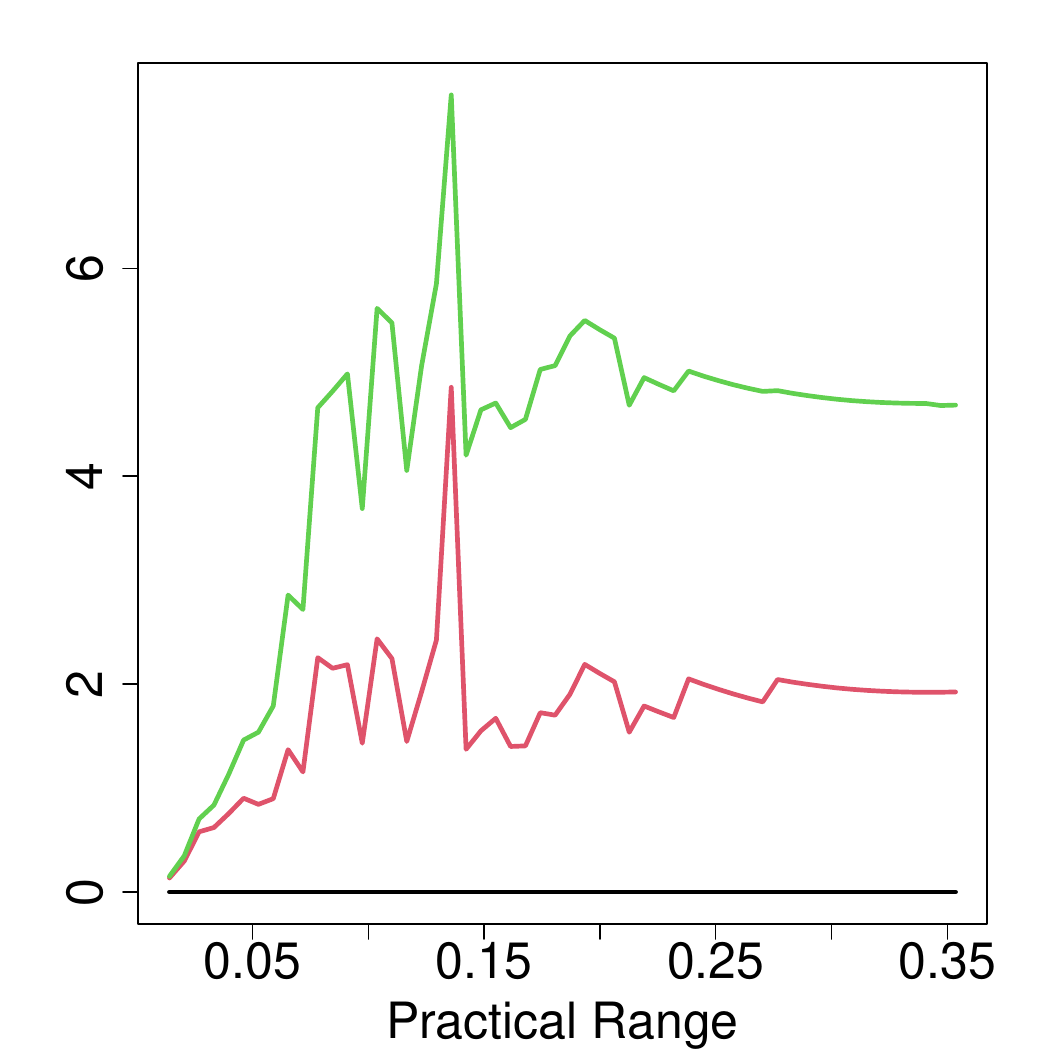}}
    \subfigure[Matérn -- AN]{\includegraphics[width=0.1925\linewidth]{figs/Variance_ratio_matern_Average.pdf}}
    \subfigure[Matérn -- DN]{\includegraphics[width=0.1925\linewidth]{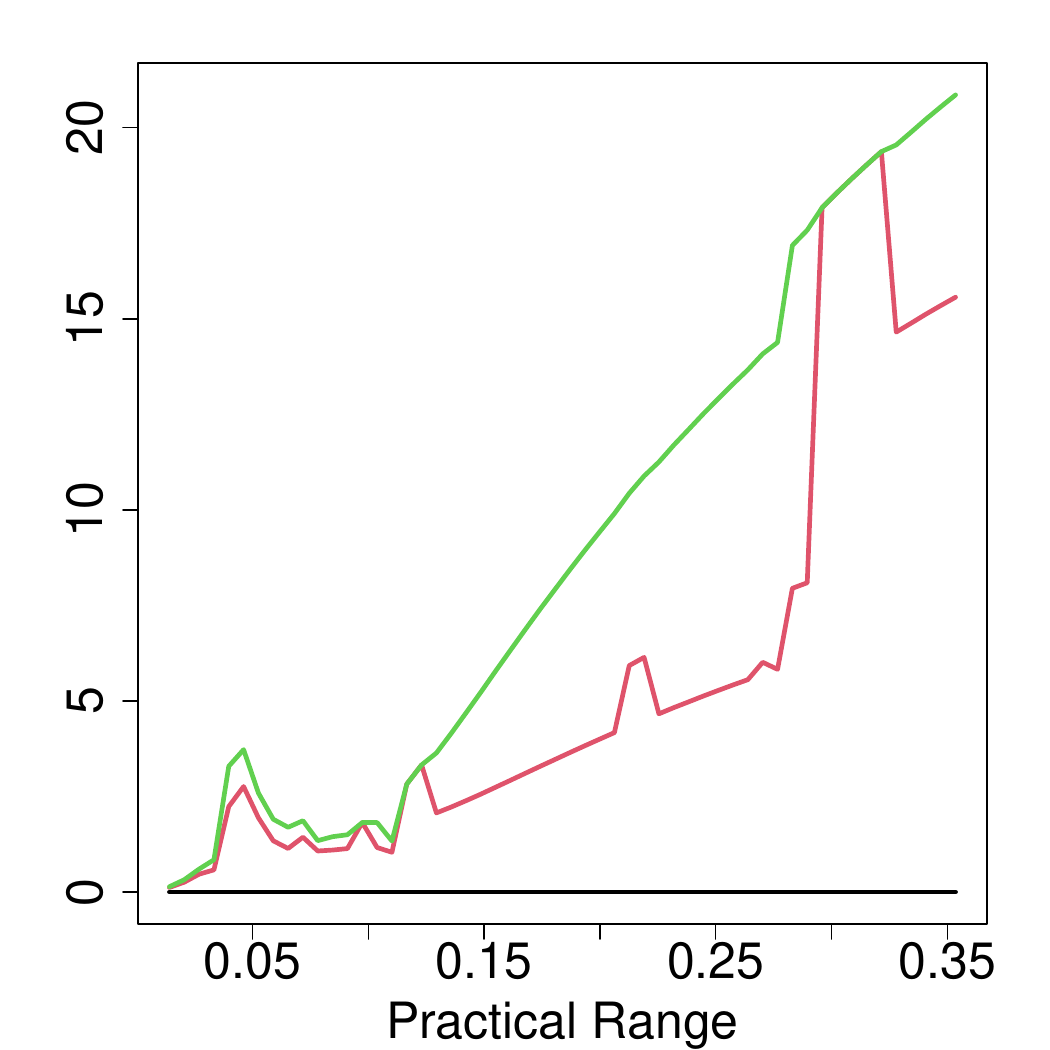}}
    \subfigure[Matérn -- SN]{\includegraphics[width=0.1925\linewidth]{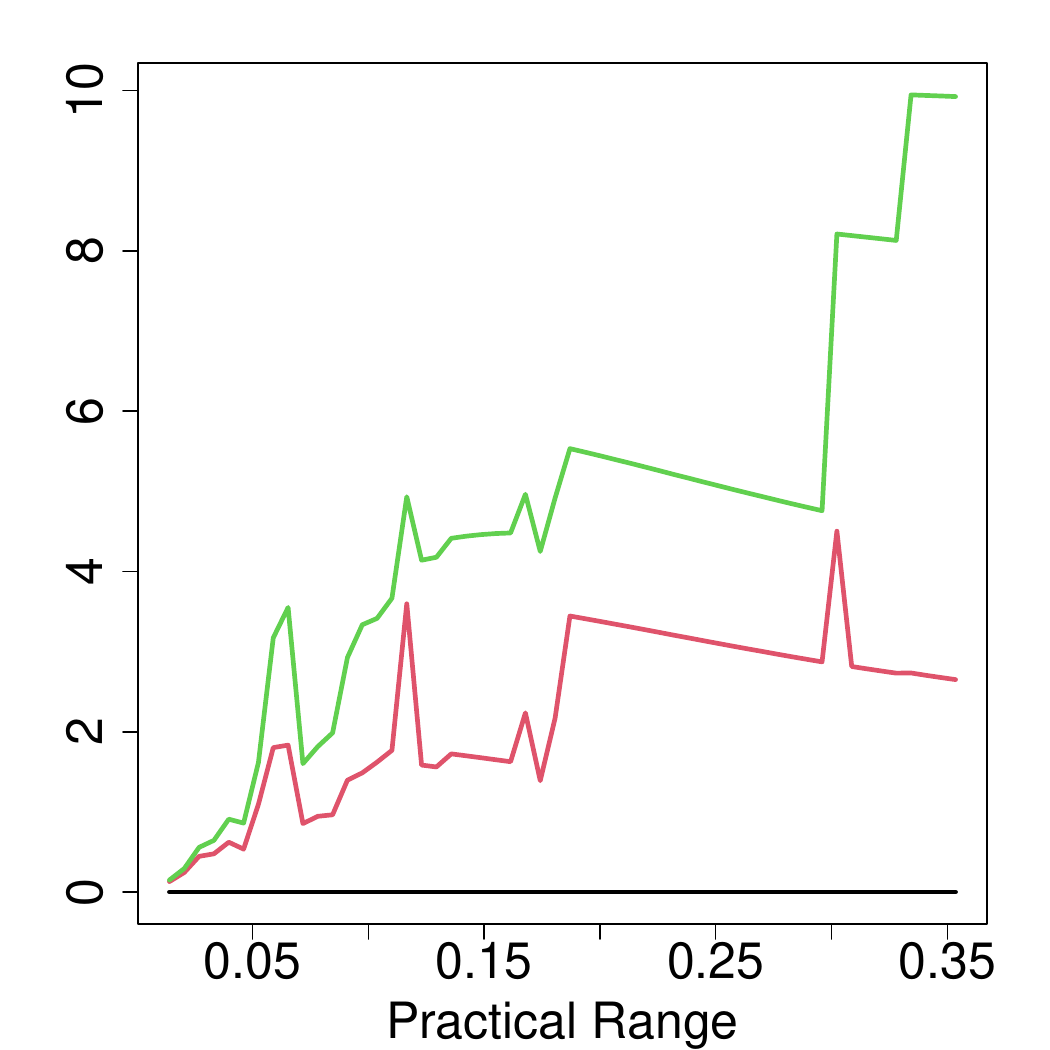}}
    \subfigure[Matérn -- CN]{\includegraphics[width=0.1925\linewidth]{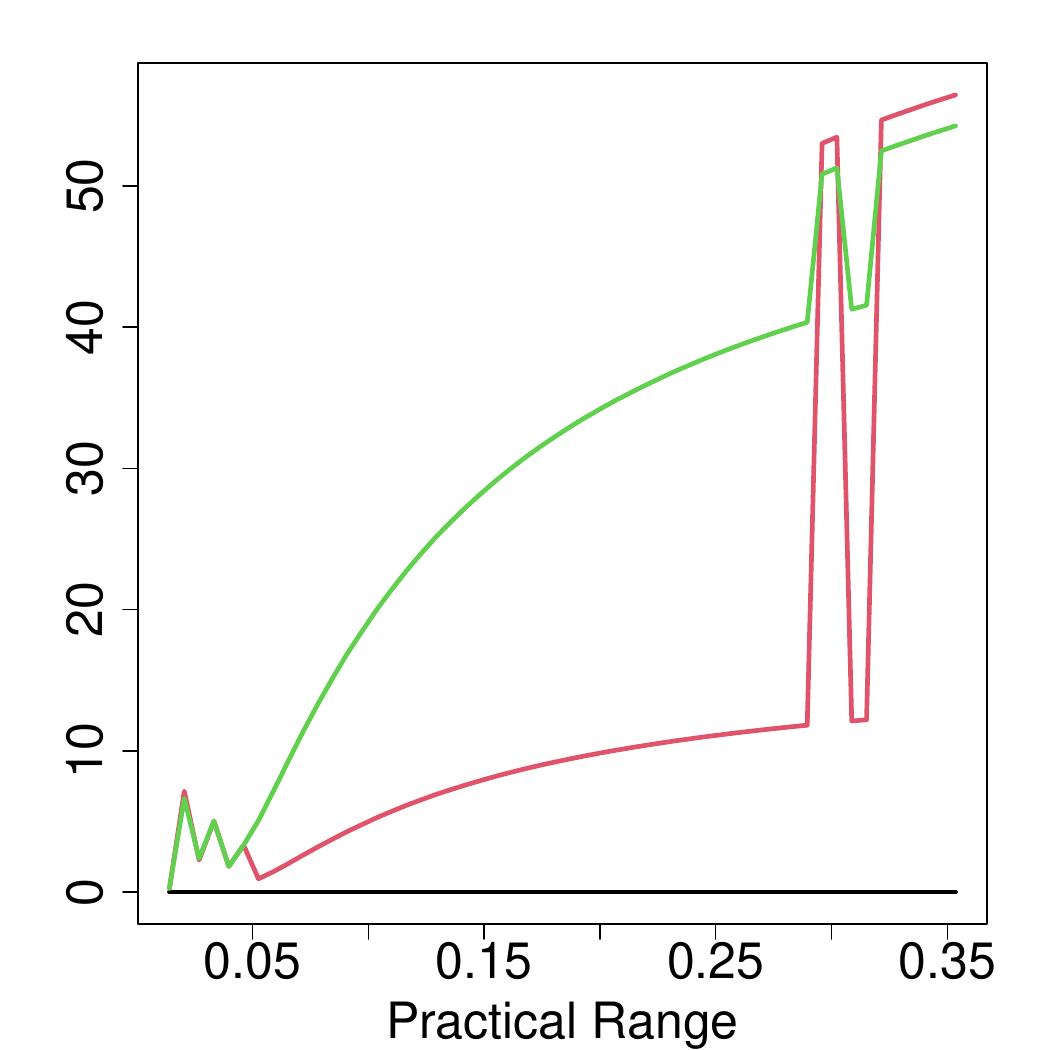}}

    \vspace{0.3em}

    \subfigure[Spherical -- FN]{\includegraphics[width=0.1925\linewidth]{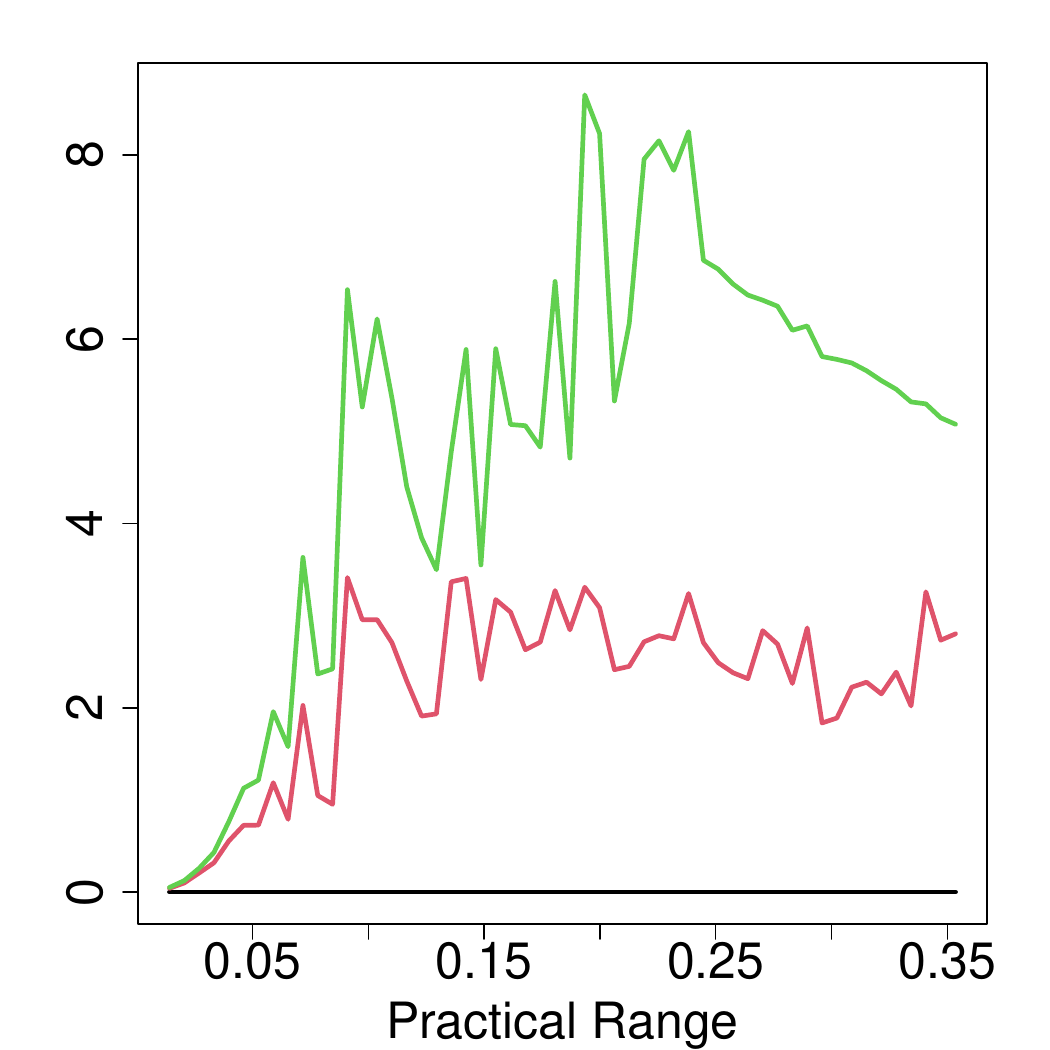}}
    \subfigure[Spherical -- AN]{\includegraphics[width=0.1925\linewidth]{figs/Variance_ratio_spherical_Average.pdf}}
    \subfigure[Spherical -- DN]{\includegraphics[width=0.1925\linewidth]{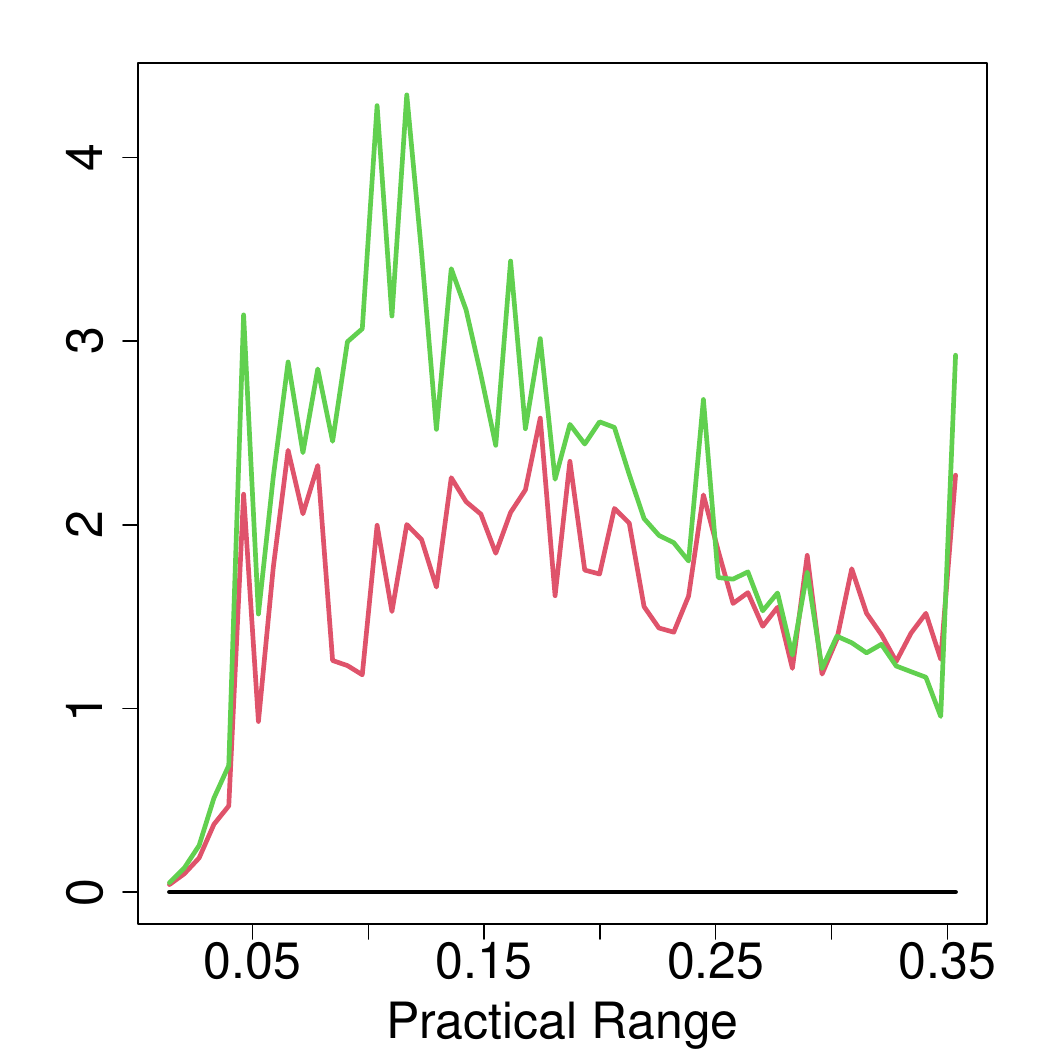}}
    \subfigure[Spherical -- SN]{\includegraphics[width=0.1925\linewidth]{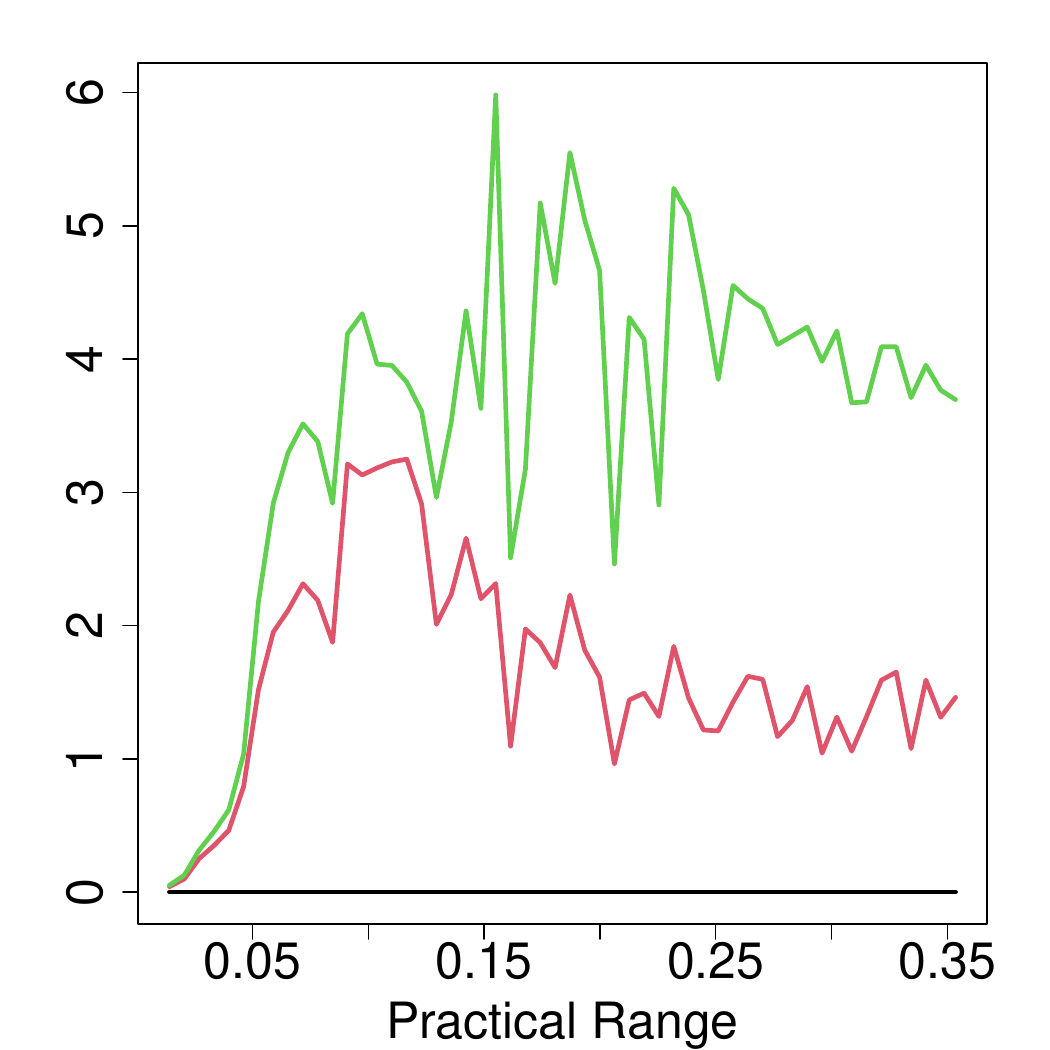}}
    \subfigure[Spherical -- CN]{\includegraphics[width=0.1925\linewidth]{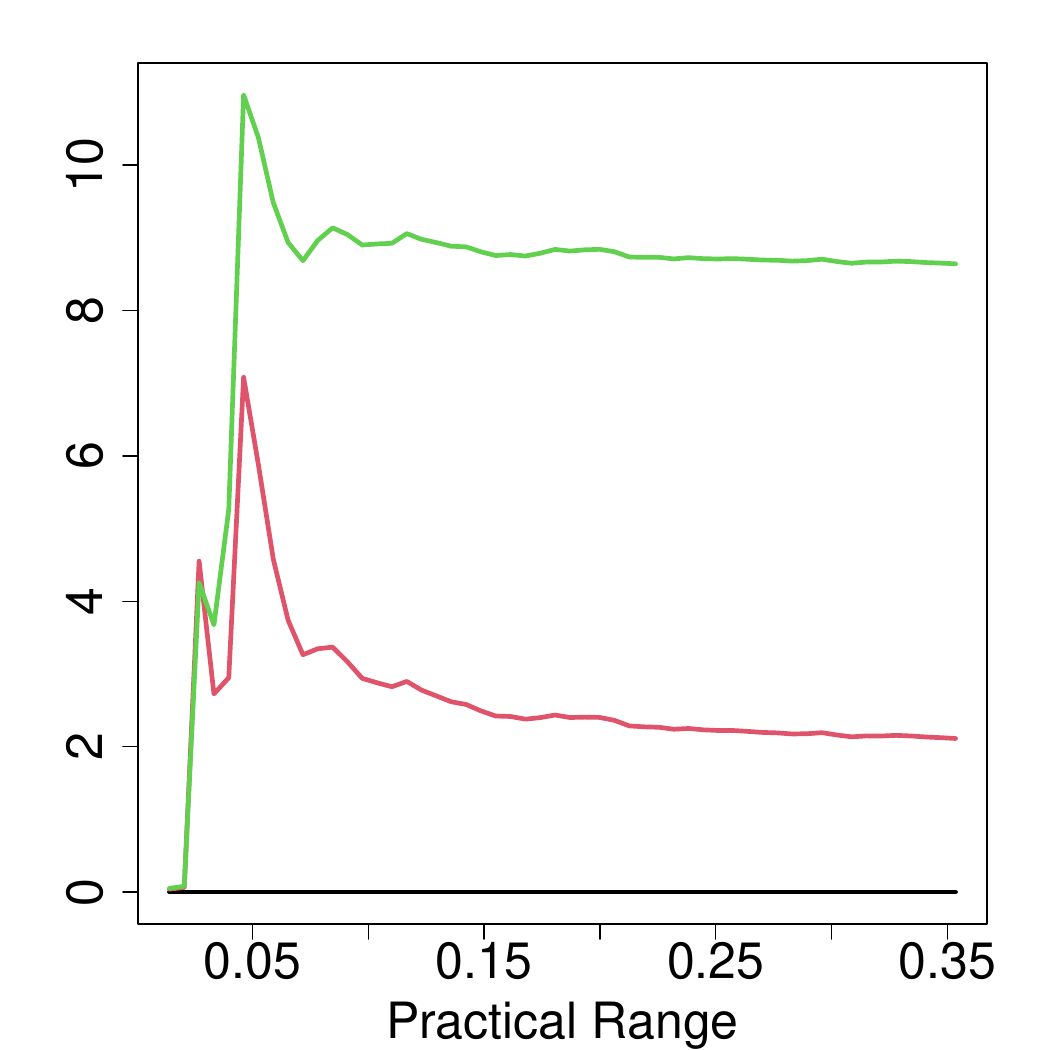}}

    \caption{\small Relative variance increase (\%) for penalized kriging (red) and local kriging using the same number of nearest neighbors (green), compared to ordinary kriging (black baseline at 0\%). 
    Rows correspond to covariance models (Exponential, Matérn $\nu=1.5$, Spherical), and columns to prediction locations (FN: farthest, AN: average, DN: densest, SN: side, CN: corner).}
    \label{fig:variance_all_combined_transposed}
\end{figure}

\section{Spatial Distribution of Kriging Coefficients}\label{appendix:spatial_coeffs}

This section presents the complete spatial coefficient distributions for all combinations of covariance model (Exponential, Matérn $\nu=1.5$, Spherical), practical range level (Low, Medium, Large), and prediction location (Farthest, Average, Densest, Corner, Side). 
Grey points indicate neighbors with zero coefficients (not selected by adaptive LASSO), 
colored points show selected neighbors with color intensity indicating coefficient magnitude, and red points mark the prediction locations. The color scale is identical across all figures to facilitate comparison.

\begin{figure}[ht!]
    \centering
    \begin{minipage}{0.9\textwidth}
        \begin{minipage}{0.08\textwidth}
            \centering
            \rotatebox{90}{\small Farthest}
        \end{minipage}
        \begin{minipage}{0.88\textwidth}
            \includegraphics[width=0.32\textwidth]{figs/Low_exponential_Farthest.pdf}
            \includegraphics[width=0.32\textwidth]{figs/Med_exponential_Farthest.pdf}
            \includegraphics[width=0.32\textwidth]{figs/Lar_exponential_Farthest.pdf}
        \end{minipage}
        
        \begin{minipage}{0.08\textwidth}
            \centering
            \rotatebox{90}{\small Average}
        \end{minipage}
        \begin{minipage}{0.88\textwidth}
            \includegraphics[width=0.32\textwidth]{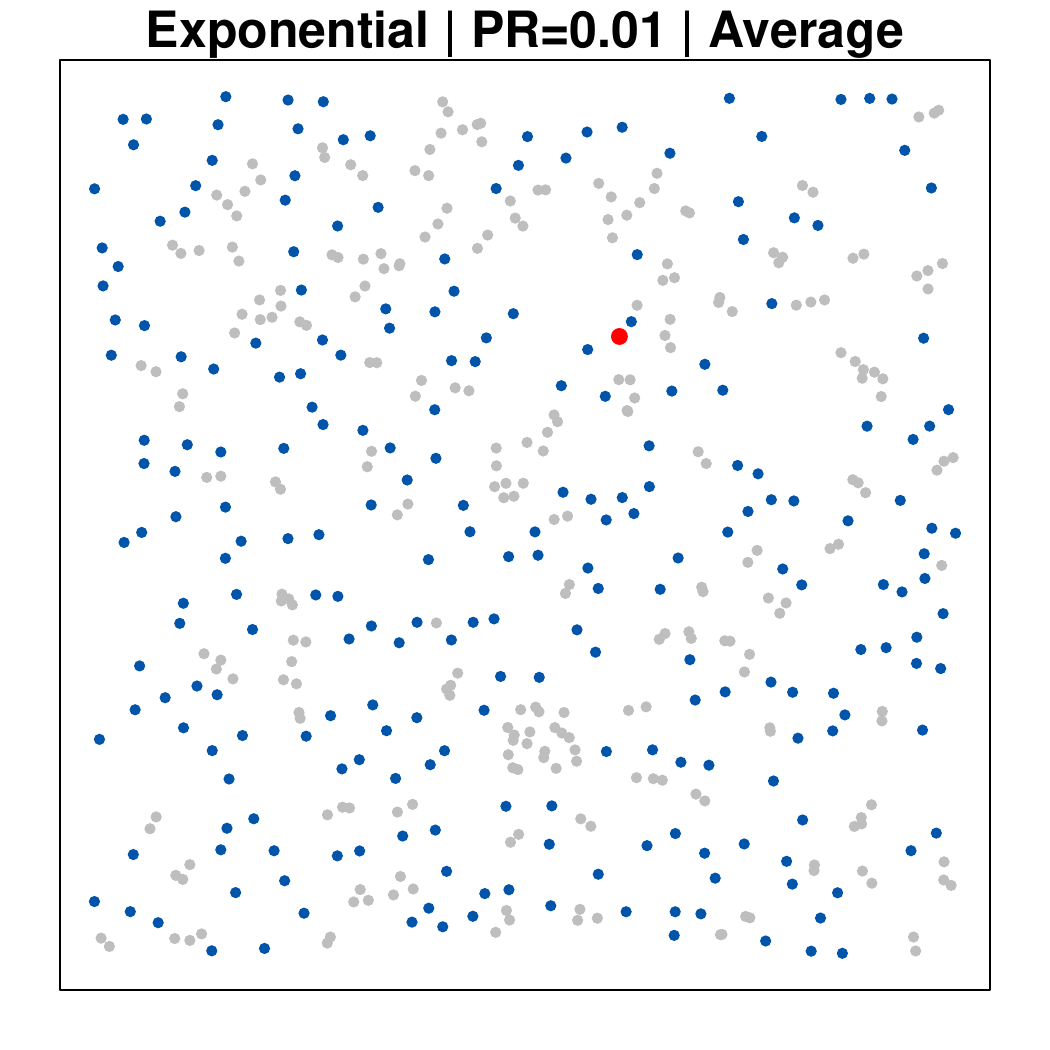}
            \includegraphics[width=0.32\textwidth]{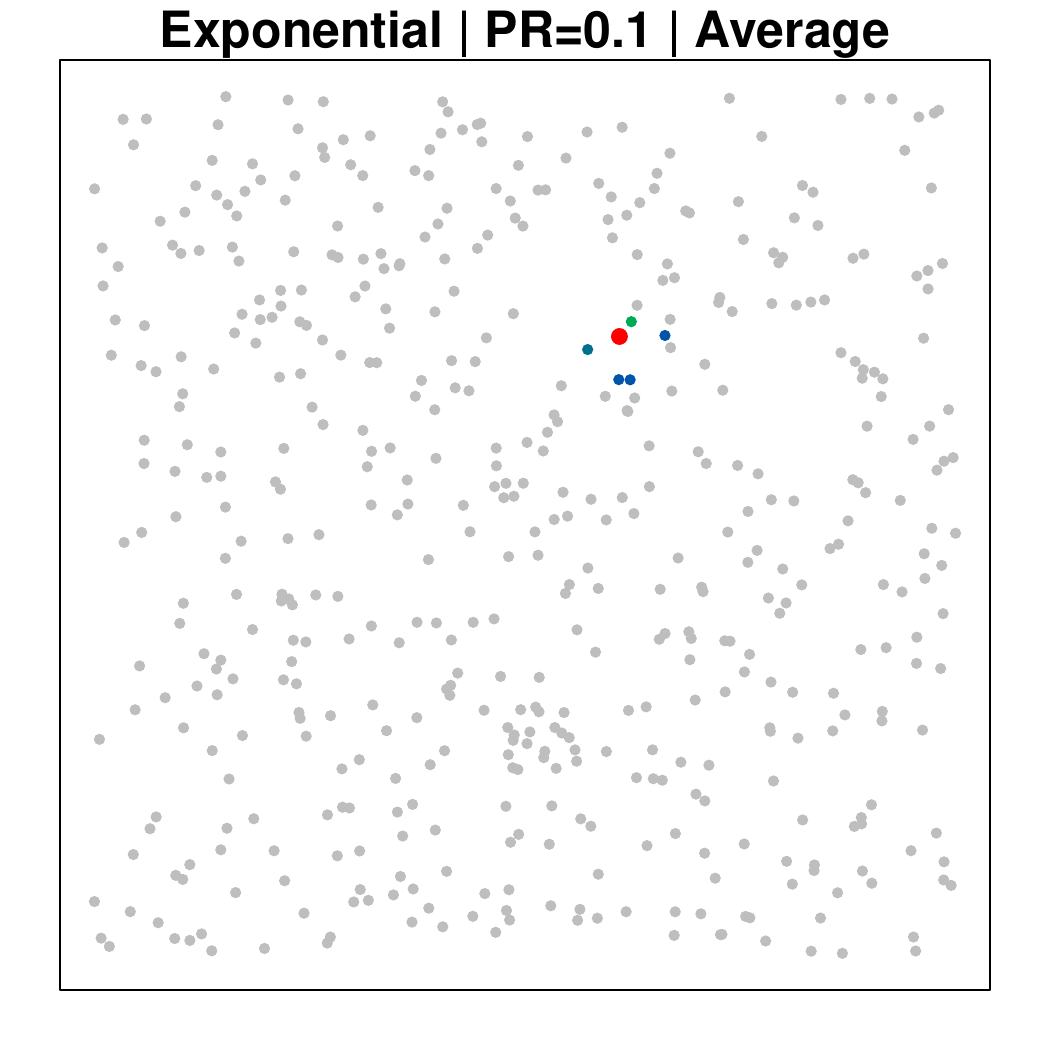}
            \includegraphics[width=0.32\textwidth]{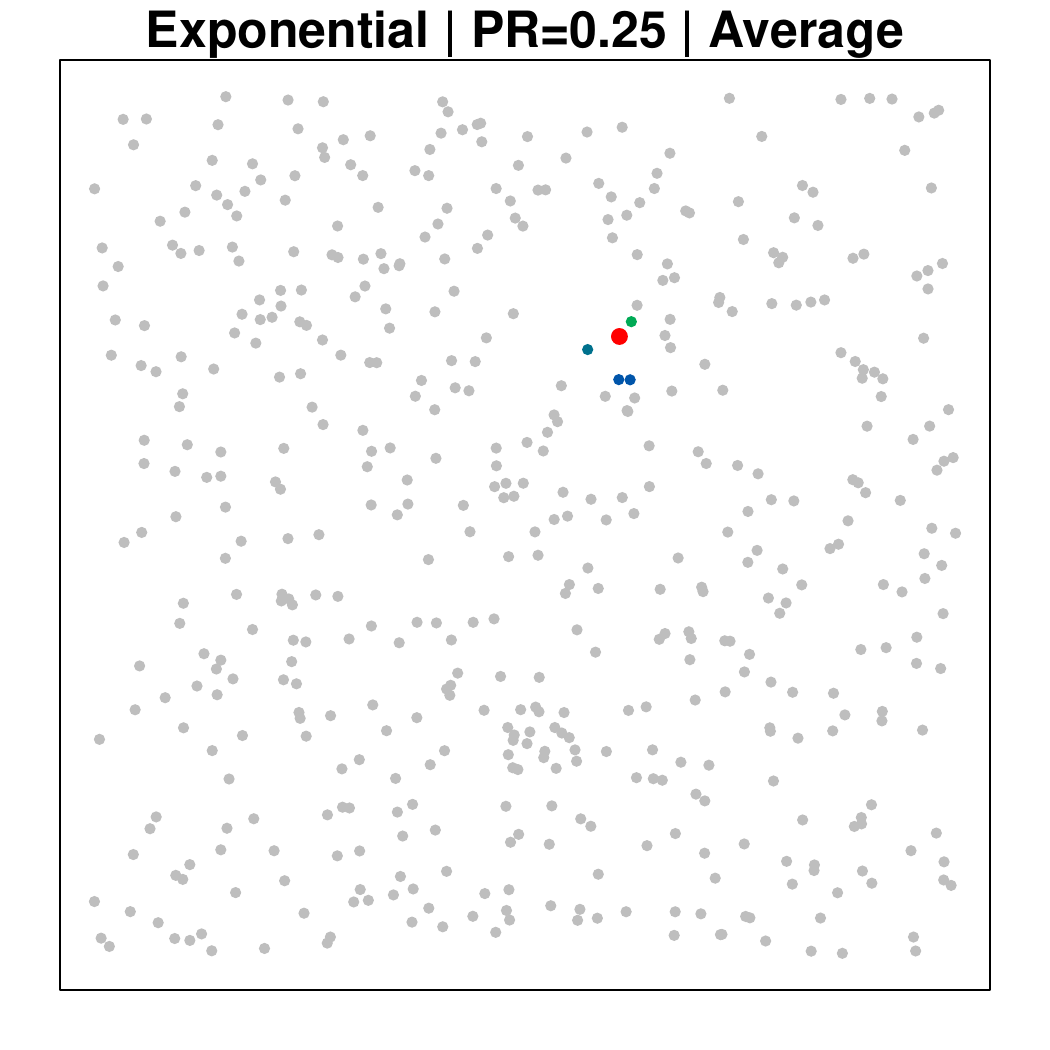}
        \end{minipage}
        
        \begin{minipage}{0.08\textwidth}
            \centering
            \rotatebox{90}{\small Densest}
        \end{minipage}
        \begin{minipage}{0.88\textwidth}
            \includegraphics[width=0.32\textwidth]{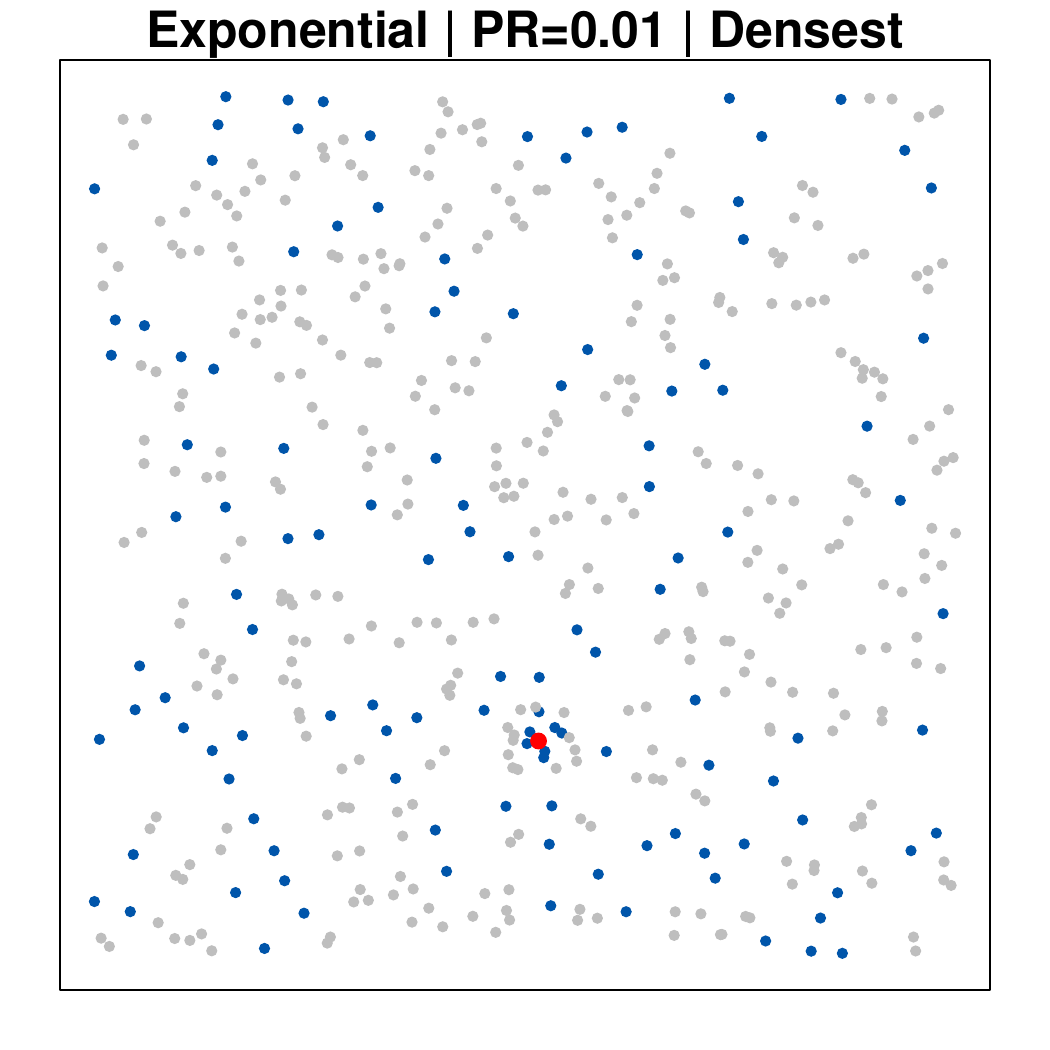}
            \includegraphics[width=0.32\textwidth]{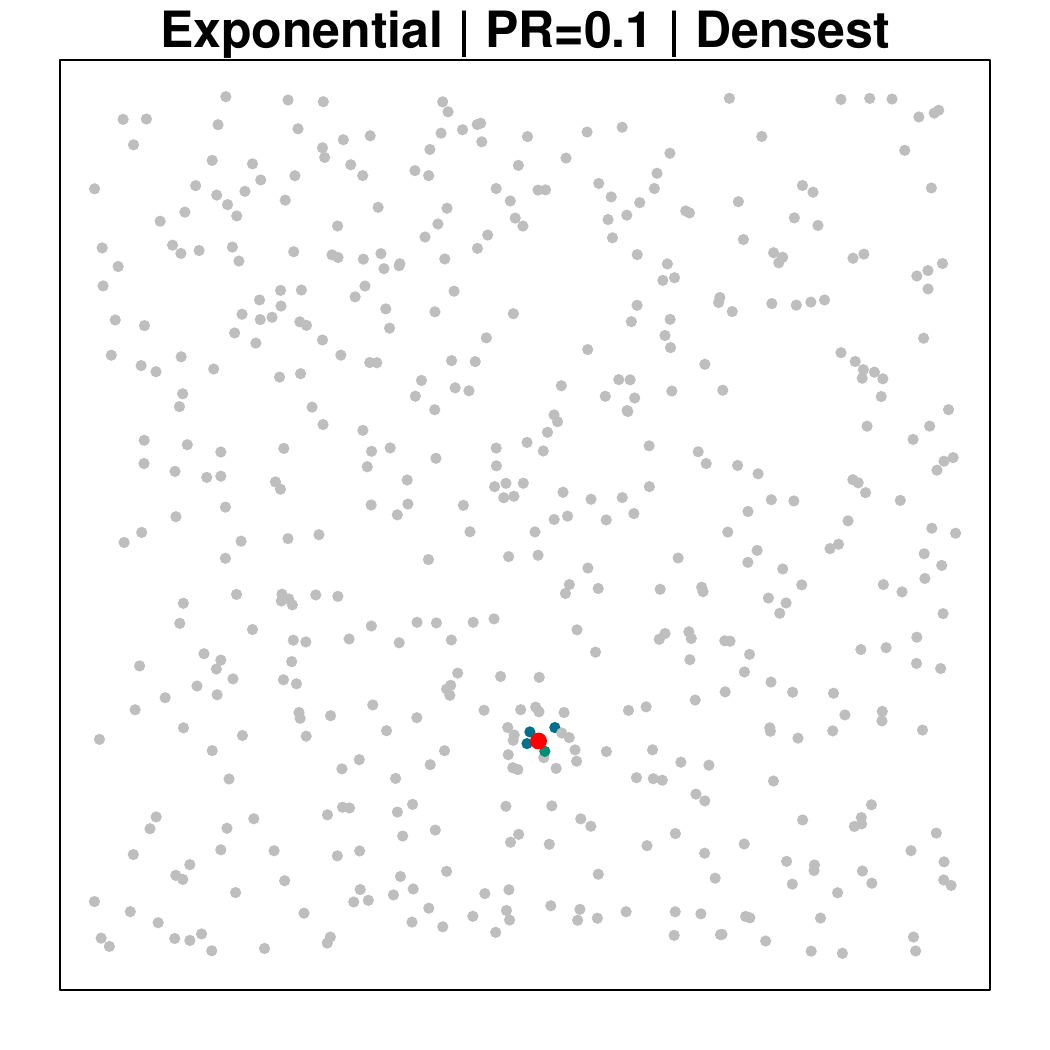}
            \includegraphics[width=0.32\textwidth]{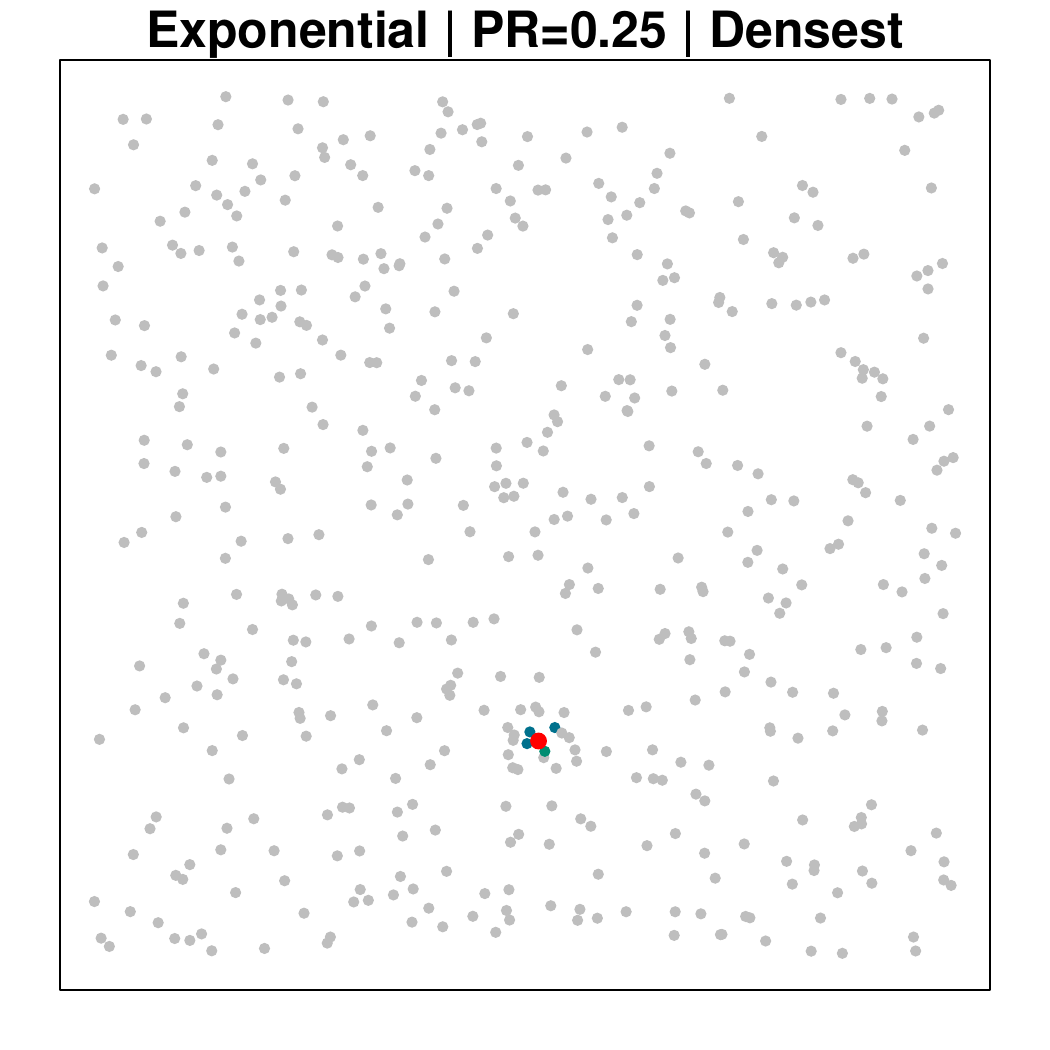}
        \end{minipage}
        
        \begin{minipage}{0.08\textwidth}
            \centering
            \rotatebox{90}{\small Corner}
        \end{minipage}
        \begin{minipage}{0.88\textwidth}
            \includegraphics[width=0.32\textwidth]{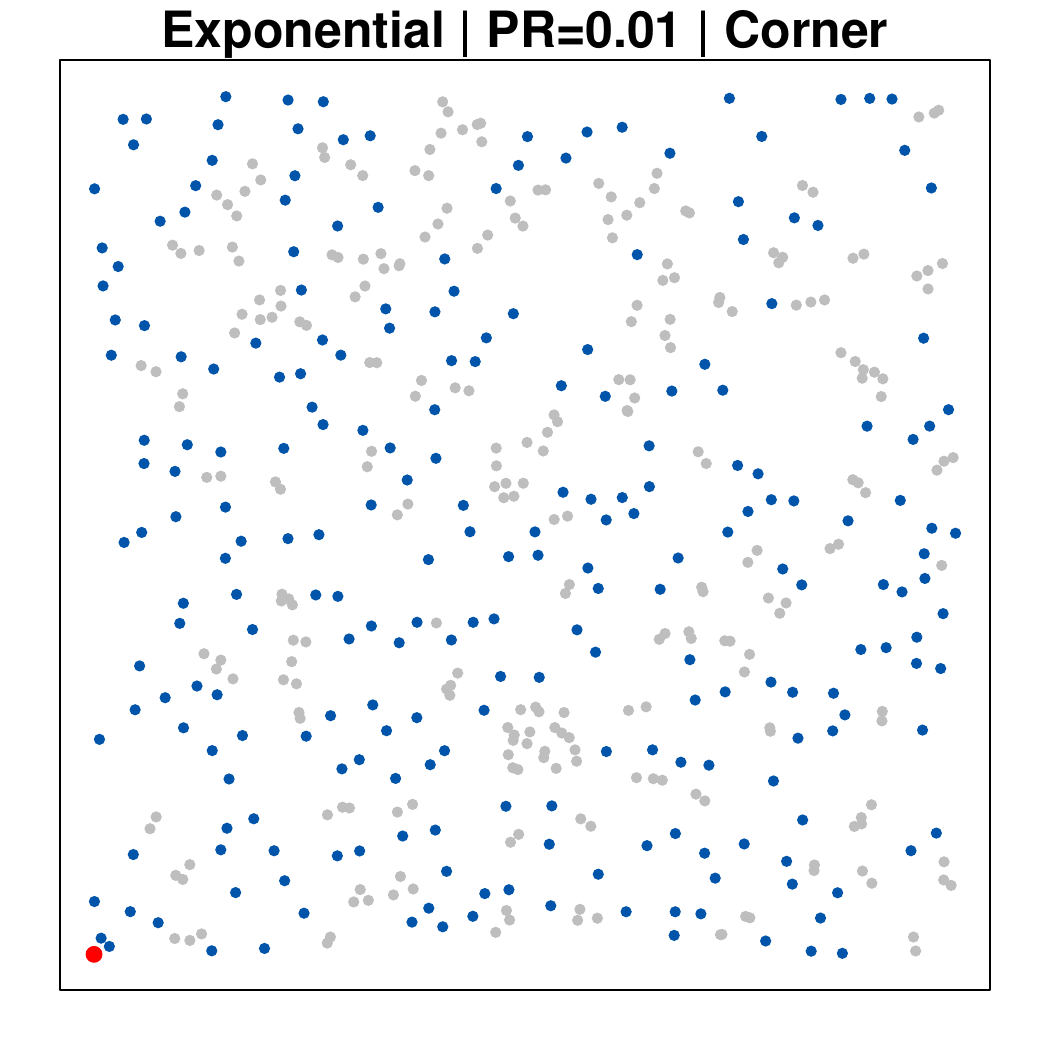}
            \includegraphics[width=0.32\textwidth]{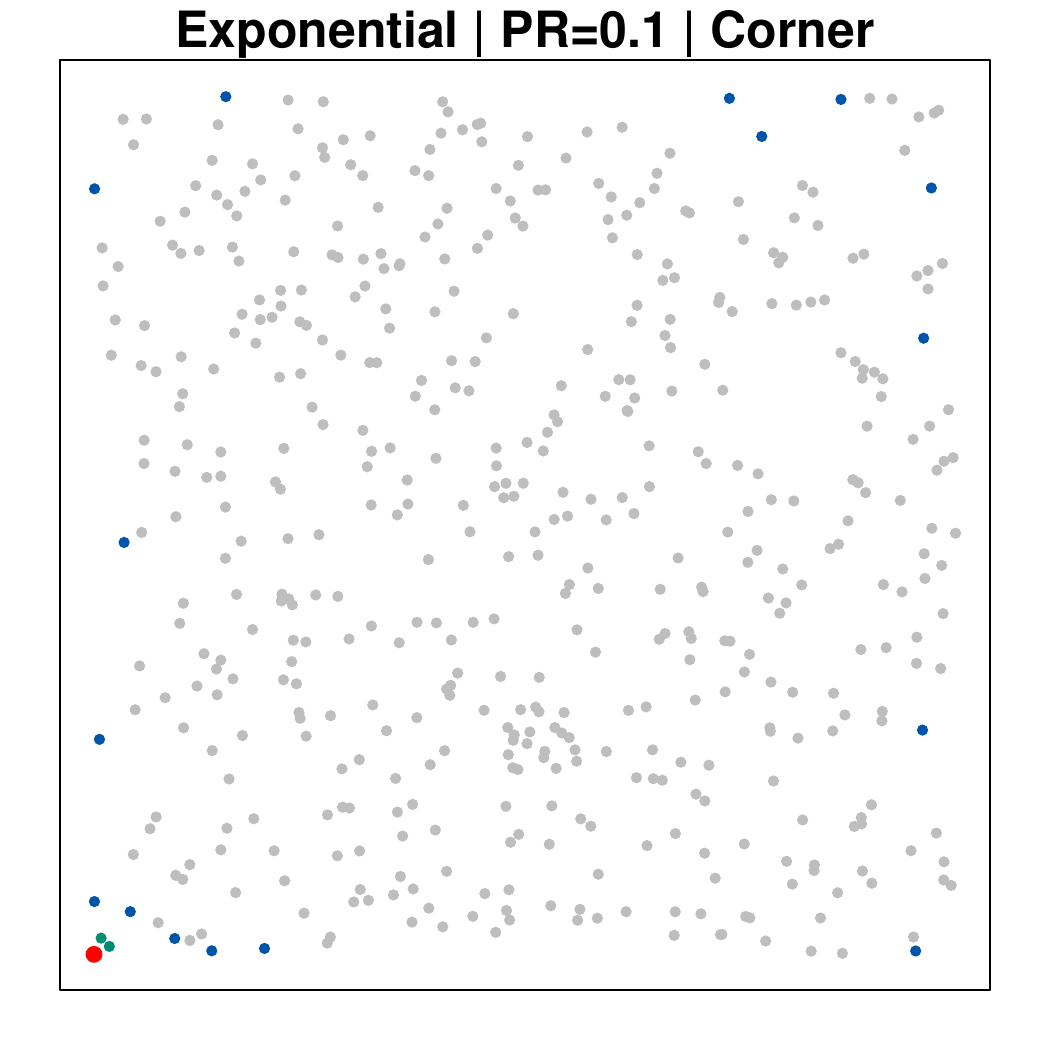}
            \includegraphics[width=0.32\textwidth]{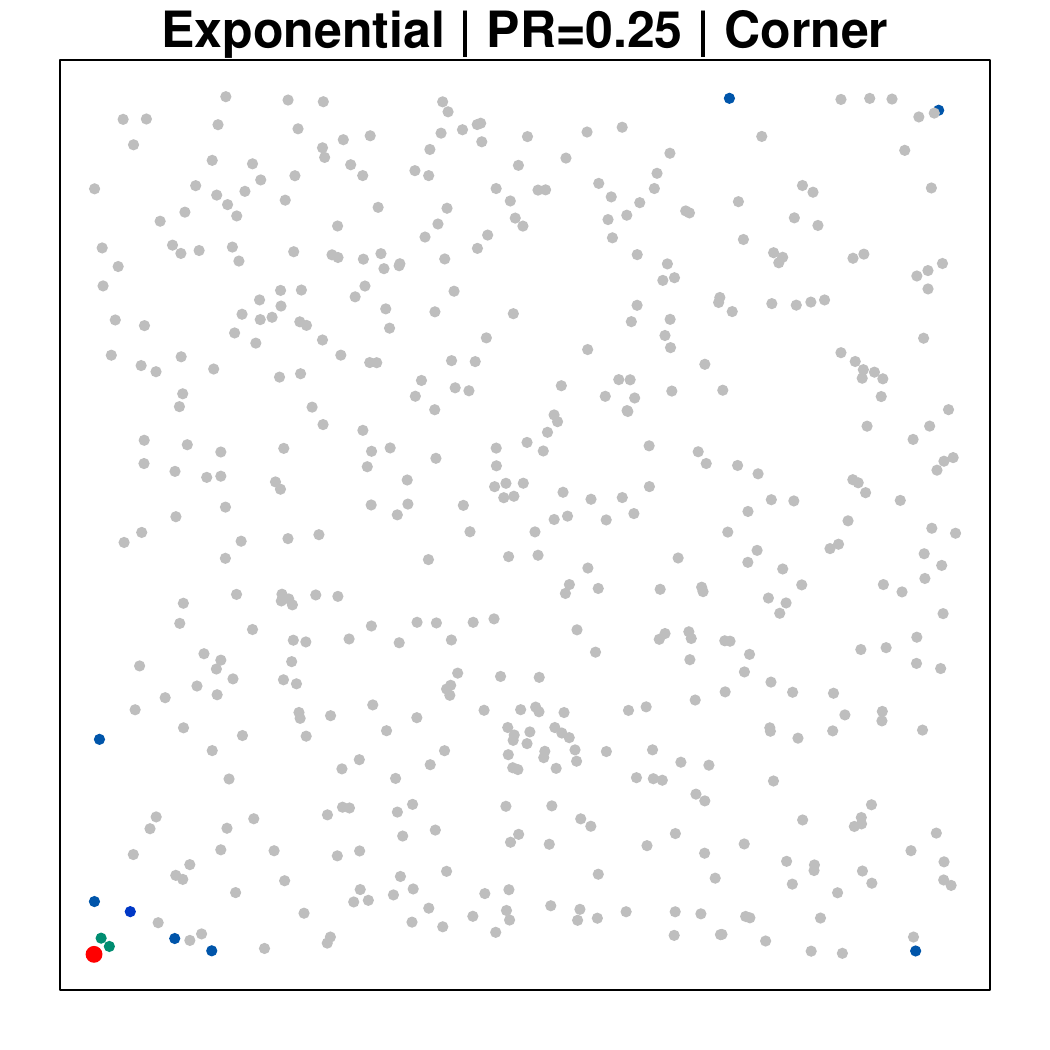}
        \end{minipage}
        
        \begin{minipage}{0.08\textwidth}
            \centering
            \rotatebox{90}{\small Side}
        \end{minipage}
        \begin{minipage}{0.88\textwidth}
            \includegraphics[width=0.32\textwidth]{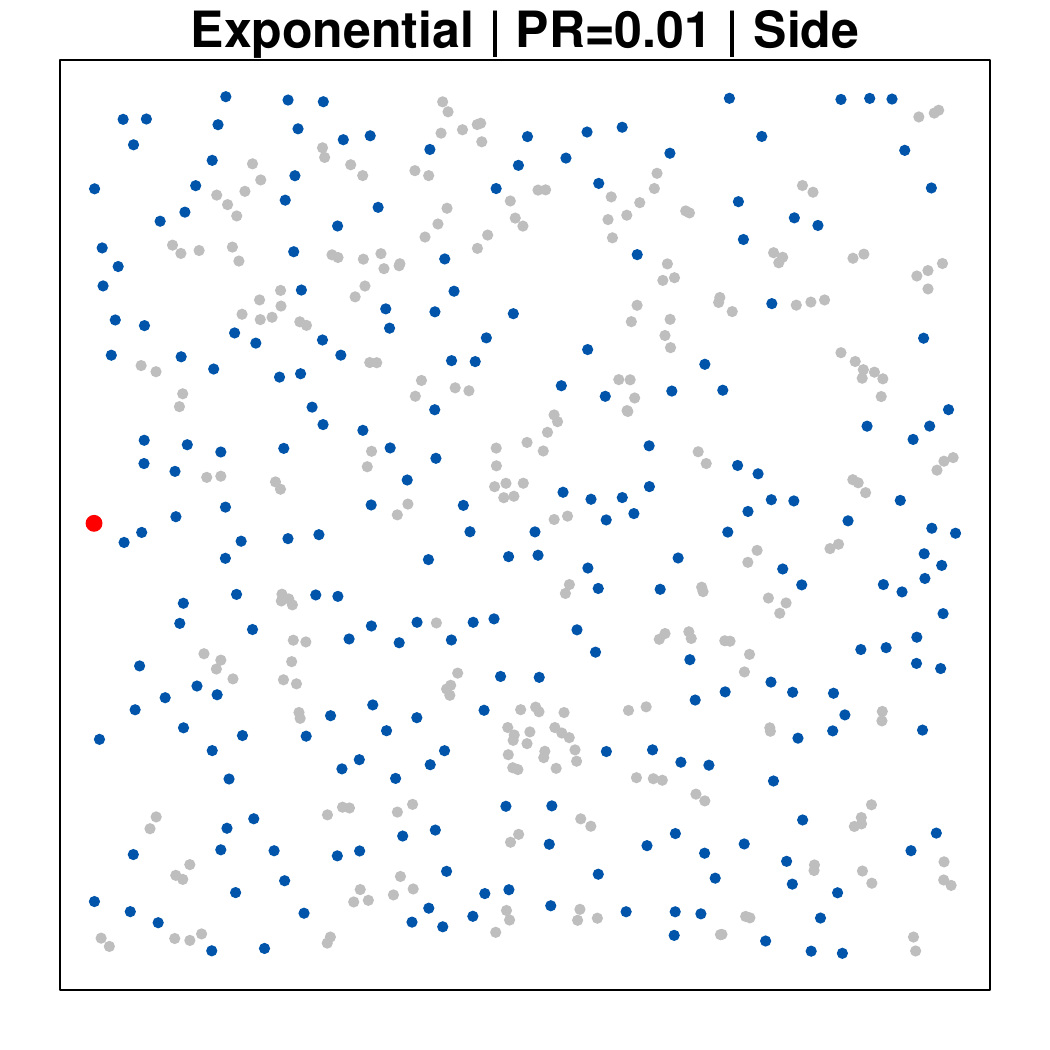}
            \includegraphics[width=0.32\textwidth]{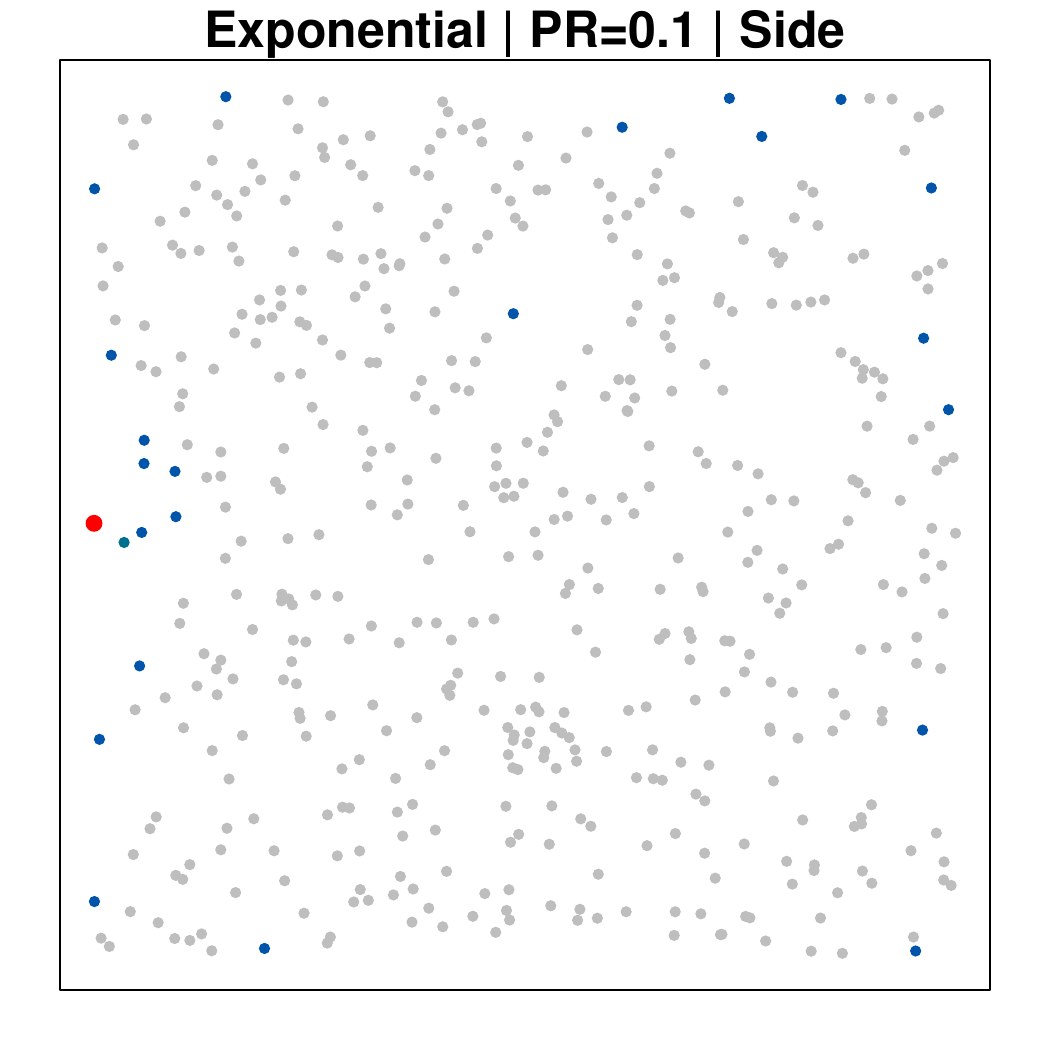}
            \includegraphics[width=0.32\textwidth]{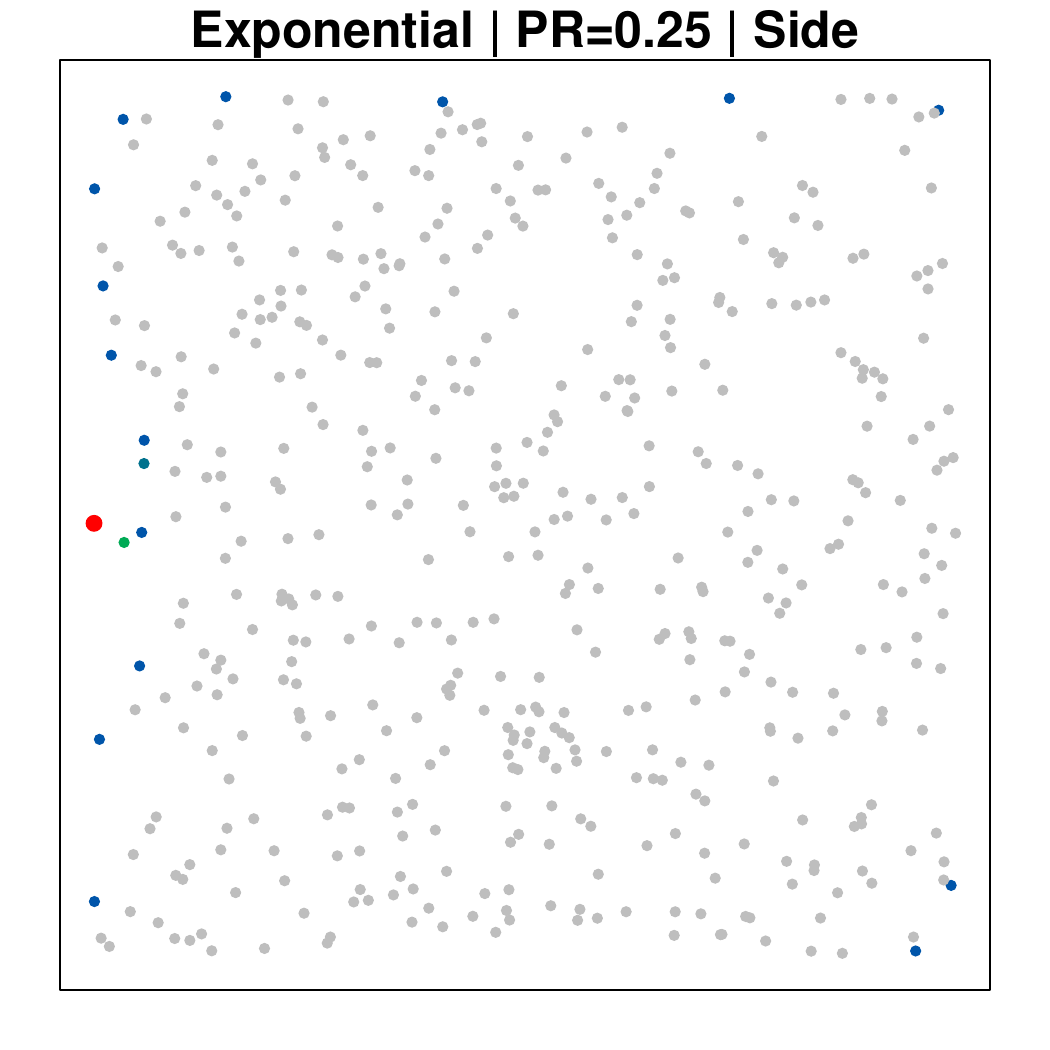}
        \end{minipage}
        
        \vspace{0.3cm}
        \begin{minipage}{0.08\textwidth}
        \end{minipage}
        \begin{minipage}{0.88\textwidth}
            \centering
            \textbf{Low PR} \hspace{2.5cm} \textbf{Med PR} \hspace{2.5cm} \textbf{Lar PR}
        \end{minipage}
    \end{minipage}%
    \hfill
    \begin{minipage}{0.06\textwidth}
        \centering
        \includegraphics[width=1.75\textwidth, keepaspectratio=false]{figs/Global_Legend.pdf}
    \end{minipage}
    
    \caption{\small Spatial distribution of adaptive LASSO kriging coefficients for exponential covariance across five prediction locations (rows) and three practical range levels (columns). Grey points indicate neighbors with zero coefficients (not selected). Colored points show selected neighbors, with color intensity indicating coefficient magnitude (blue = small, green = large). The red point marks the prediction location $\bm{s}_0$.}
    \label{fig:spatial_coefficients_exponential}
\end{figure}

\begin{figure}[ht!]
    \centering
    \begin{minipage}{0.9\textwidth}
        \begin{minipage}{0.08\textwidth}
            \centering
            \rotatebox{90}{\small Farthest}
        \end{minipage}
        \begin{minipage}{0.88\textwidth}
            \includegraphics[width=0.32\textwidth]{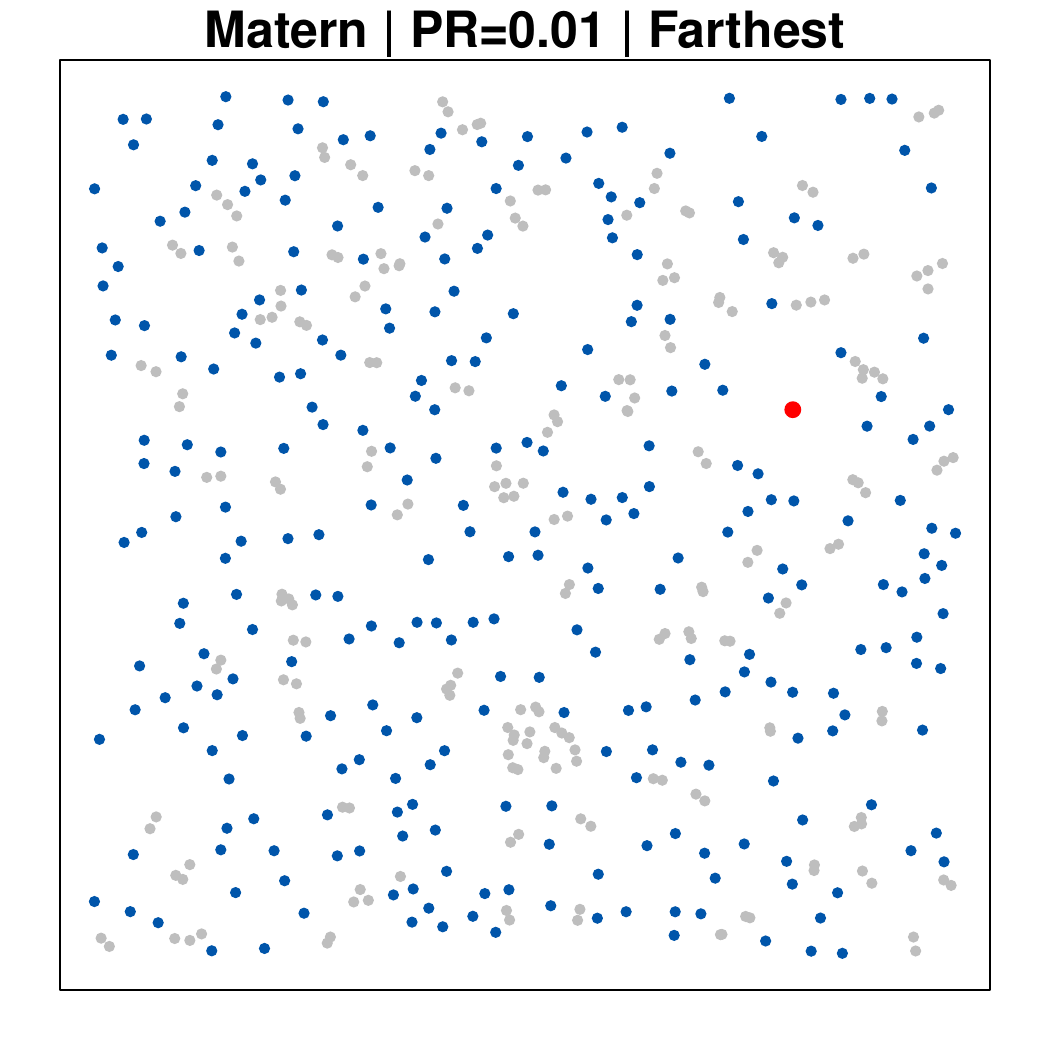}
            \includegraphics[width=0.32\textwidth]{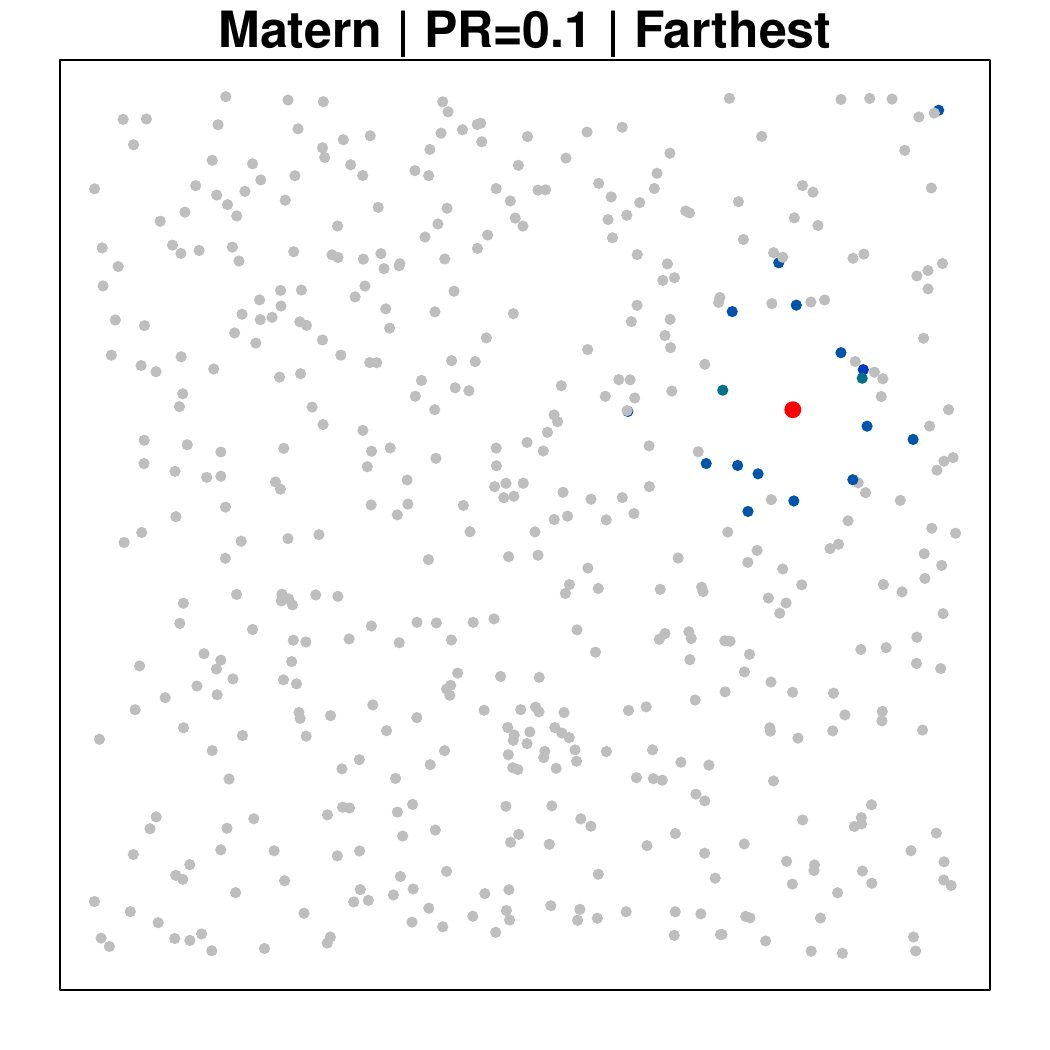}
            \includegraphics[width=0.32\textwidth]{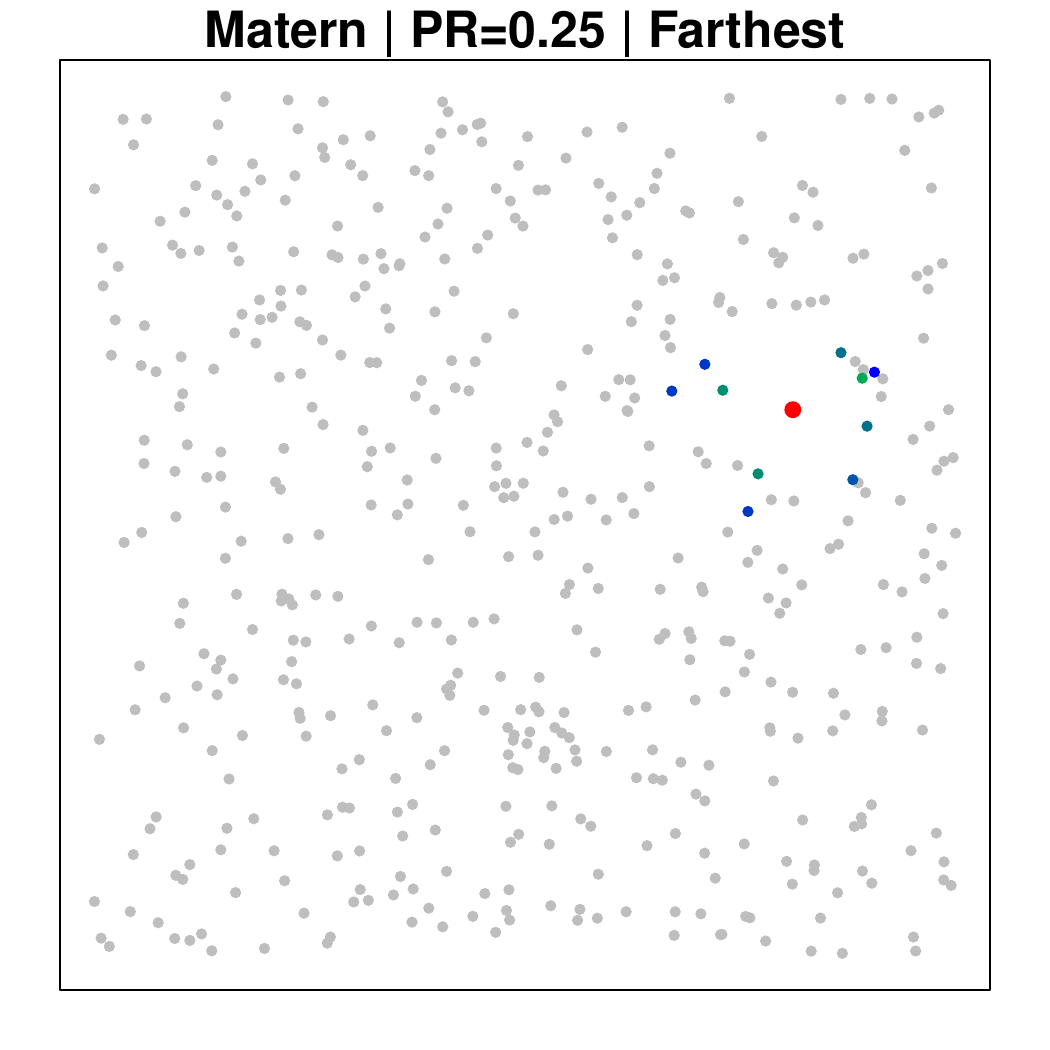}
        \end{minipage}
        
        \begin{minipage}{0.08\textwidth}
            \centering
            \rotatebox{90}{\small Average}
        \end{minipage}
        \begin{minipage}{0.88\textwidth}
            \includegraphics[width=0.32\textwidth]{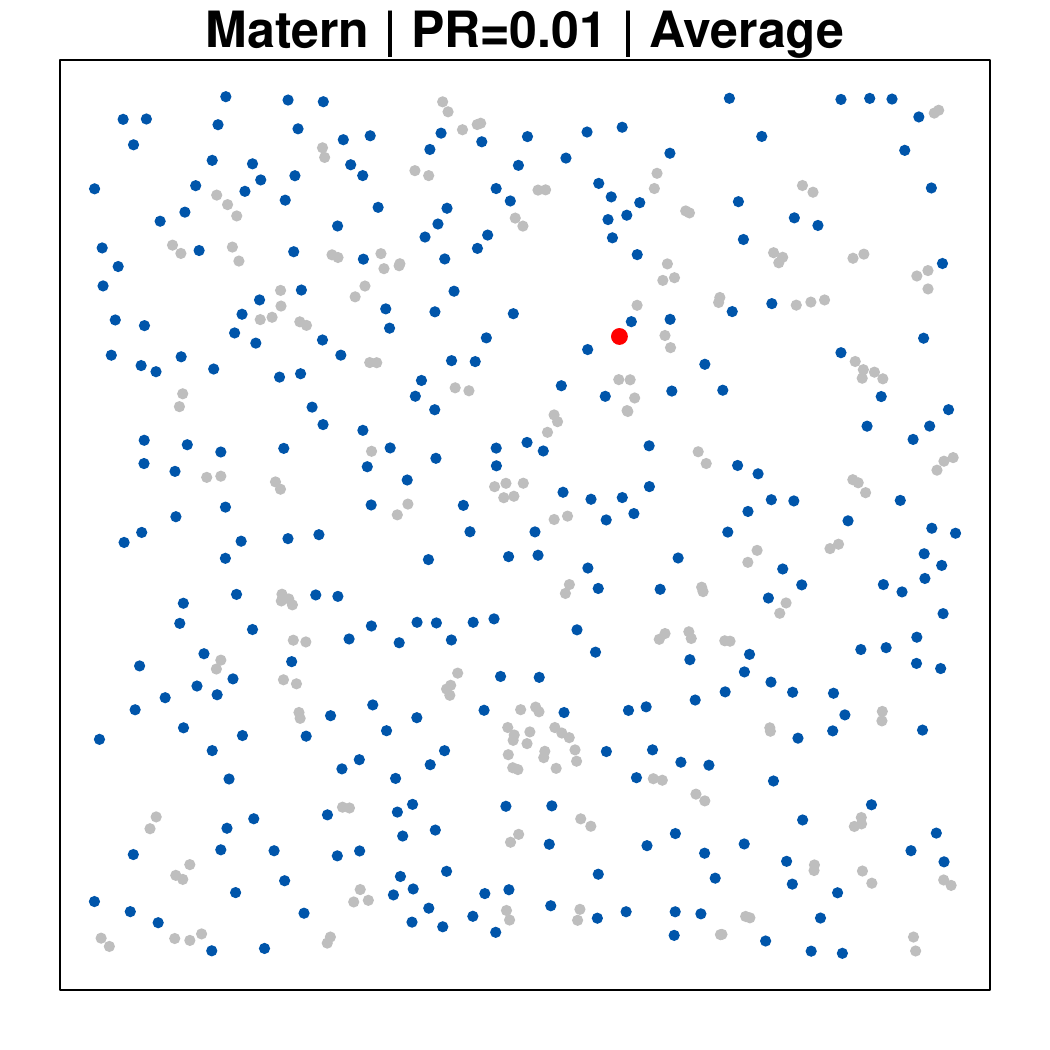}
            \includegraphics[width=0.32\textwidth]{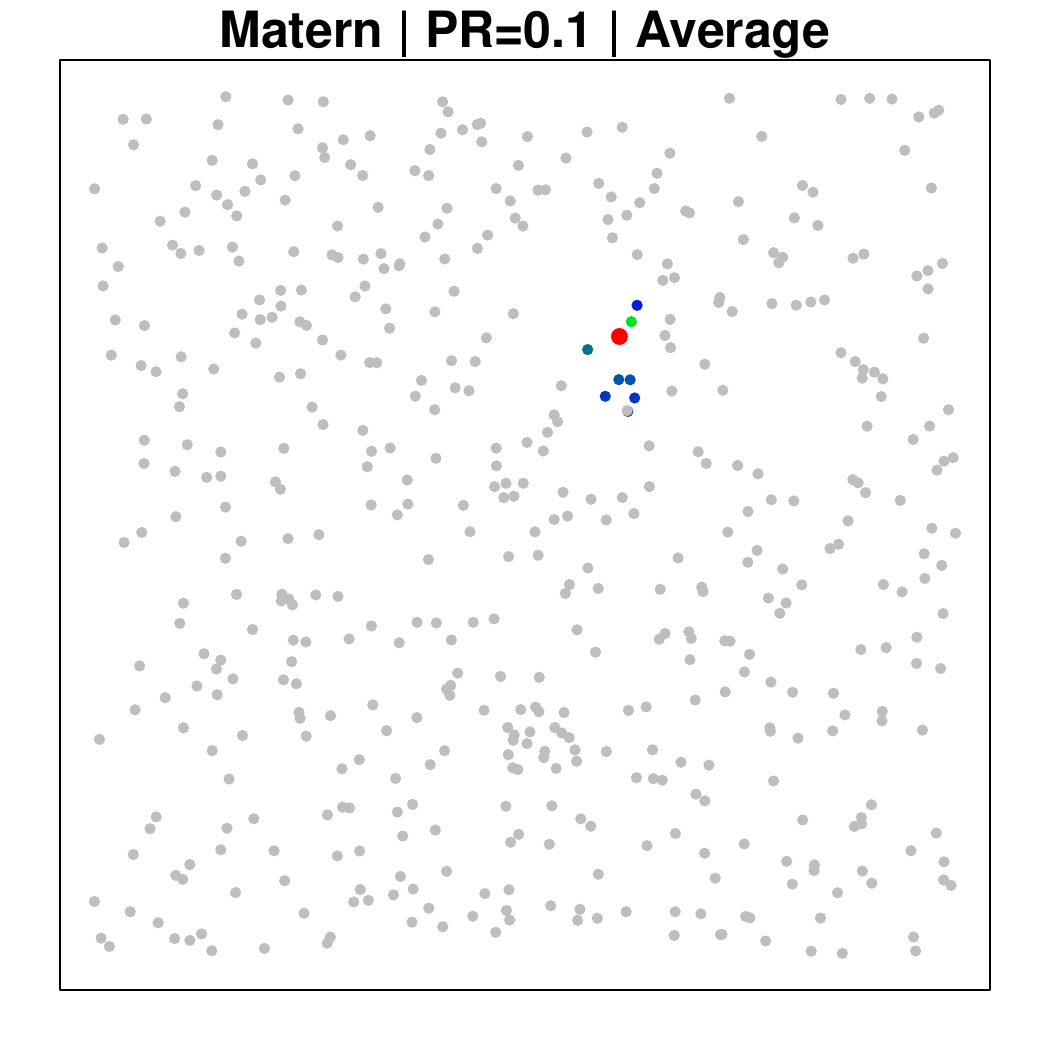}
            \includegraphics[width=0.32\textwidth]{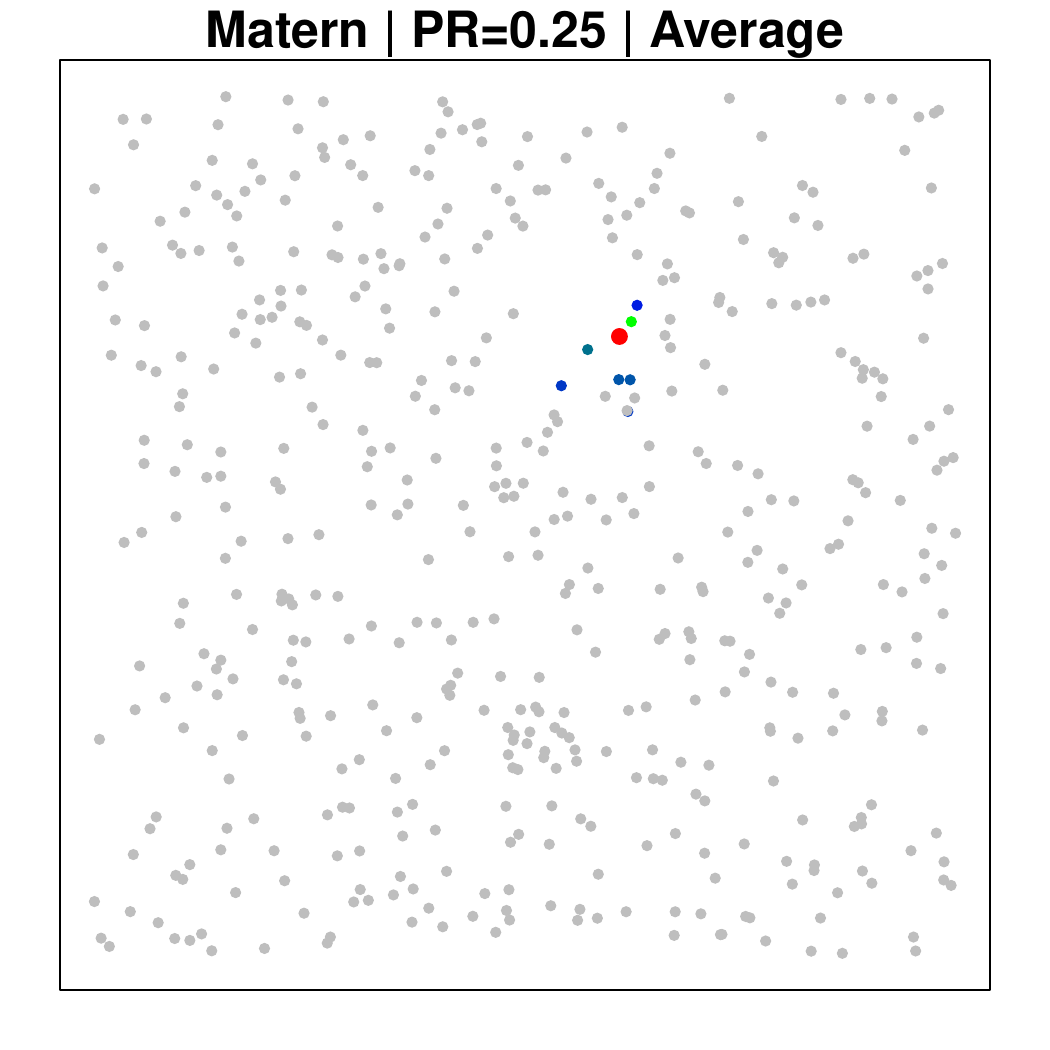}
        \end{minipage}
        
        \begin{minipage}{0.08\textwidth}
            \centering
            \rotatebox{90}{\small Densest}
        \end{minipage}
        \begin{minipage}{0.88\textwidth}
            \includegraphics[width=0.32\textwidth]{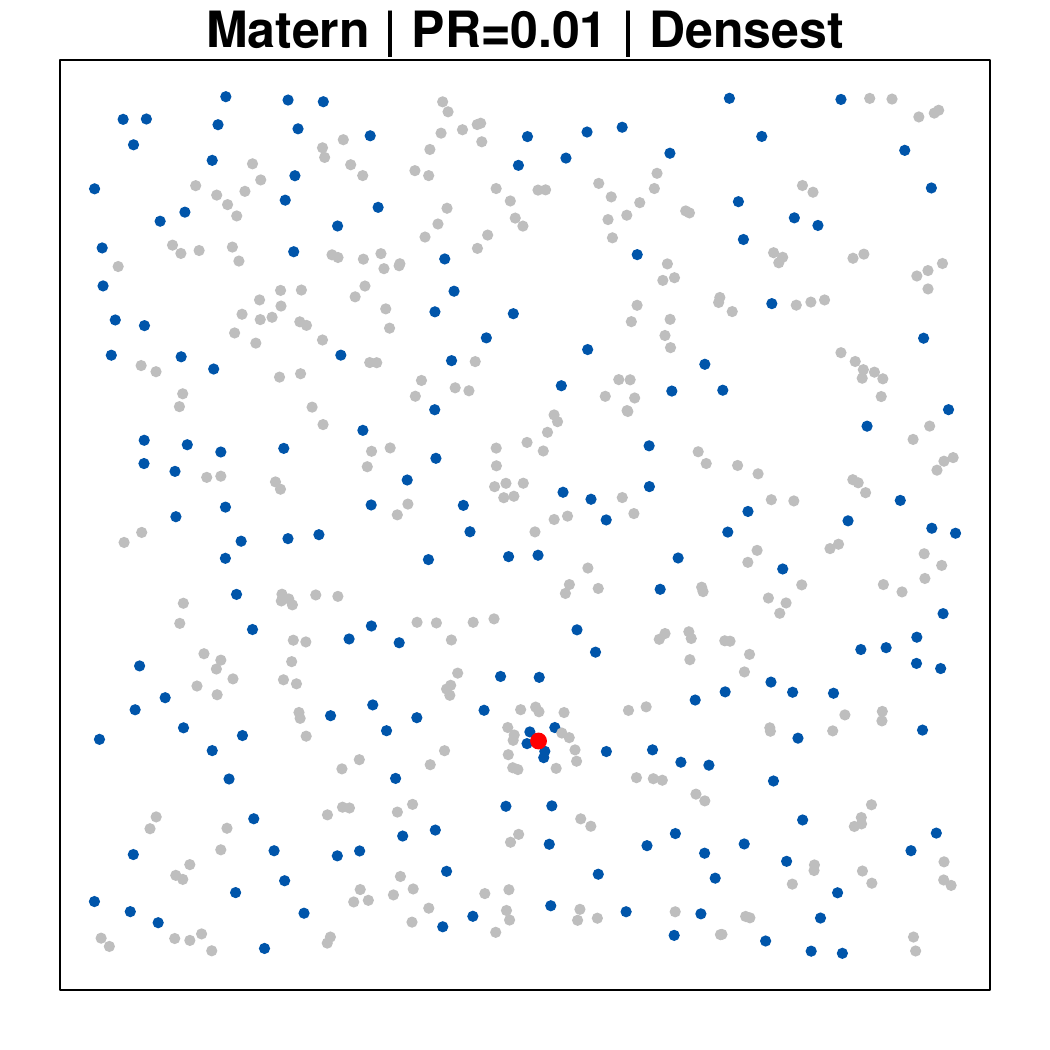}
            \includegraphics[width=0.32\textwidth]{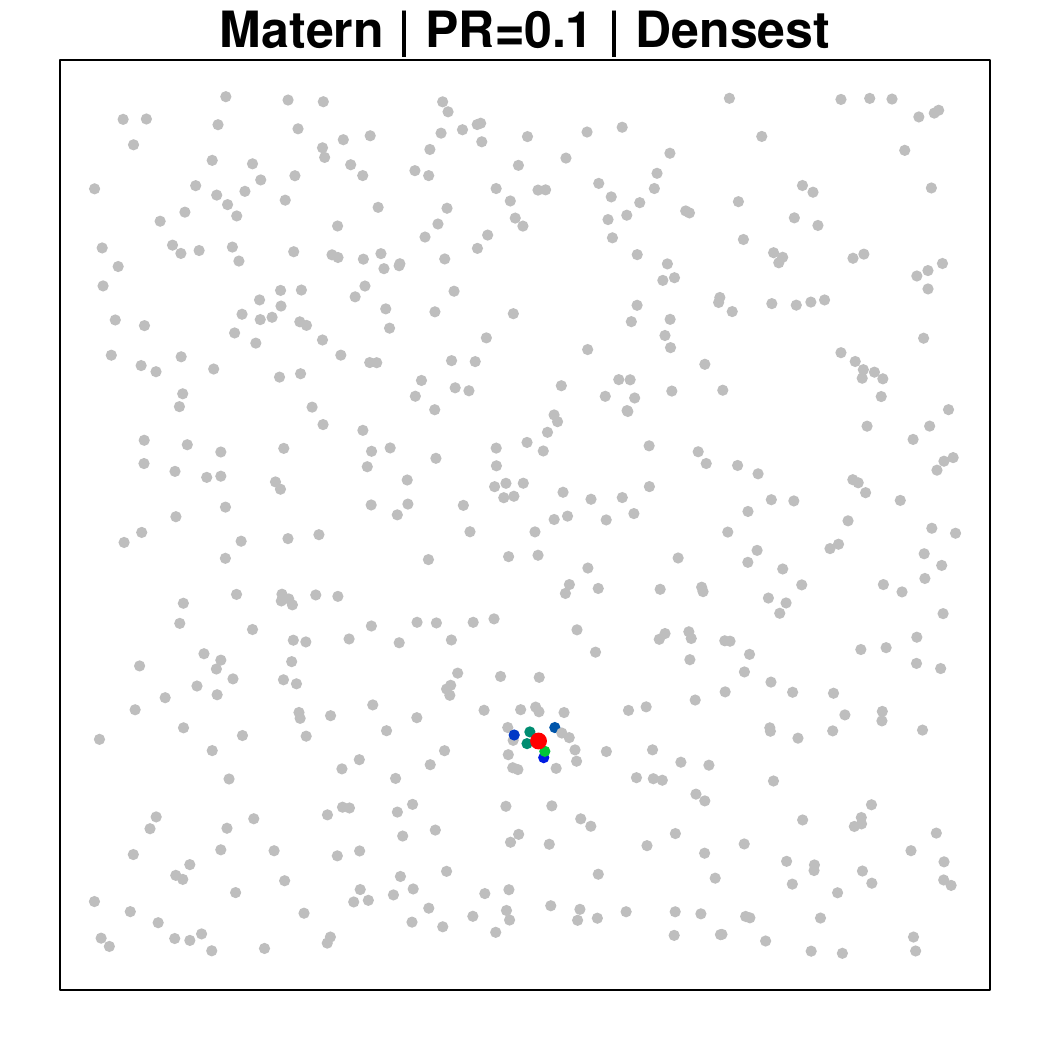}
            \includegraphics[width=0.32\textwidth]{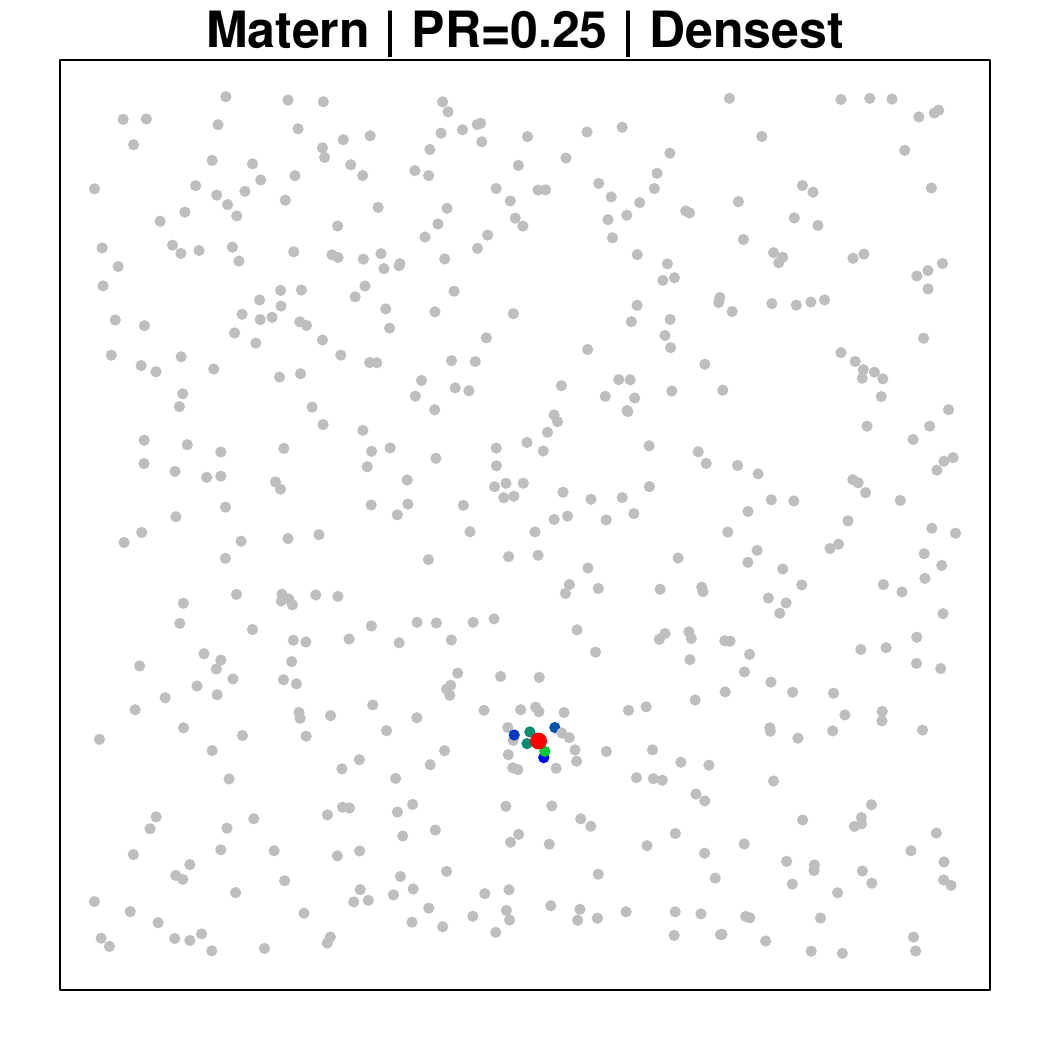}
        \end{minipage}
        
        \begin{minipage}{0.08\textwidth}
            \centering
            \rotatebox{90}{\small Corner}
        \end{minipage}
        \begin{minipage}{0.88\textwidth}
            \includegraphics[width=0.32\textwidth]{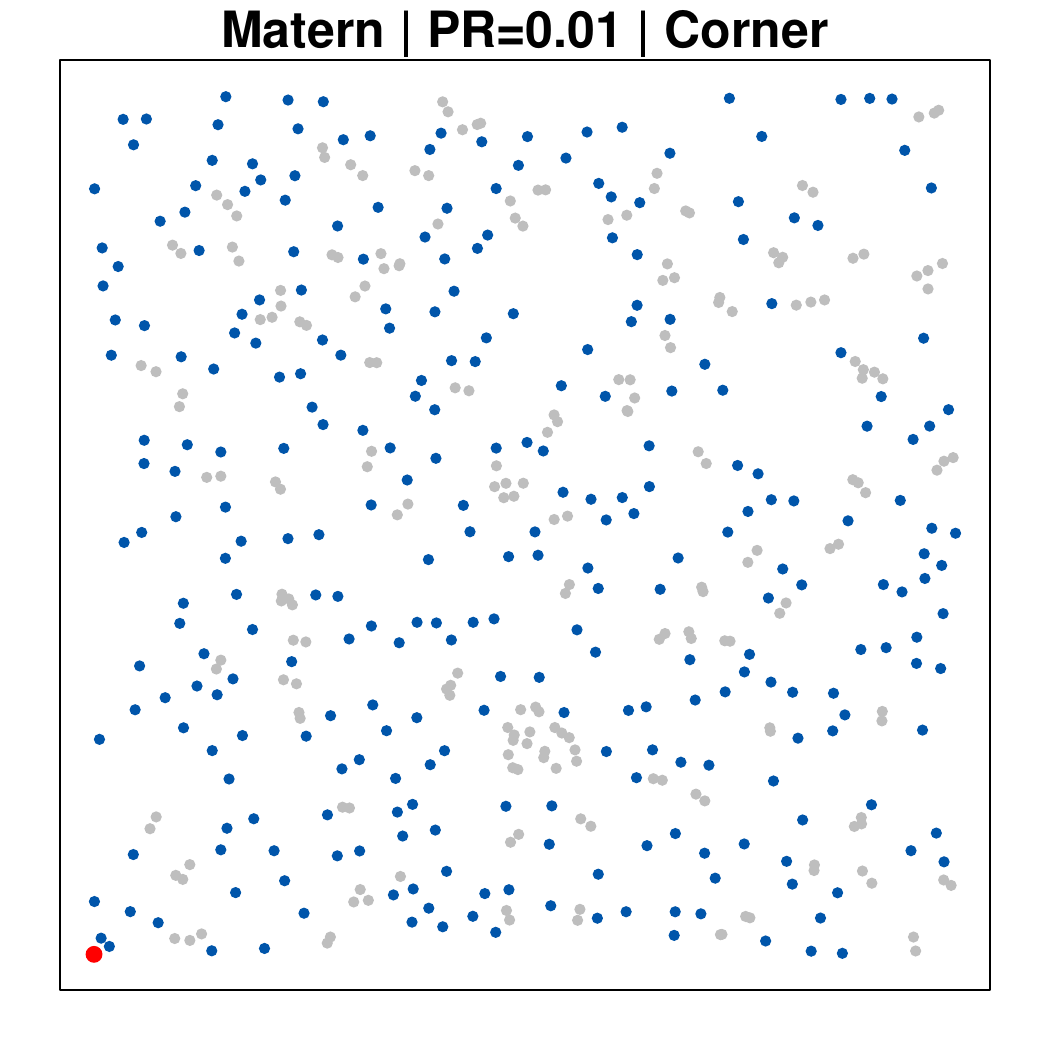}
            \includegraphics[width=0.32\textwidth]{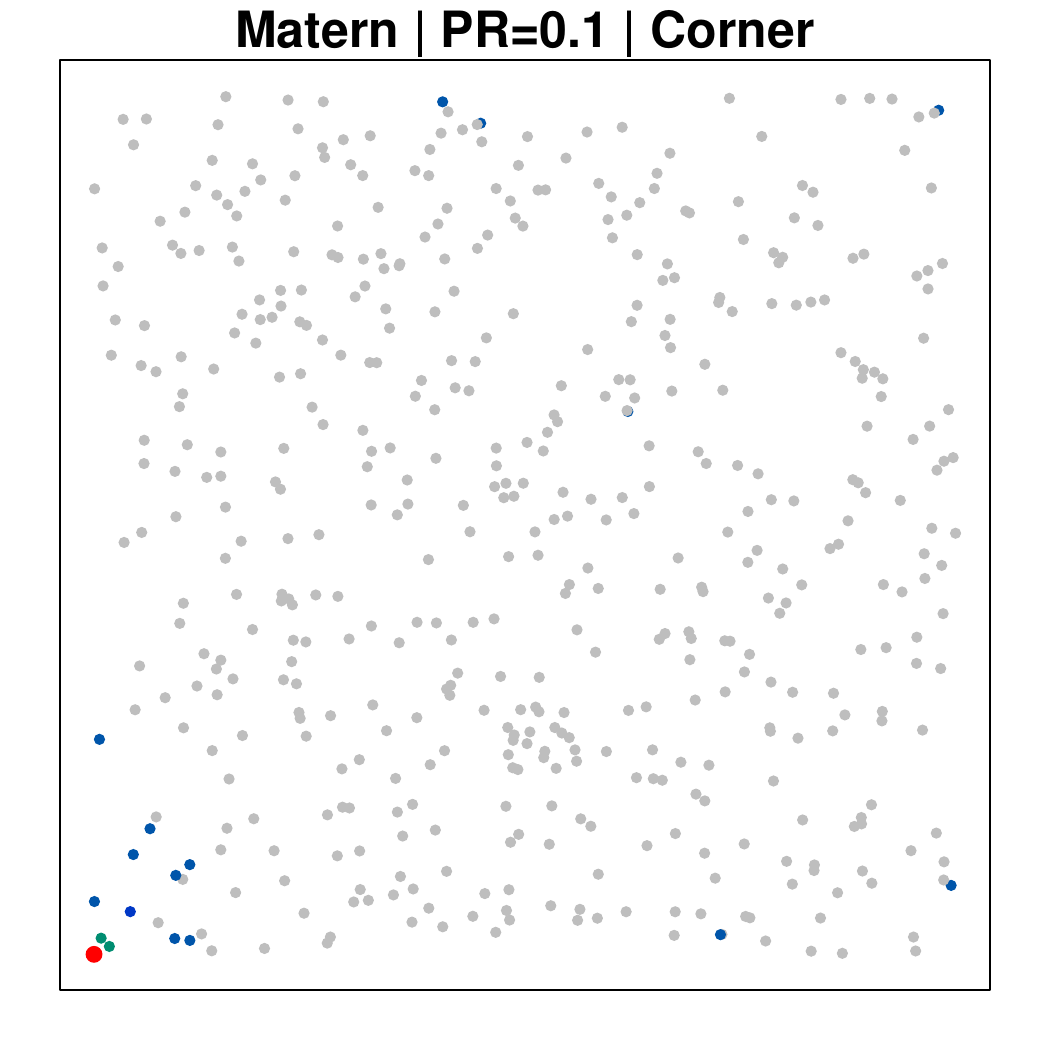}
            \includegraphics[width=0.32\textwidth]{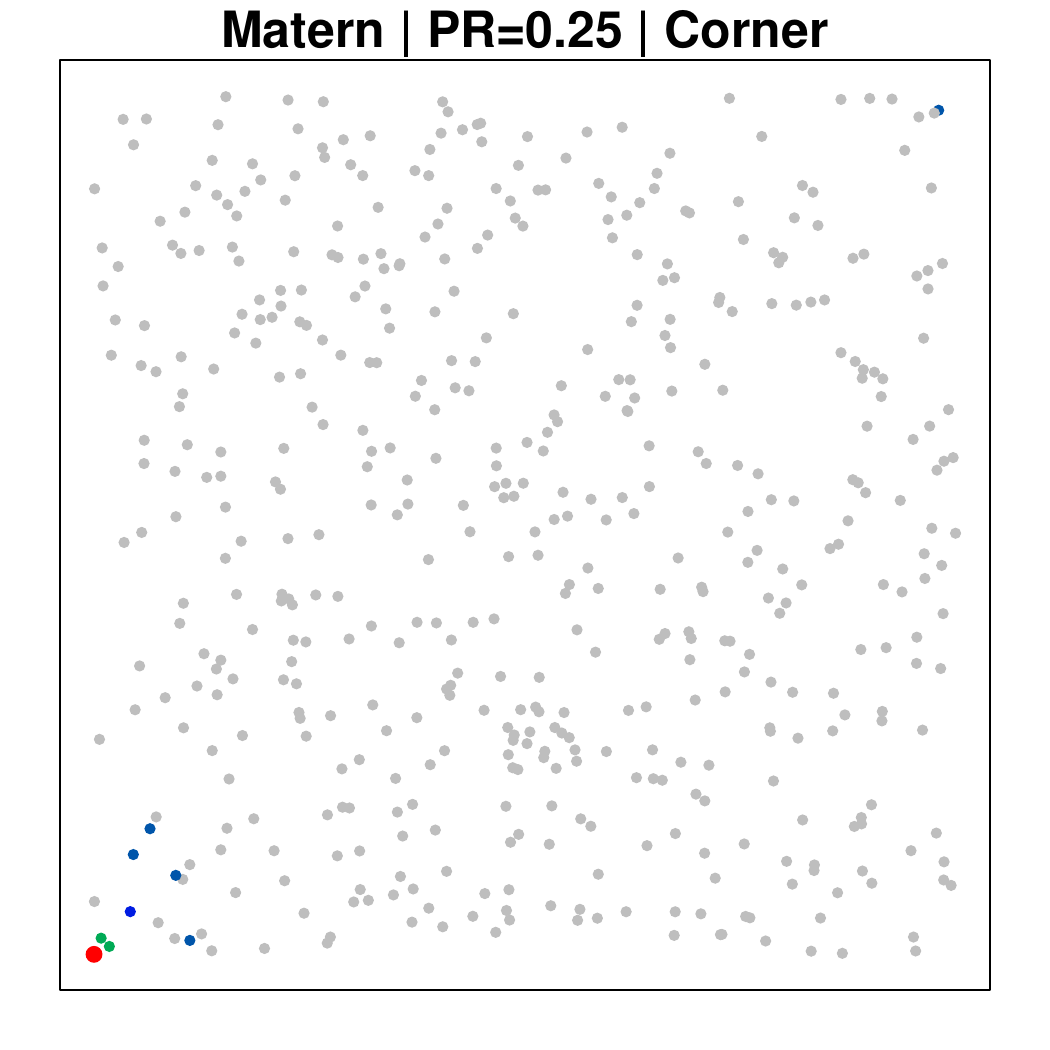}
        \end{minipage}
        
        \begin{minipage}{0.08\textwidth}
            \centering
            \rotatebox{90}{\small Side}
        \end{minipage}
        \begin{minipage}{0.88\textwidth}
            \includegraphics[width=0.32\textwidth]{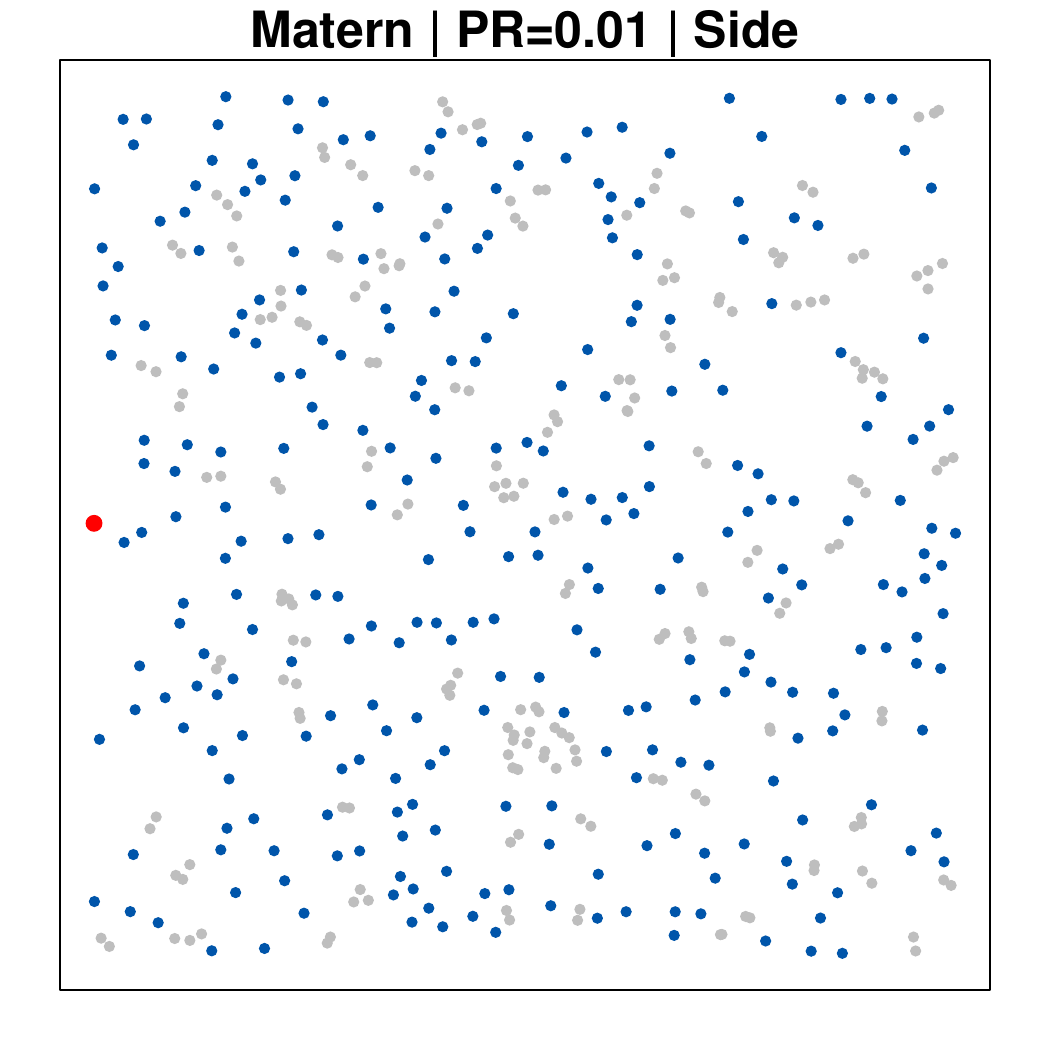}
            \includegraphics[width=0.32\textwidth]{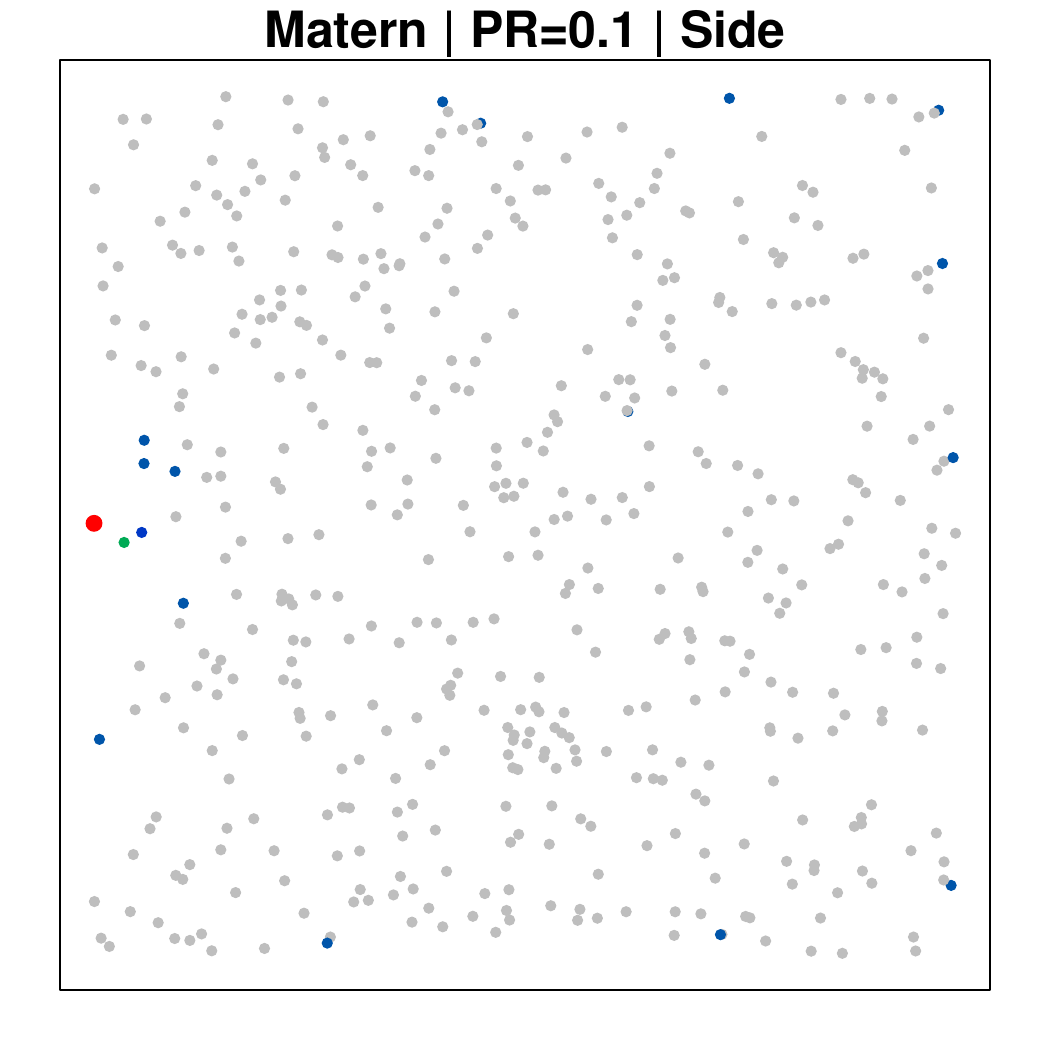}
            \includegraphics[width=0.32\textwidth]{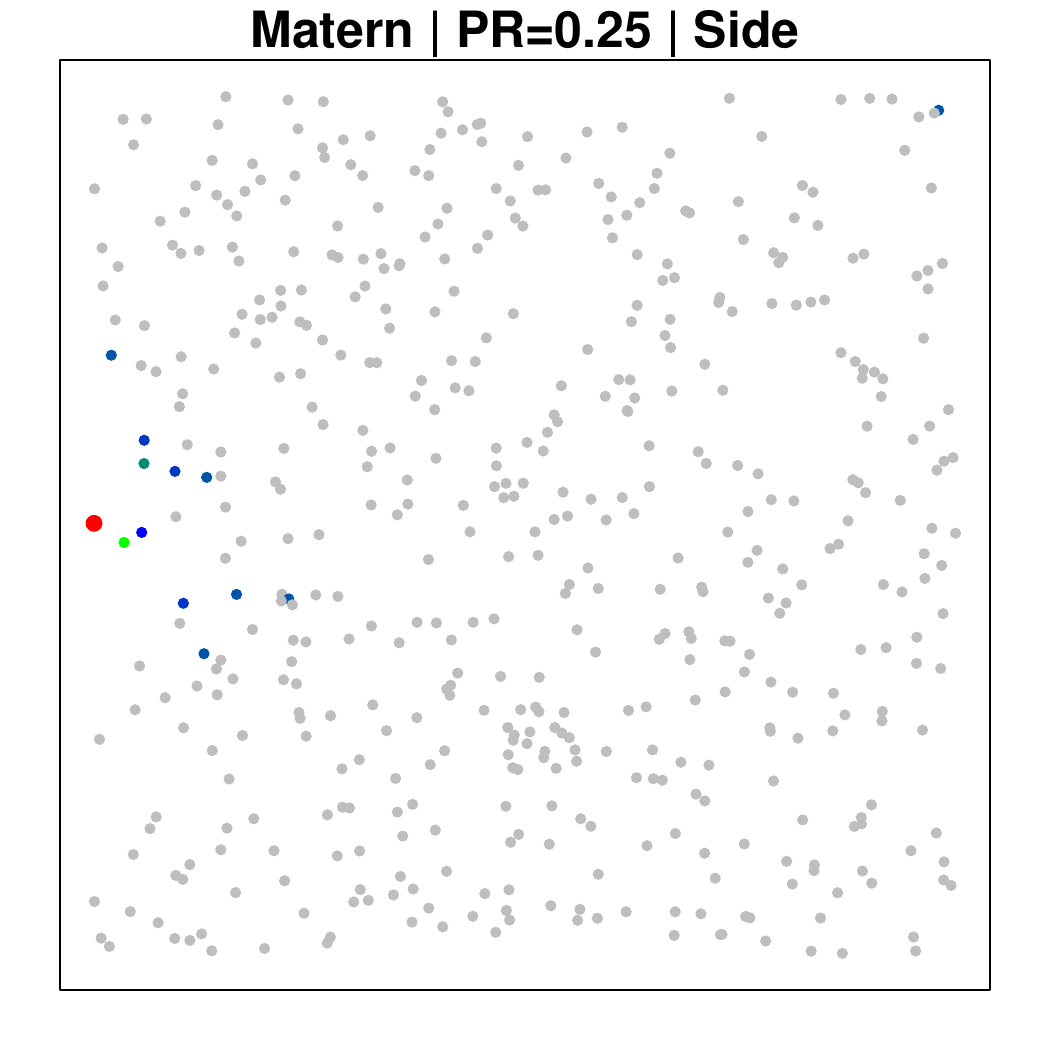}
        \end{minipage}
        
        \vspace{0.3cm}
        \begin{minipage}{0.08\textwidth}
        \end{minipage}
        \begin{minipage}{0.88\textwidth}
            \centering
            \textbf{Low PR} \hspace{2.5cm} \textbf{Med PR} \hspace{2.5cm} \textbf{Lar PR}
        \end{minipage}
    \end{minipage}%
    \hfill
    \begin{minipage}{0.06\textwidth}
        \centering
        \includegraphics[width=1.75\textwidth, keepaspectratio=false]{figs/Global_Legend.pdf}
    \end{minipage}
    
     \caption{\small Spatial distribution of adaptive LASSO kriging coefficients for Matérn covariance ($\nu=1.5$) across five prediction locations (rows) and three practical range levels (columns). Grey points indicate neighbors with zero coefficients (not selected). Colored points show selected neighbors, with color intensity indicating coefficient magnitude (blue = small, green = large). The red point marks the prediction location $\bm{s}_0$.}
    \label{fig:spatial_coefficients_matern}
\end{figure}

\begin{figure}[ht!]
    \centering
    \begin{minipage}{0.9\textwidth}
        \begin{minipage}{0.08\textwidth}
            \centering
            \rotatebox{90}{\small Farthest}
        \end{minipage}
        \begin{minipage}{0.88\textwidth}
            \includegraphics[width=0.32\textwidth]{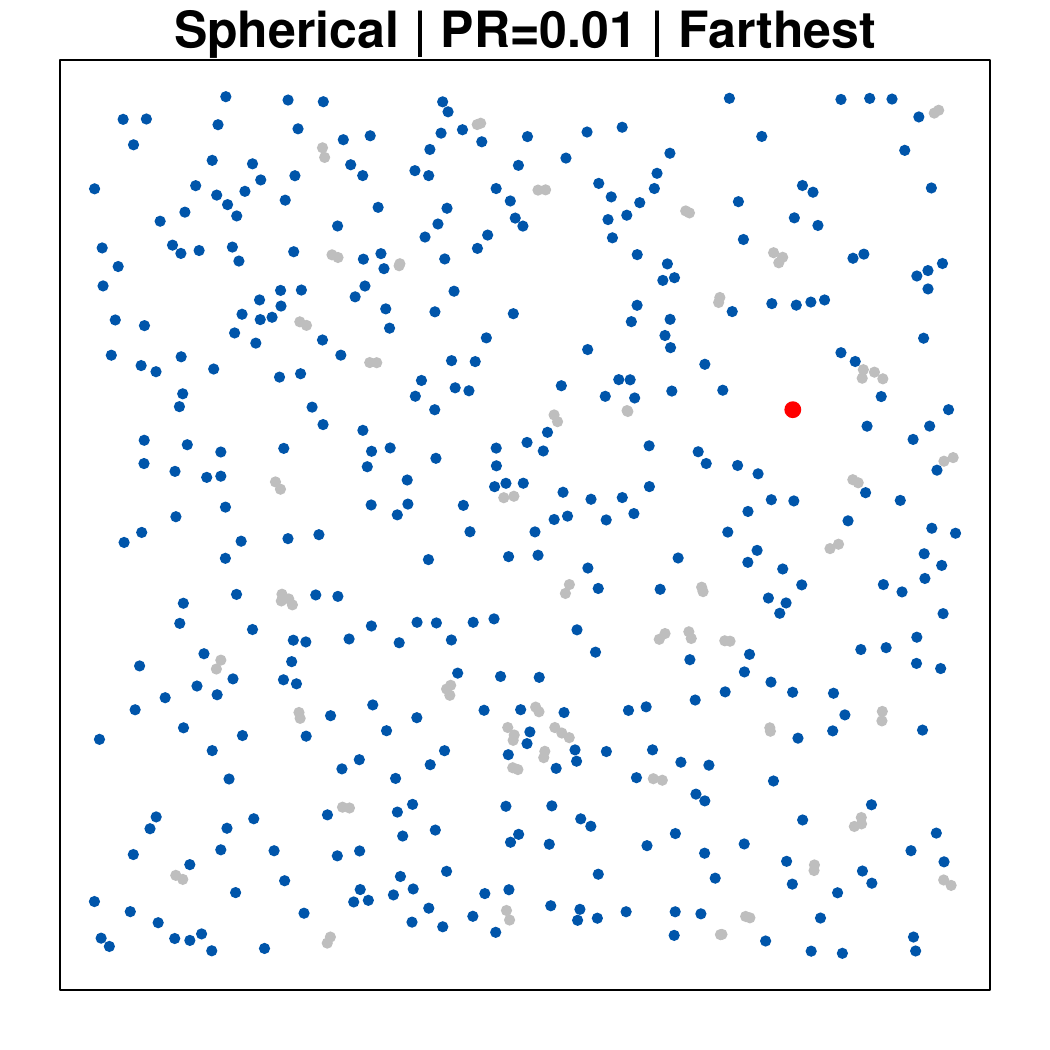}
            \includegraphics[width=0.32\textwidth]{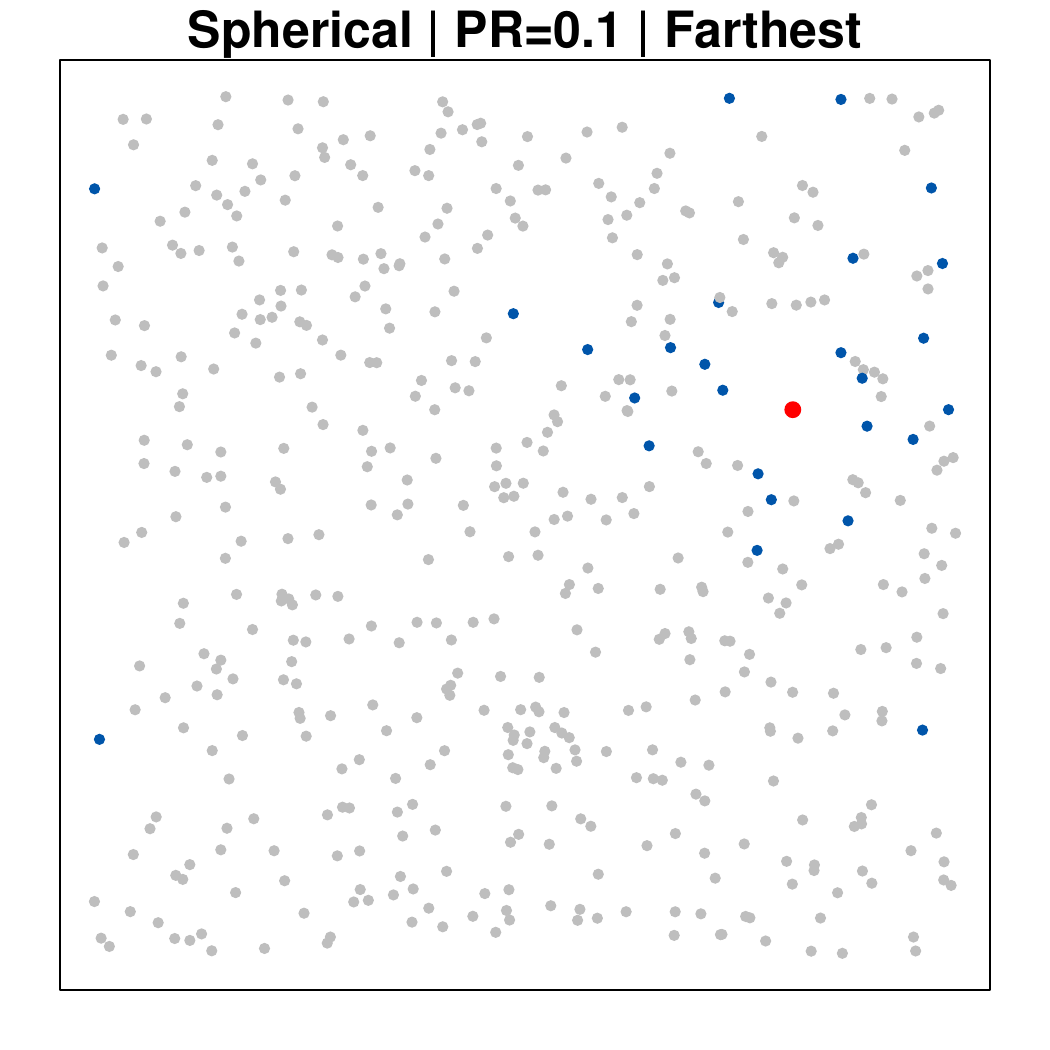}
            \includegraphics[width=0.32\textwidth]{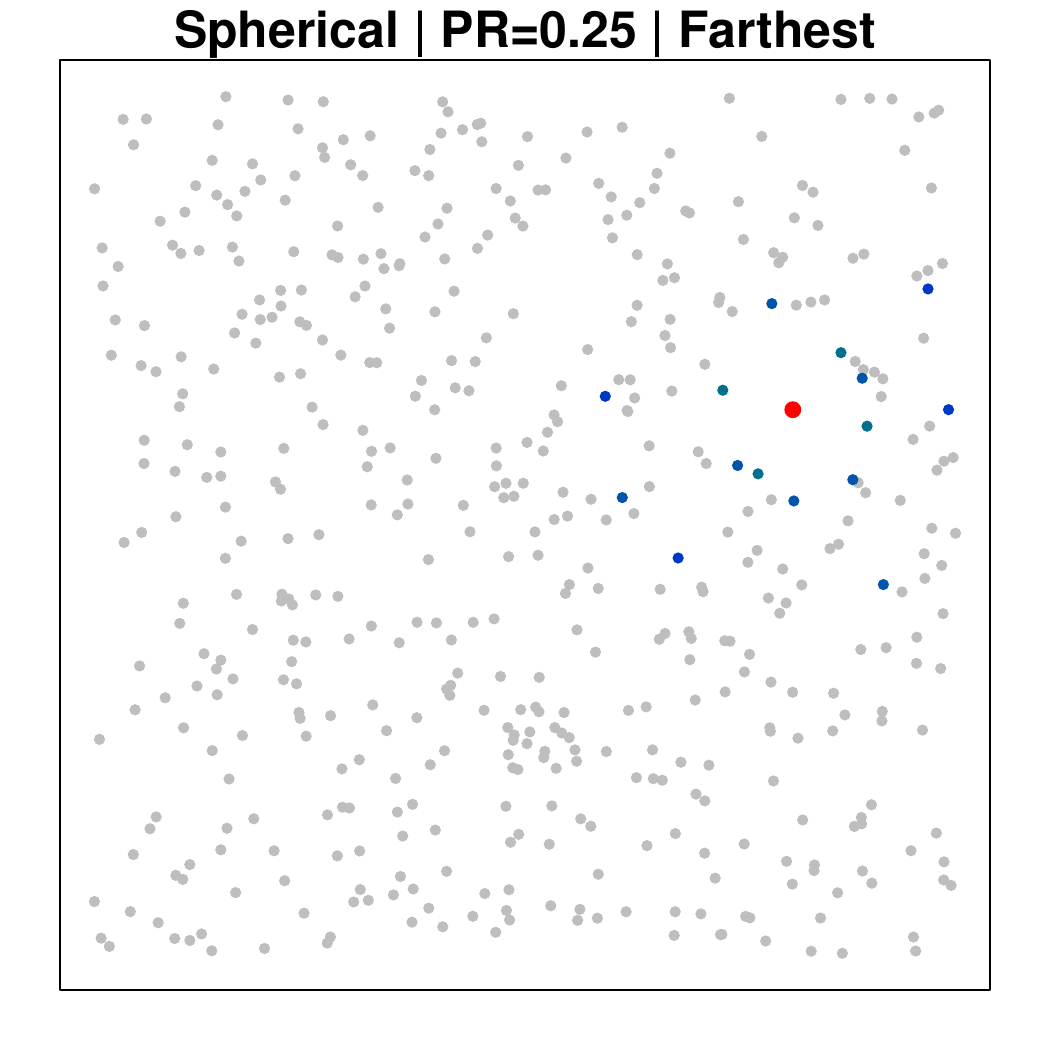}
        \end{minipage}
        
        \begin{minipage}{0.08\textwidth}
            \centering
            \rotatebox{90}{\small Average}
        \end{minipage}
        \begin{minipage}{0.88\textwidth}
            \includegraphics[width=0.32\textwidth]{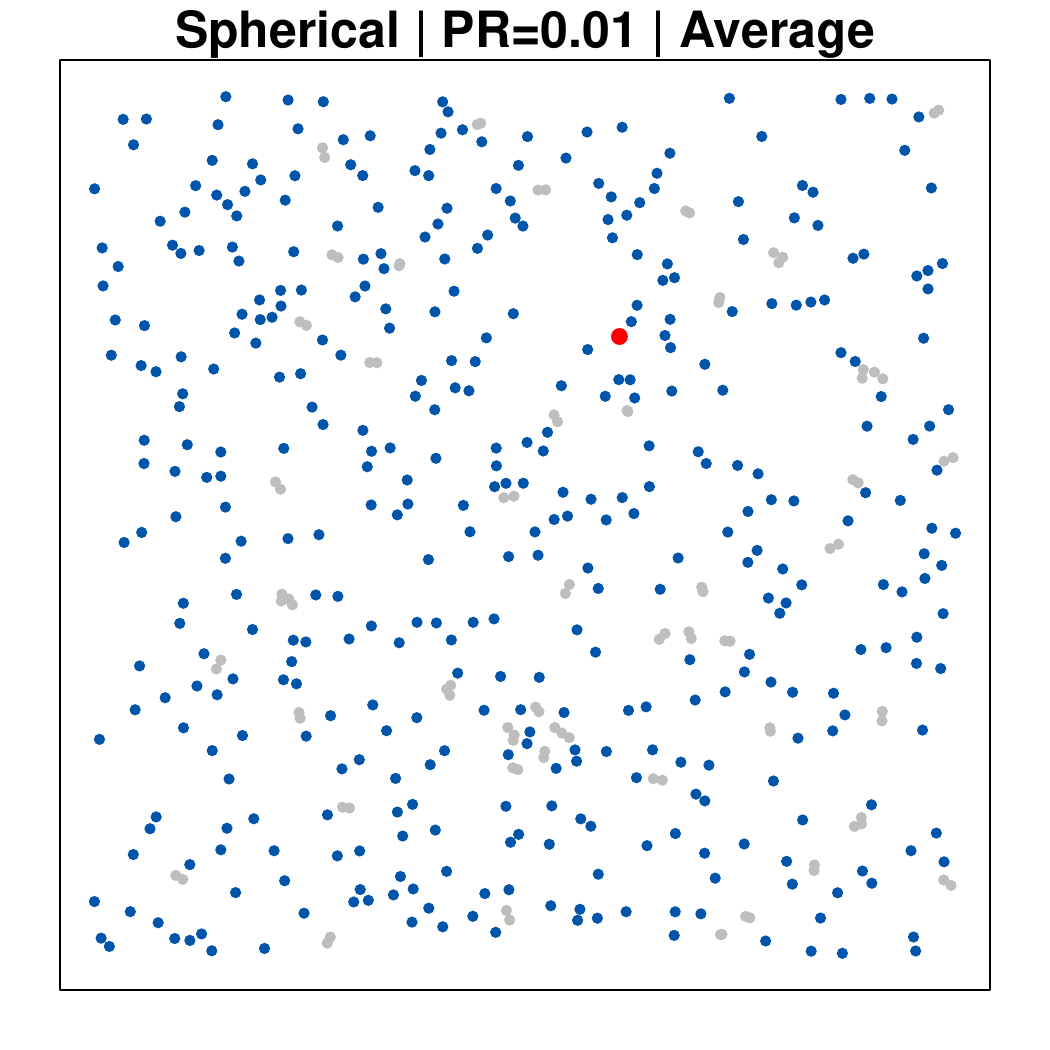}
            \includegraphics[width=0.32\textwidth]{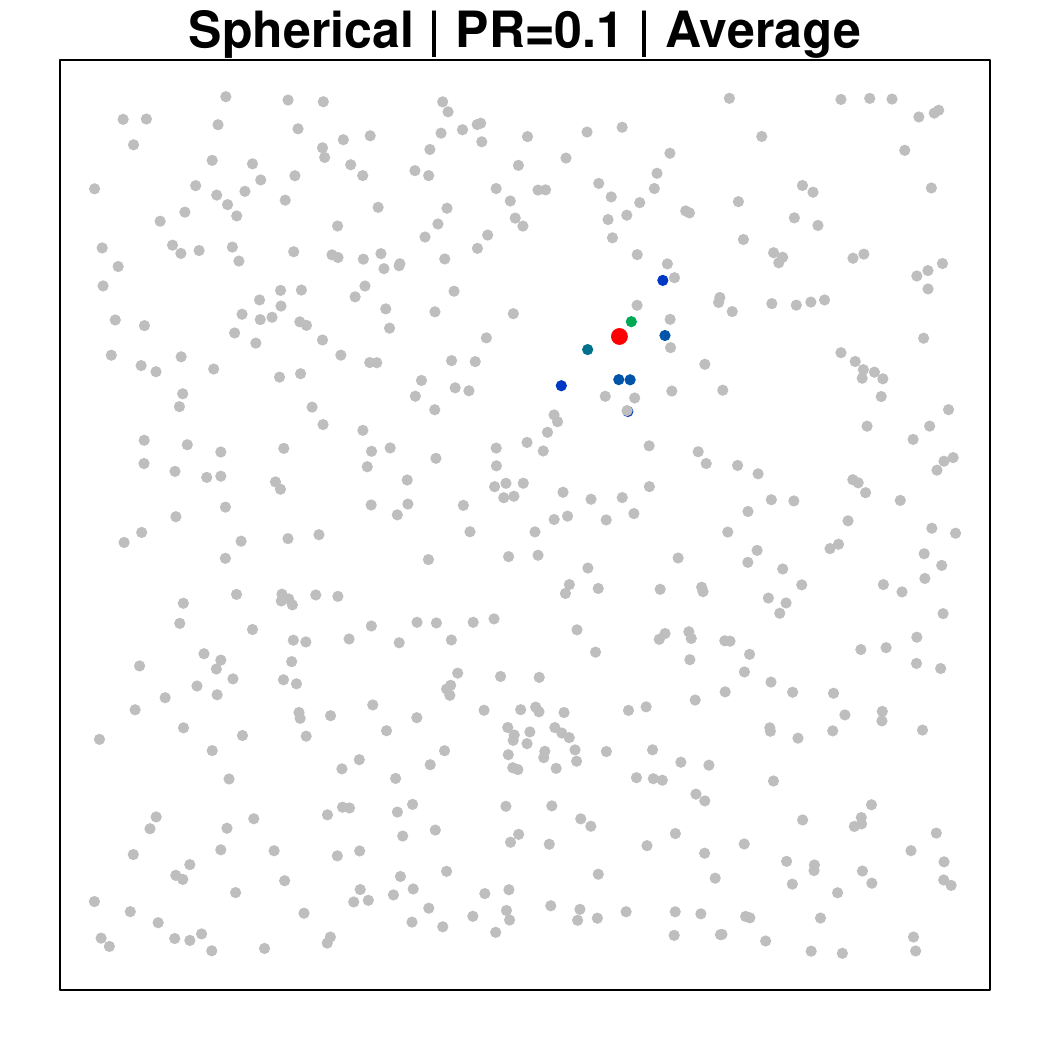}
            \includegraphics[width=0.32\textwidth]{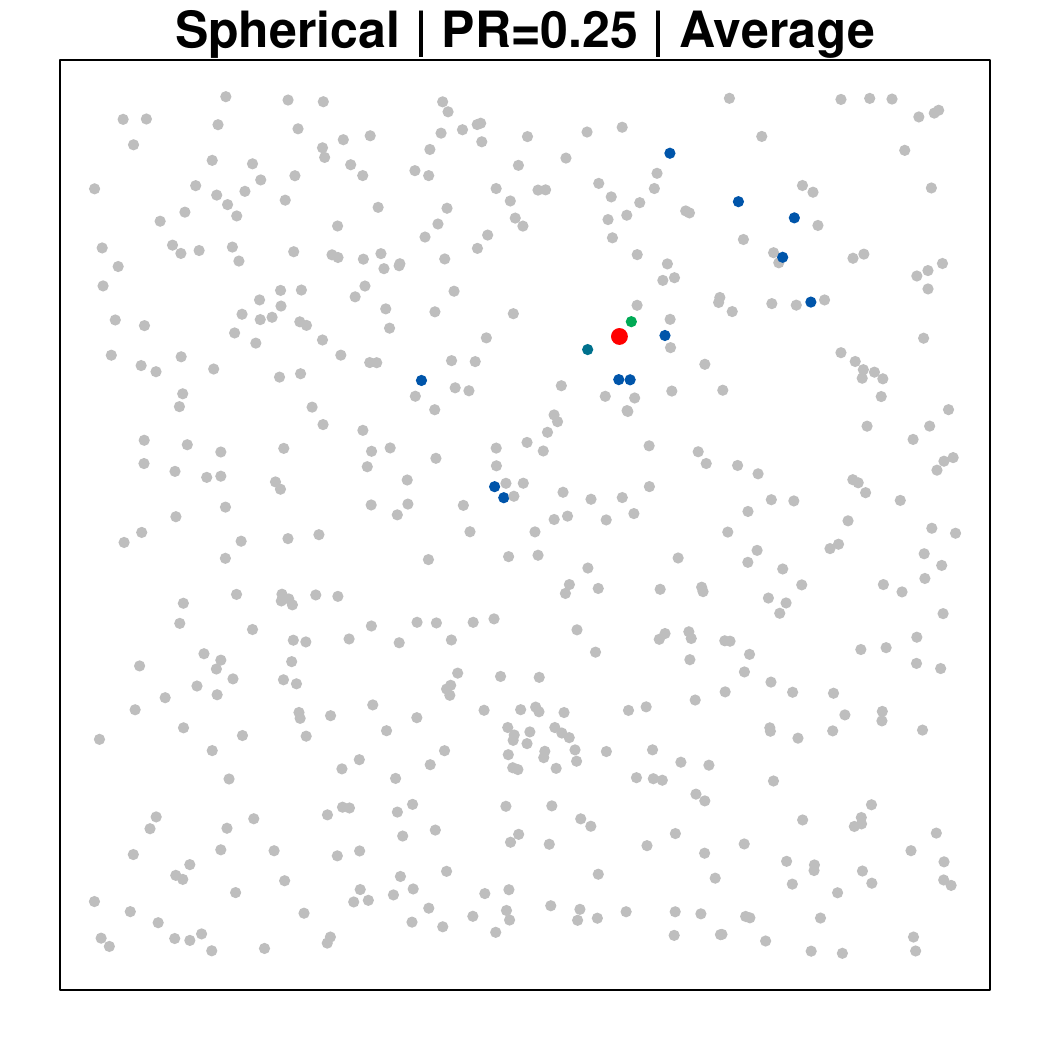}
        \end{minipage}
        
        \begin{minipage}{0.08\textwidth}
            \centering
            \rotatebox{90}{\small Densest}
        \end{minipage}
        \begin{minipage}{0.88\textwidth}
            \includegraphics[width=0.32\textwidth]{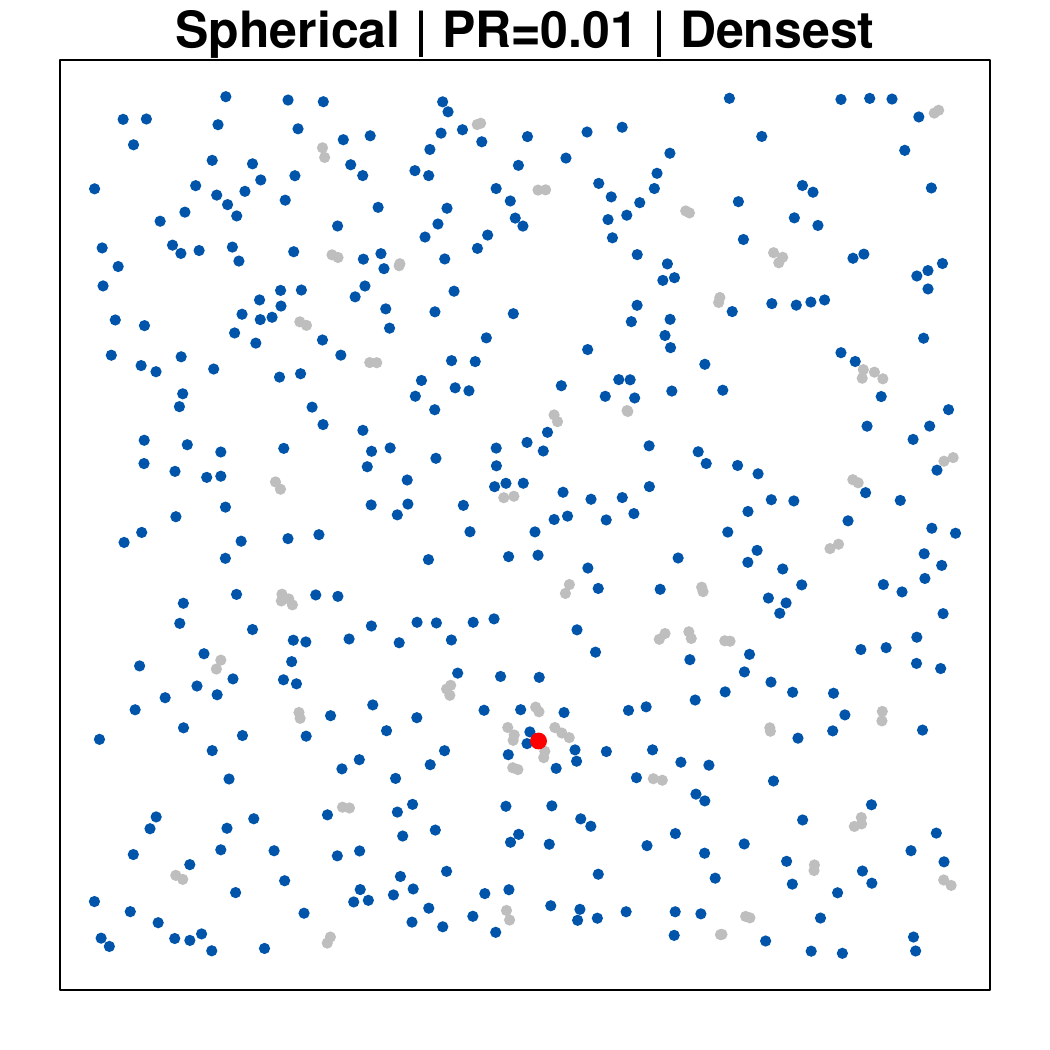}
            \includegraphics[width=0.32\textwidth]{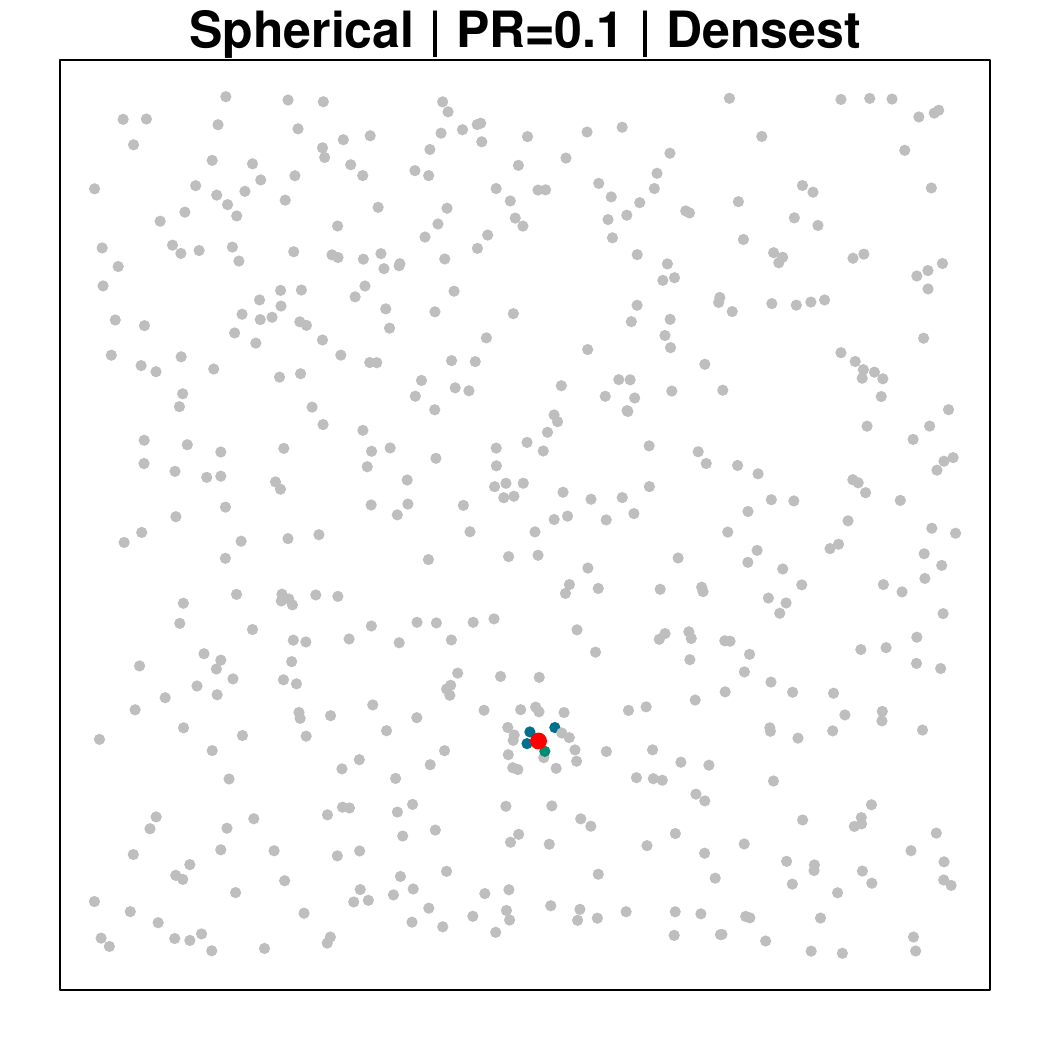}
            \includegraphics[width=0.32\textwidth]{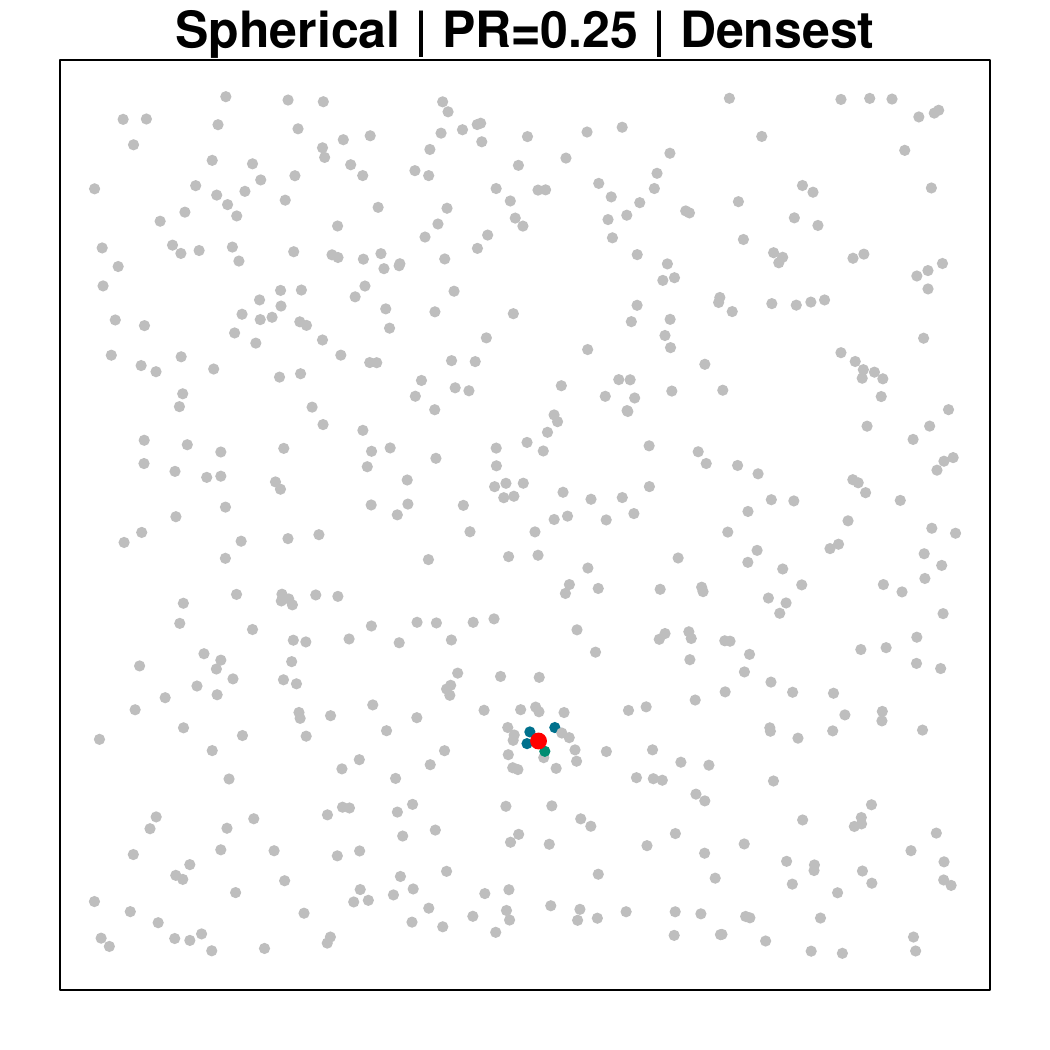}
        \end{minipage}
        
        \begin{minipage}{0.08\textwidth}
            \centering
            \rotatebox{90}{\small Corner}
        \end{minipage}
        \begin{minipage}{0.88\textwidth}
            \includegraphics[width=0.32\textwidth]{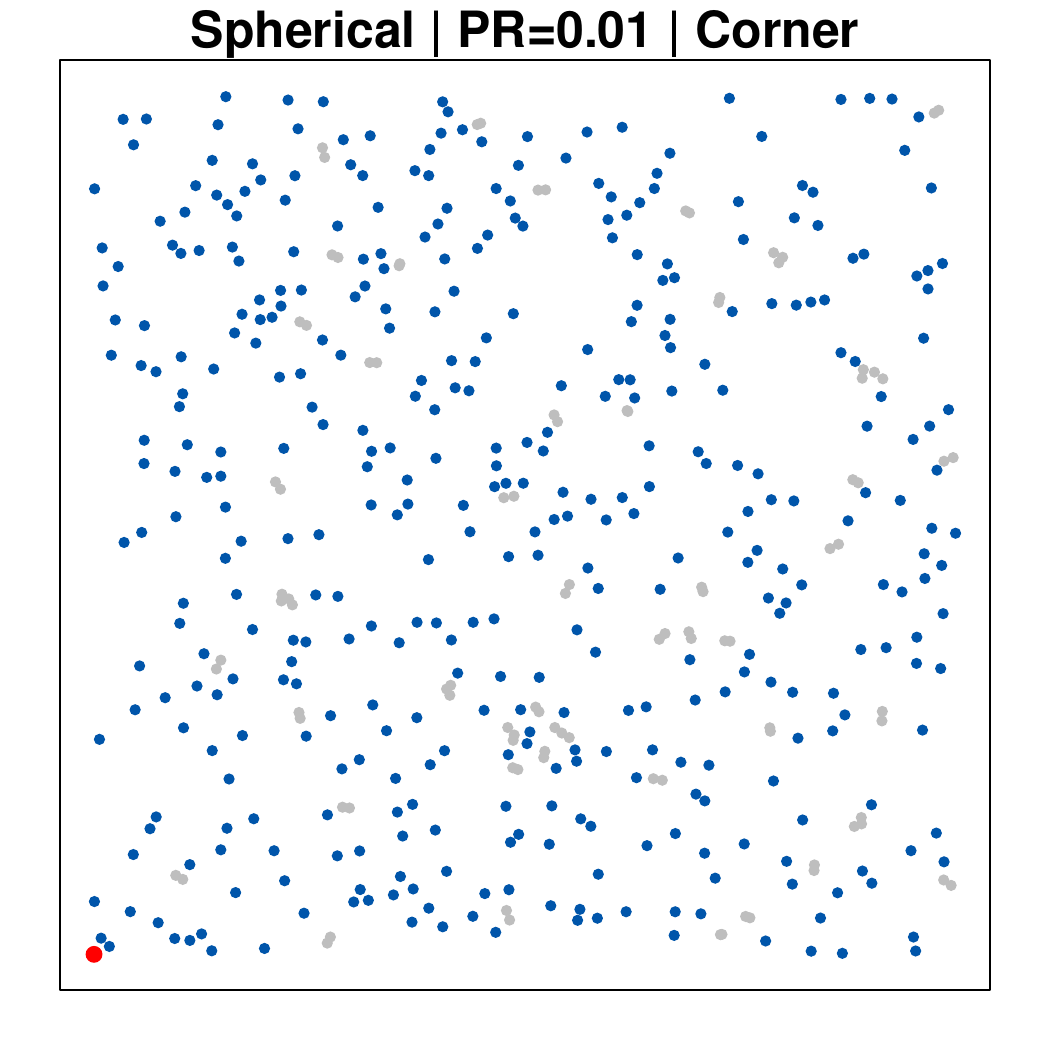}
            \includegraphics[width=0.32\textwidth]{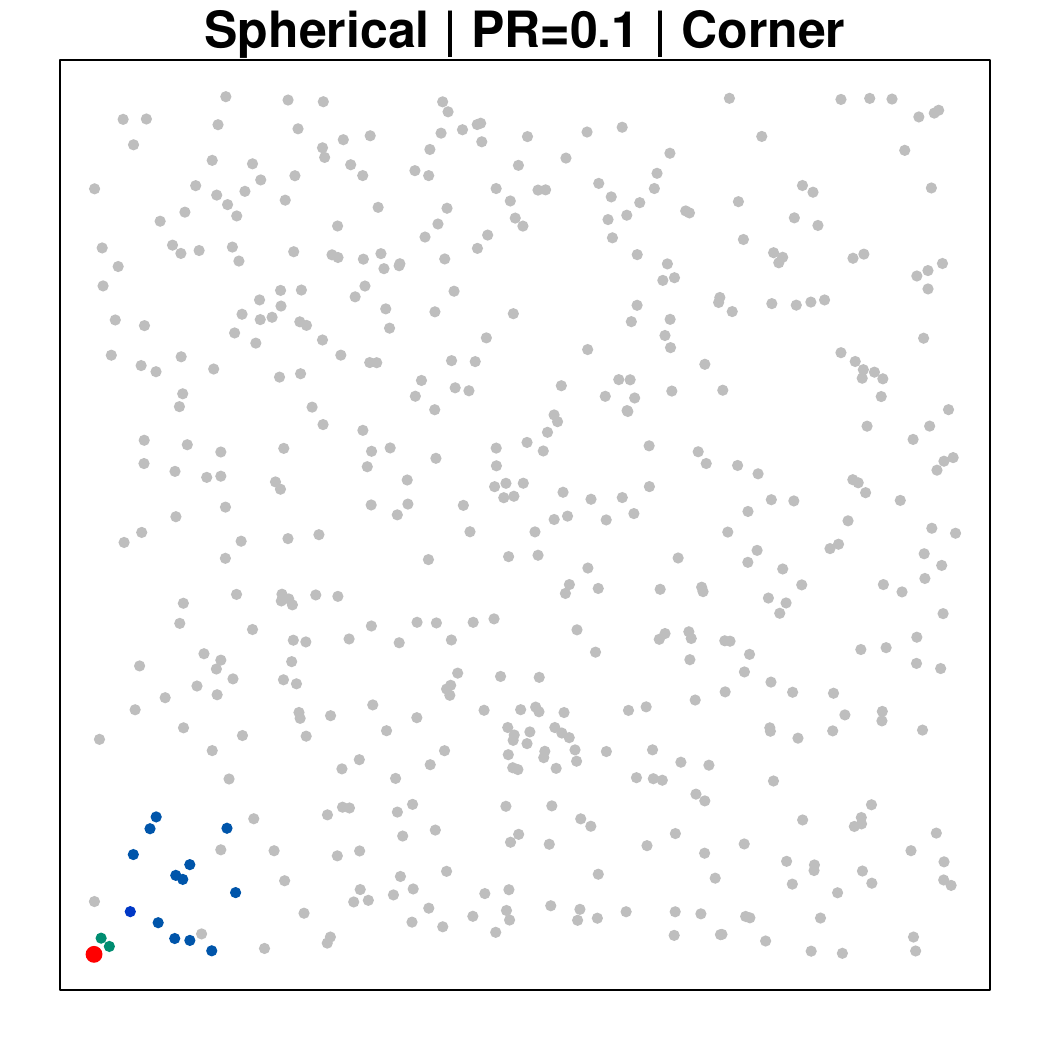}
            \includegraphics[width=0.32\textwidth]{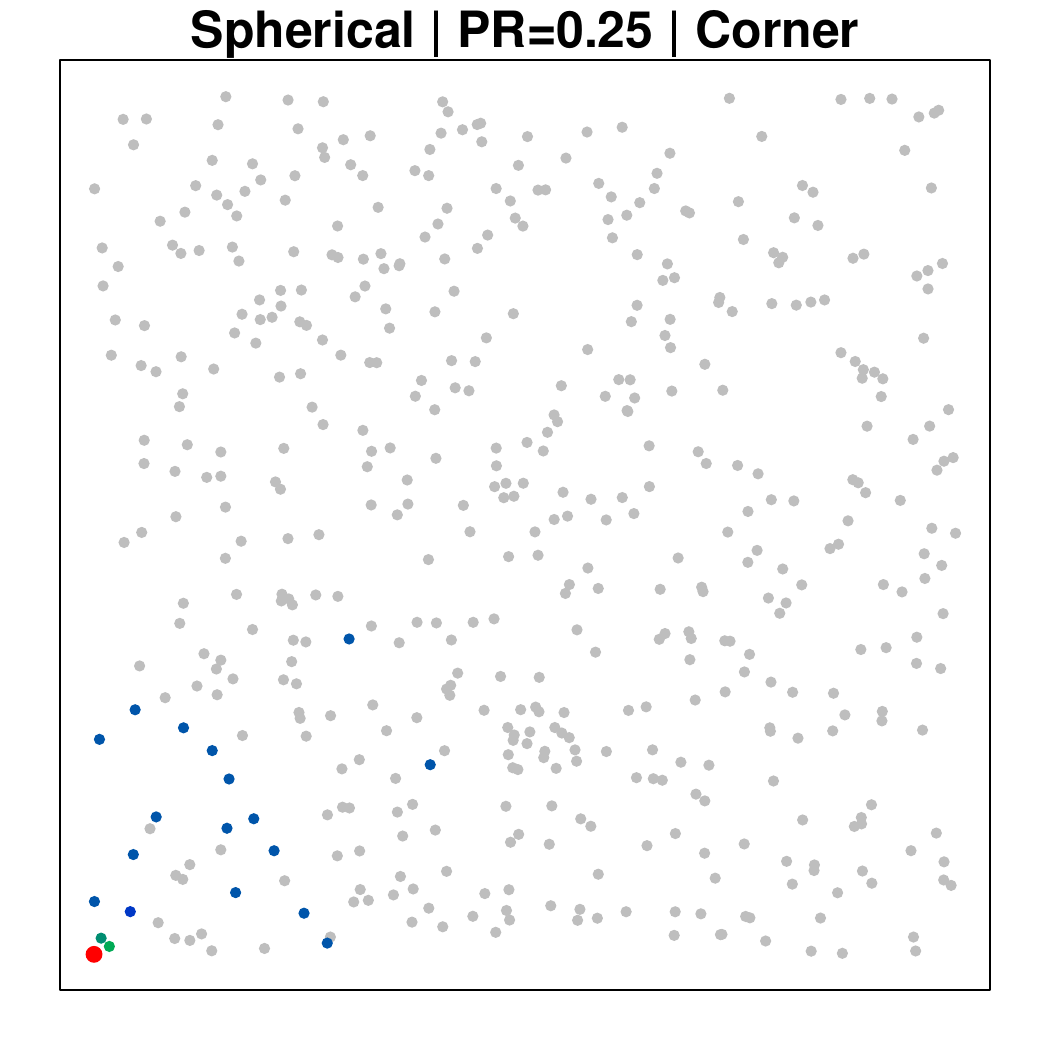}
        \end{minipage}
        
        \begin{minipage}{0.08\textwidth}
            \centering
            \rotatebox{90}{\small Side}
        \end{minipage}
        \begin{minipage}{0.88\textwidth}
            \includegraphics[width=0.32\textwidth]{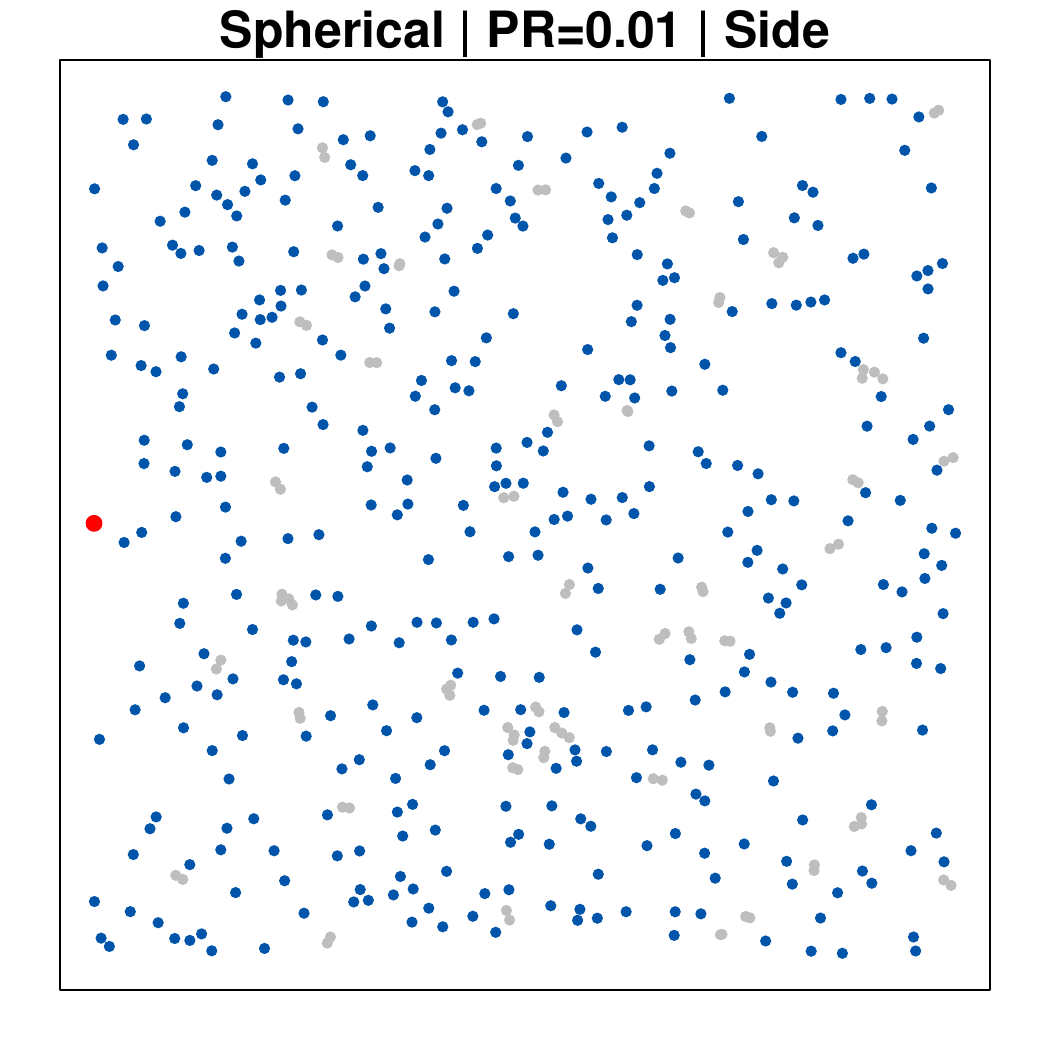}
            \includegraphics[width=0.32\textwidth]{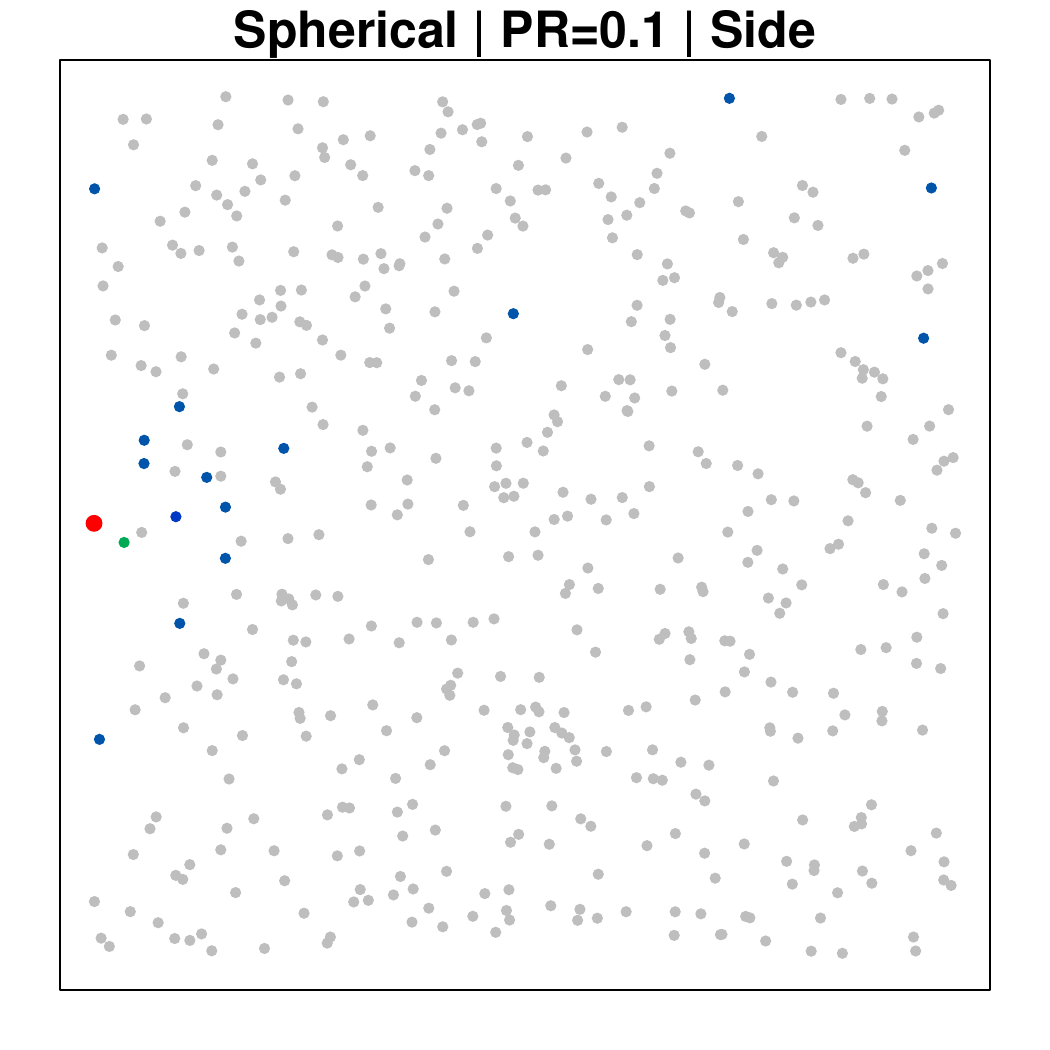}
            \includegraphics[width=0.32\textwidth]{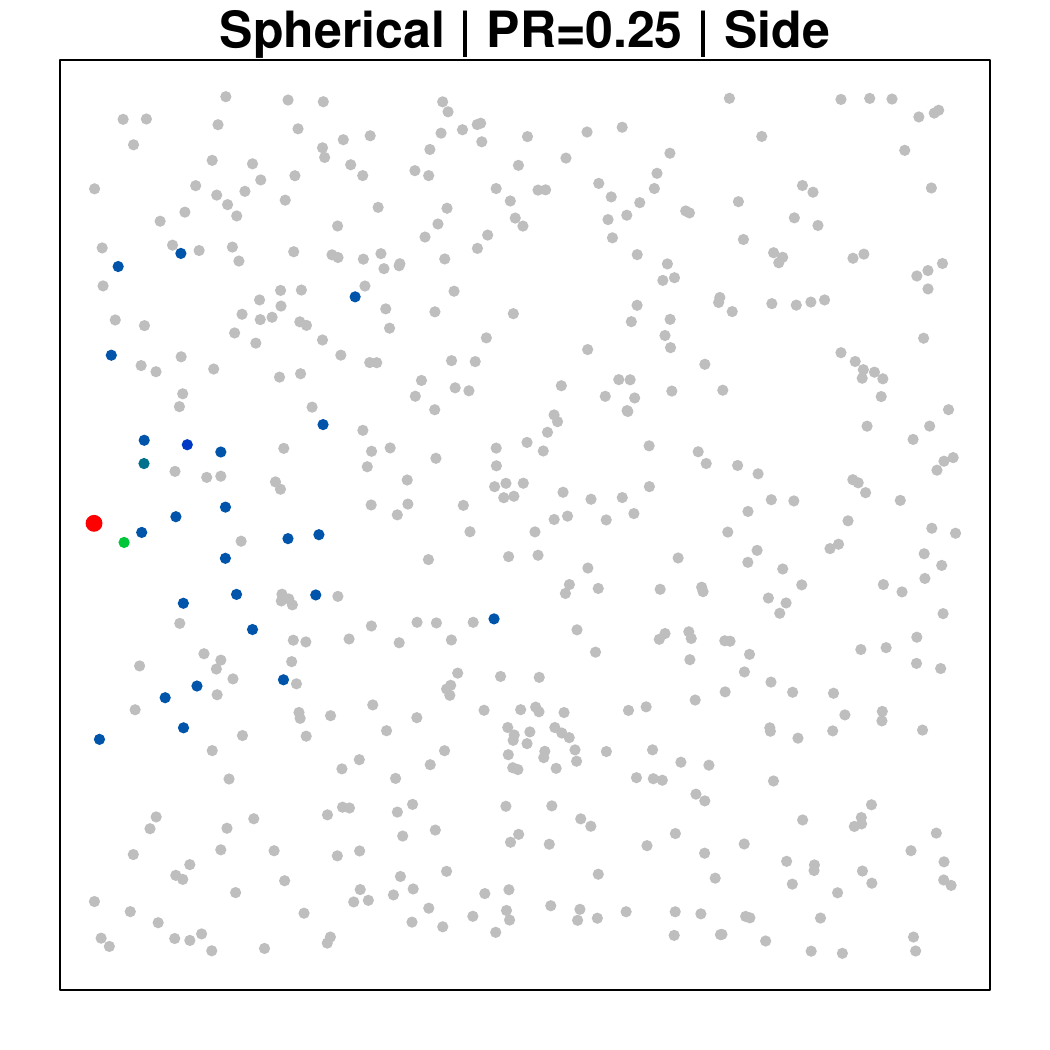}
        \end{minipage}
        
        \vspace{0.3cm}
        \begin{minipage}{0.08\textwidth}
        \end{minipage}
        \begin{minipage}{0.88\textwidth}
            \centering
            \textbf{Low PR} \hspace{2.5cm} \textbf{Med PR} \hspace{2.5cm} \textbf{Lar PR}
        \end{minipage}
    \end{minipage}%
    \hfill
    \begin{minipage}{0.06\textwidth}
        \centering
        \includegraphics[width=1.75\textwidth,  keepaspectratio=false]{figs/Global_Legend.pdf}
    \end{minipage}
    
     \caption{\small Spatial distribution of adaptive LASSO kriging coefficients for spherical covariance across five prediction locations (rows) and three practical range levels (columns). Grey points indicate neighbors with zero coefficients (not selected). Colored points show selected neighbors, with color intensity indicating coefficient magnitude (blue = small, green = large). The red point marks the prediction location $\bm{s}_0$.}
    \label{fig:spatial_coefficients_spherical}
\end{figure}

\section{Gradient Descent Algorithm}\label{app:gradient_descent}

We solve the penalized kriging problem \eqref{eq:proposed_lasso_problem_mod} using proximal gradient descent. The constraint $\bm{X} \bm{\lambda} = \bm{x}_{0}$ allows us to express $\bm{\lambda}_{p} = \bm{X}_{p}^{-1}(\bm{x}_{0} - \bm{X}_{-p}\bm{\lambda}_{-p})$, reducing the problem to an unconstrained optimization over $\bm{\lambda}_{-p} \in \mathbb{R}^{N-p}$.

Let $Q(\bm{\lambda}_{-p}) = \bm{\lambda}^{\top} \Sigma \bm{\lambda} - 2\bm{\lambda}^{\top} \bm{c}_{0}$. Partitioning $\Sigma = [\Sigma_{p,p}, \Sigma_{p,-p}; \Sigma_{-p,p}, \Sigma_{-p,-p}]$ and $\bm{c}_0 = [\bm{c}_{0p}, \bm{c}_{0-p}]^{\top}$ conformably, the gradient is:
\begin{align*}
\nabla Q(\bm{\lambda}_{-p}) = &-2\bm{X}_{-p}^{\top}(\bm{X}_{p}^{-\top}\Sigma_{p,p} \bm{X}_{p}^{-1})(\bm{x}_{0} - \bm{X}_{-p}\bm{\lambda}_{-p}) \\
&- 2\bm{X}_{-p}^{\top}\bm{X}_{p}^{-\top}\Sigma_{p,-p}\bm{\lambda}_{-p} + 2\Sigma_{-p,p}\bm{X}_{p}^{-1}(\bm{x}_{0} - \bm{X}_{-p}\bm{\lambda}_{-p}) \\
&+ 2\Sigma_{-p,-p}\bm{\lambda}_{-p} + 2\bm{X}_{-p}^{\top}\bm{X}_{p}^{-\top}\bm{c}_{0p} - 2\bm{c}_{0-p}
\end{align*}

The algorithm alternates between gradient descent and soft-thresholding:

\begin{algorithm}[H]
\caption{Gradient Descent for Penalized Kriging}\label{alg:GD}
\begin{algorithmic}[1]
\Require $\Sigma$, $\bm{c}_{0}$, $\eta \geq 0$, $\bm{\lambda}_{-p}^{(0)}$, $\epsilon > 0$, $\alpha > 0$
\State $k \gets 0$
\While{$\| \bm{\lambda}_{-p}^{(k+1)} - \bm{\lambda}_{-p}^{(k)} \|_2 > \epsilon$ and $k < 1000$}
    \State $\bm{g} \gets \nabla Q(\bm{\lambda}_{-p}^{(k)})$
    \State $\bm{\lambda}_{-p}^{(k+1)} \gets \bm{\lambda}_{-p}^{(k)} - \alpha \bm{g}$
    \State $\bm{\lambda}_{-p}^{(k+1)} \gets \text{sign}(\bm{\lambda}_{-p}^{(k+1)}) \odot (|\bm{\lambda}_{-p}^{(k+1)}| - \eta\alpha)_+$
    \State $k \gets k + 1$
\EndWhile
\State \Return $\widehat{\bm{\lambda}} = [\bm{X}_{p}^{-1}(\bm{x}_{0} - \bm{X}_{-p}\bm{\lambda}_{-p}^{(k)}), \bm{\lambda}_{-p}^{(k)}]^{\top}$
\end{algorithmic}
\end{algorithm}

Here $(a)_+ = \max(a, 0)$ and $\odot$ denotes element-wise multiplication. For adaptive LASSO, replace $\eta\alpha$ with $\eta\alpha w_i$ for each component $i$, where $w_i = 1/|\tilde{\lambda}_i|$ are the adaptive weights. We use $\alpha = 0.01$ and warm start from ordinary kriging weights when available.

\end{document}